\documentclass[a4paper,11pt]{article}
\pdfoutput=1
\usepackage{jheppub}
\usepackage[utf8]{inputenc}

\usepackage[vcentermath]{youngtab}
\usepackage{pdflscape}
\usepackage{tikz}
\usepackage{ marvosym }
\usetikzlibrary{positioning}
\usetikzlibrary{arrows}
\usetikzlibrary{arrows.meta}
\usetikzlibrary{decorations.pathreplacing}
\usepackage{subcaption}
\usepackage{xcolor,colortbl}
\usepackage{mathtools}
\usepackage{ragged2e}
\newlength\ubwidth
\newcommand\parunderbrace[2]{\settowidth\ubwidth{$#1$}\underbrace{#1}_{\parbox{\ubwidth}{\scriptsize\RaggedRight#2}}}
\usepackage{booktabs}

\usepackage{afterpage}

\usepackage{paralist}
\tikzset{Reddotted/.style={dashed,color=red}}
\usetikzlibrary{arrows,fit,decorations.pathreplacing, shapes.misc,decorations.pathmorphing}
\usepackage[noabbrev,capitalise]{cleveref}
\usepackage{xcolor}
\usepackage{soul}
\usepackage{colortbl}
\definecolor{Gray}{gray}{0.9}
\usepackage{multirow}
\usepackage[colorinlistoftodos]{todonotes}
\usepackage{amsthm}
\usepackage{amsmath}
\usepackage{amsthm}
\usepackage{changepage}
\usepackage{tabularx}
\usepackage{tikz}
\usepackage{multirow}
\usepackage{longtable}
\usepackage{multirow}
\usepackage{amsmath}
\usepackage{amssymb}
\usepackage{mathtools}
\usepackage{adjustbox}
\usetikzlibrary{shapes,arrows}
\usepackage[english]{babel}
\newcolumntype{P}[1]{>{\centering\arraybackslash}p{#1}}
\usepackage{tabularx}

\usepackage{color}
\usepackage{subcaption}
\usepackage{setspace}
\usepackage{pdflscape}

\usepackage[all]{xy}
\newcommand{\slice}[1]{\mathcal{S}_{\mathcal N,#1} }
\newcommand{\orbit}[1]{\overline{\mathcal{O}}_{#1} }
\newcommand{\ghs}[1]{g_{HS}^{#1}}
\newcommand{\bv}[1]{d_{BV}(#1)}
\newcommand{\pal}{\ldots \text{palindrome}\ldots}
\newcommand{\RK}[1]{\textcolor{red}{[RK: #1]}}

\usetikzlibrary{shapes.misc}
\tikzset{gauge1/.style={draw=none,minimum size=0.6cm,fill=white,circle, draw}}
\tikzset{gauge3/.style={draw=none,minimum size=0.35cm,fill=white,circle, draw}}
\tikzset{gauge5/.style={draw=none,minimum size=0.35cm,fill=white,circle, draw}}
\tikzset{oranger/.style={draw=orange,minimum size=0.35cm,fill=none,circle, draw}}
\tikzset{crosses/.style={draw,circle,cross out}}
\tikzset{orangecrosses/.style={draw=orange,circle,cross out}}
\tikzset{blank/.style={draw=none,minimum size=0.4cm,fill=none,circle, draw}}
\tikzset{flavour2/.style={draw=none,minimum size=0.4cm,fill=white,regular polygon sides=4,draw}}
\tikzset{flavour2/.style={draw=none,minimum size=0.4cm,fill=white,regular polygon sides=4,draw}}
\tikzset{flavourBlue/.style={draw=none,minimum size=0.4cm,fill=blue,regular polygon sides=4,draw}}
\tikzset{flavourRed/.style={draw=none,minimum size=0.4cm,fill=red,regular polygon sides=4,draw}}
\tikzset{none/.style={draw=none}}
\tikzset{flavourBlue/.style={draw=none,minimum size=0.4cm,fill=blue,regular polygon sides=4,draw}}
\tikzset{flavourRed/.style={draw=none,minimum size=0.4cm,fill=red,regular polygon sides=4,draw}}
\tikzset{none/.style={draw=none}}
\tikzset{redgauge/.style={draw=none,minimum size=0.4cm,fill=red,circle, draw}}
\tikzset{miniU/.style={draw=none,minimum size=0.1cm,fill=red,circle, draw}}
\tikzset{miniBlue/.style={draw=none,minimum size=0.1cm,fill=blue,circle, draw}}
\tikzset{gauge2/.style={draw=none,minimum size=0.35mm,fill=red,circle, draw}}
\tikzset{bluegauge/.style={draw=none,minimum size=0.4cm,fill=blue,circle, draw}}
\tikzset{o5/.style={green, line width=0.05cm}}
\tikzset{new edge style 0/.style={magenta}}
\tikzset{o3tildeplus/.style={dashed}}
\tikzset{o3plus/.style={dotted}}
\tikzset{new edge style 1/.style={dashed}}
\tikzset{brace/.style={decorate,decoration={brace,amplitude=10pt}}}
\tikzset{brace1/.style={decorate,decoration={brace,amplitude=10pt}}}
\pgfdeclarelayer{edgelayer}
\pgfdeclarelayer{nodelayer}
\usetikzlibrary{decorations.pathreplacing}
\pgfsetlayers{edgelayer,nodelayer,main}

\tikzset{wd/.style={circle, draw,inner sep=3.5pt}}
\tikzset{bd/.style={circle, draw,inner sep=3.5pt, fill=black}}

\tikzset{gaw/.style={inner sep=1mm,draw=none,fill=white,minimum size=2mm,circle, draw}}
\tikzset{gar/.style={inner sep=1mm,draw=none,fill=red,minimum size=2mm,circle, draw}}
\tikzset{gab/.style={inner sep=1mm,draw=none,fill=blue,minimum size=2mm,circle, draw}}
\tikzset{flaw/.style={draw=none,minimum size=0.2mm,fill=white, regular polygon,regular polygon sides=4,draw}}
\tikzset{flar/.style={draw=none,minimum size=0.2mm,fill=red, regular polygon,regular polygon sides=4,draw}}
\tikzset{flab/.style={draw=none,minimum size=0.2mm,fill=blue, regular polygon,regular polygon sides=4,draw}}
\newcommand{\NS}{\scalebox{0.3}{\begin{tikzpicture}
    \draw[very thick] (0,0) circle (0.5cm);
    \draw[very thick] (-0.3,0.3)--(0.3,-0.3) (-0.3,-0.3)--(0.3,0.3);
\end{tikzpicture}}}
\newcommand{\ONp}{\scalebox{0.3}{\begin{tikzpicture}
    \draw[orange,very thick] (0,0) circle (0.5cm);
    \draw[orange,very thick] (-0.3,0.3)--(0.3,-0.3) (-0.3,-0.3)--(0.3,0.3);
\end{tikzpicture}}}

\allowdisplaybreaks

\preprint{Imperial/TP/21/AH/03}

\title{Folding Orthosymplectic Quivers}

\author[a]{Antoine Bourget,}
\author[a]{Julius F.\ Grimminger,}
\author[a] {Amihay Hanany,}
\author[a]{Rudolph Kalveks,}
\author[b]{Marcus Sperling,}
\author[a]{and Zhenghao Zhong}
\affiliation[a]{Theoretical Physics Group, The Blackett Laboratory, Imperial College London,\\ Prince Consort Road
London, SW7 2AZ, UK}
\affiliation[b]{Yau Mathematical Sciences Center, Tsinghua University,\\ Haidian District, Beijing, 100084, China}
\emailAdd{a.bourget@imperial.ac.uk}
\emailAdd{julius.grimminger17@imperial.ac.uk}
\newcommand{\JG}[1]{\textcolor{magenta}{[XXX JG: #1]}}
\emailAdd{a.hanany@imperial.ac.uk}
\emailAdd{rudolph.kalveks09@imperial.ac.uk}
\emailAdd{marcus.sperling@univie.ac.at}
\newcommand{\MS}[1]{\textcolor{orange}{[XXX MS: #1]}}
\emailAdd{zhenghao.zhong14@imperial.ac.uk}
\newcommand{\ZZ}[1]{\textcolor{cyan}{[XXX ZZ: #1]}}

\abstract{Folding identical legs of a simply-laced quiver creates a quiver with a non-simply laced edge. So far, this has been explored for quivers containing unitary gauge groups. In this paper, orthosymplectic quivers are folded, giving rise to a new family of quivers. This is realised by intersecting orientifolds in the brane system. The monopole formula for these non-simply laced orthosymplectic quivers is introduced. Some of the folded quivers have Coulomb branches that are closures of minimal nilpotent orbits of exceptional algebras, thus providing a new construction of these fundamental moduli spaces.
Moreover, a general family of folded orthosymplectic quivers is shown to be a new magnetic quiver realisation of Higgs branches of 4d $\mathcal{N}=2$ theories. The Hasse (phase) diagrams of certain families are derived via quiver subtraction as well as Kraft-Procesi transitions in the brane system.}

\begin{document} 
\maketitle

\section{Introduction}

Amongst $3$d $\mathcal{N}=4$ quiver gauge theories, a natural set of theories to consider is that of affine Dynkin quivers with unitary gauge nodes. The Coulomb branches of affine ADE Dynkin quivers are moduli spaces of instantons which can be identified with the closures of minimal nilpotent orbits of the $A_n$, $D_n$, and $E_{6,7,8}$ algebras.
These are examples of \emph{simply-laced quivers}. The Coulomb branches of affine Dynkin quivers of BCFG-type, explored in \cite{Cremonesi:2014xha}, are the closures of minimal nilpotent orbits of the $B_n$, $C_n$, $F_4$, and $G_2$ algebras. The corresponding quivers are termed \emph{non-simply laced} quivers. 

The concept of \emph{folding}, in the sense that identical legs of simply-laced quivers are folded into a quiver with a non-simply laced edge, has been studied recently in the context of the Coulomb branch Hilbert series of 3d $\mathcal{N}=4$ quivers \cite{Hanany:2012dm,Dey:2016qqp,Nakajima:2019olw, Bourget:2020bxh}. In particular, the non-simply laced edge within the quiver implies that the theory has no obvious path integral formulation. 
Nonetheless, the Coulomb branch Hilbert series can be readily computed via the \emph{monopole formula} \cite{Cremonesi:2013lqa,Cremonesi:2014xha}. This allows one to study non-simply laced quivers that are more general than affine Dynkin type quivers.  One purpose of this paper is to demonstrate that these quivers can provide new magnetic quiver constructions of known moduli spaces and in many cases lead to new interesting moduli spaces. Other approaches to folding include \cite{Cecotti:2012gh,Kimura:2017hez,Haouzi:2017vec,Chen:2018ntf}.  

So far, works on non-simply laced quivers have focused solely on quivers with unitary gauge groups. In light of the recent understanding of orthosymplectic quivers \cite{Cremonesi:2014uva,Cremonesi:2014vla,Cabrera:2017ucb,Cabrera:2019dob,Bourget:2020gzi,Akhond:2020vhc,Akhond:2021knl}, and the applicability of the monopole formula to framed and unframed orthosymplectic quivers \cite{Bourget:2020xdz}, one is finally able to extend this program to non-simply laced orthosymplectic quivers. 

Amongst framed/flavoured orthosymplectic quivers, one natural set to fold is that of magnetic quivers for nilpotent orbit closures of $\mathfrak{so}(2n)$. To be more precise, framed orthosymplectic quivers whose Coulomb branches are height 2 nilpotent orbit closures of $\mathfrak{so}(2n)$ carry a natural $\mathbb{Z}_2$ symmetry that allows us to fold the identical legs. The Coulomb branches of the resulting non-simply laced framed orthosymplectic quivers turn out to be height 2  nilpotent orbit closures of $\mathfrak{sl}(n)$. Furthermore, for framed non-simply laced orthosymplectic quivers there exist corresponding brane configurations with D3-D5-NS5 branes in the presence of O3, O5 and ON orientifold planes.

\afterpage{
\begin{landscape}
\begin{table}[]
    \centering
    \scalebox{0.9}{
    \begin{tabular}{cc|cc}
    \multicolumn{2}{c}{\Large{Before folding}} &     \multicolumn{2}{c}{\Large{After folding}} \\ 
    \multicolumn{2}{c}{ } &     \multicolumn{2}{c}{ } \\  \toprule 
        Orthosymplectic quiver & Coulomb branch & Orthosymplectic quiver & Coulomb branch \\ \midrule 
         \raisebox{-.5\height}{\scalebox{.811}{\begin{tikzpicture}
	\begin{pgfonlayer}{nodelayer}
		\node [style=miniU] (0) at (-4.5, 1) {};
		\node [style=miniBlue] (1) at (-3.5, 1) {};
		\node [style=miniBlue] (2) at (-5.5, 1) {};
		\node [style=miniBlue] (26) at (-4.5, 1) {};
		\node [style=miniU] (32) at (-5.5, 1) {};
		\node [style=miniU] (33) at (-3.5, 1) {};
		\node [style=miniU] (38) at (-4.5, 1) {};
		\node [style=miniBlue] (39) at (-3.5, 1) {};
		\node [style=miniBlue] (40) at (-5.5, 1) {};
		\node [style=miniU] (47) at (2.5, 1) {};
		\node [style=miniBlue] (48) at (3.5, 1) {};
		\node [style=miniU] (53) at (-5.5, 1) {};
		\node [style=miniU] (54) at (3.5, 1) {};
		\node [style=miniBlue] (55) at (-4.5, 1) {};
		\node [style=miniBlue] (56) at (2.5, 1) {};
		\node [style=miniU] (57) at (-3.5, 1) {};
		\node [style=miniU] (58) at (1.5, 1) {};
		\node [style=flavourRed] (59) at (-4.5, 2) {};
		\node [style=flavourRed] (60) at (2.5, 2) {};
		\node [style=none] (61) at (2.5, 2.5) {$1$};
		\node [style=none] (62) at (-4.5, 2.5) {$1$};
		\node [style=none] (63) at (-2.75, 1) {\dots};
		\node [style=none] (64) at (-3, 1) {};
		\node [style=none] (65) at (-2.5, 1) {};
		\node [style=none] (66) at (-5.5, 0.5) {$2$};
		\node [style=none] (67) at (-4.5, 0.5) {$2$};
		\node [style=none] (68) at (2.5, 0.5) {$2$};
		\node [style=none] (69) at (3.5, 0.5) {$2$};
		\node [style=none] (70) at (1.5, 0.5) {$3$};
		\node [style=none] (71) at (-3.5, 0.5) {$3$};
		\node [style=none] (72) at (-5.5, 0.25) {};
		\node [style=none] (73) at (3.5, 0.25) {};
		\node [style=none] (74) at (-1, -0.4) {$4n-3$ nodes};
		\node [style=miniBlue] (78) at (-2, 1) {};
		\node [style=none] (79) at (0.75, 1) {\dots};
		\node [style=none] (80) at (1, 1) {};
		\node [style=none] (81) at (0.5, 1) {};
		\node [style=none] (82) at (-2, 0.5) {$2$};
		\node [style=miniBlue] (83) at (0, 1) {};
		\node [style=miniU] (84) at (-1, 1) {};
		\node [style=none] (85) at (-1, 0.5) {$3$};
		\node [style=none] (86) at (0, 0.5) {$2$};
	\end{pgfonlayer}
	\begin{pgfonlayer}{edgelayer}
		\draw (2) to (0);
		\draw (0) to (1);
		\draw (48) to (47);
		\draw (55) to (59);
		\draw (60) to (56);
		\draw (57) to (64.center);
		\draw (58) to (56);
		\draw [style=brace1] (73.center) to (72.center);
		\draw (78) to (65.center);
		\draw (83) to (81.center);
		\draw (58) to (80.center);
		\draw (84) to (78);
		\draw (84) to (83);
	\end{pgfonlayer}
\end{tikzpicture}}
}
&  $\overline{\mathcal{O}}^{\mathfrak{so}(4n)}_{\text{min}} = d_{2n}$   
& \scalebox{0.8}{
\begin{tabular}{cc}
\raisebox{-.5\height}{
    \begin{tikzpicture}
	\begin{pgfonlayer}{nodelayer}
		\node [style=miniU] (0) at (0, 0) {};
		\node [style=flavourBlue] (1) at (-1, 0) {};
		\node [style=none] (2) at (0, -0.5) {2};
		\node [style=none] (3) at (-1, -0.5) {2};
		\node at (0,.5) {};
	\end{pgfonlayer}
	\begin{pgfonlayer}{edgelayer}
		\draw (0) to (1);
	\end{pgfonlayer}
\end{tikzpicture}
} & for $n=1$
\\
 \raisebox{-.5\height}{ \begin{tikzpicture}
	\begin{pgfonlayer}{nodelayer}
		\node [style=miniBlue] (1) at (1, -1.25) {};
		\node [style=none] (18) at (1, -1.75) {$2$};
		\node [style=miniU] (27) at (0, -1.25) {};
		\node [style=none] (28) at (0, -1.15) {};
		\node [style=none] (29) at (1, -1.15) {};
		\node [style=none] (30) at (0, -1.35) {};
		\node [style=none] (31) at (1, -1.35) {};
		\node [style=none] (32) at (0.175, -0.825) {};
		\node [style=none] (33) at (0.675, -1.25) {};
		\node [style=none] (34) at (0.175, -1.675) {};
		\node [style=none] (36) at (0, -1.75) {$3$};
		\node [style=miniBlue] (37) at (1, -1.25) {};
		\node [style=miniU] (46) at (2, -1.25) {};
		\node [style=none] (47) at (2, -1.75) {$3$};
		\node [style=none] (48) at (2.75, -1.25) {\dots};
		\node [style=none] (49) at (2.5, -1.25) {};
		\node [style=none] (50) at (3, -1.25) {};
		\node [style=miniBlue] (51) at (3.5, -1.25) {};
		\node [style=miniU] (52) at (4.5, -1.25) {};
		\node [style=none] (53) at (3.5, -1.75) {$2$};
		\node [style=none] (54) at (4.5, -1.75) {$3$};
		\node [style=miniBlue] (55) at (5.5, -1.25) {};
		\node [style=none] (56) at (5.5, -1.75) {$2$};
		\node [style=flavourRed] (57) at (5.5, -0.25) {};
		\node [style=none] (58) at (5.5, 0.25) {$1$};
		\node [style=miniU] (59) at (6.5, -1.25) {};
		\node [style=none] (60) at (6.5, -1.75) {$2$};
		\node [style=none] (61) at (0, -2) {};
		\node [style=none] (62) at (6.5, -2) {};
		\node [style=none] (63) at (3.2, -2.75) {$2n-1$ nodes};
	\end{pgfonlayer}
	\begin{pgfonlayer}{edgelayer}
		\draw (28.center) to (29.center);
		\draw (30.center) to (31.center);
		\draw (32.center) to (33.center);
		\draw (33.center) to (34.center);
		\draw (46) to (37);
		\draw (49.center) to (46);
		\draw (51) to (50.center);
		\draw (51) to (52);
		\draw (55) to (52);
		\draw (57) to (55);
		\draw (55) to (59);
		\draw [style=brace1] (62.center) to (61.center);
	\end{pgfonlayer}
\end{tikzpicture}} & for $n>1$
\end{tabular}
}
&  $\overline{\mathcal{O}}^{\mathfrak{sl}(2n)}_{\text{min}} = a_{2n-1}$ \\ \midrule 
         \scalebox{0.8}{  
         \raisebox{-.5\height}{\begin{tikzpicture}
	\begin{pgfonlayer}{nodelayer}
		\node [style=miniU] (0) at (-4.5, 1) {};
		\node [style=miniBlue] (1) at (-3.5, 1) {};
		\node [style=miniBlue] (2) at (-5.5, 1) {};
		\node [style=miniBlue] (26) at (-4.5, 1) {};
		\node [style=miniU] (32) at (-5.5, 1) {};
		\node [style=miniU] (33) at (-3.5, 1) {};
		\node [style=miniU] (38) at (-4.5, 1) {};
		\node [style=miniBlue] (39) at (-3.5, 1) {};
		\node [style=miniBlue] (40) at (-5.5, 1) {};
		\node [style=miniU] (47) at (2.5, 1) {};
		\node [style=miniBlue] (48) at (3.5, 1) {};
		\node [style=miniU] (53) at (-5.5, 1) {};
		\node [style=miniU] (54) at (3.5, 1) {};
		\node [style=miniBlue] (55) at (-4.5, 1) {};
		\node [style=miniBlue] (56) at (2.5, 1) {};
		\node [style=miniU] (57) at (-3.5, 1) {};
		\node [style=miniU] (58) at (1.5, 1) {};
		\node [style=flavourRed] (59) at (-4.5, 2) {};
		\node [style=flavourRed] (60) at (2.5, 2) {};
		\node [style=none] (61) at (2.5, 2.5) {$1$};
		\node [style=none] (62) at (-4.5, 2.5) {$1$};
		\node [style=none] (63) at (-2.75, 1) {\dots};
		\node [style=none] (64) at (-3, 1) {};
		\node [style=none] (65) at (-2.5, 1) {};
		\node [style=none] (66) at (-5.5, 0.5) {$2$};
		\node [style=none] (67) at (-4.5, 0.5) {$2$};
		\node [style=none] (68) at (2.5, 0.5) {$2$};
		\node [style=none] (69) at (3.5, 0.5) {$2$};
		\node [style=none] (70) at (1.5, 0.5) {$3$};
		\node [style=none] (71) at (-3.5, 0.5) {$3$};
		\node [style=none] (72) at (-5.5, 0.25) {};
		\node [style=none] (73) at (3.5, 0.25) {};
		\node [style=none] (74) at (-1, -0.4) {$4n-1$ nodes};
		\node [style=miniBlue] (78) at (-2, 1) {};
		\node [style=none] (79) at (0.75, 1) {\dots};
		\node [style=none] (80) at (1, 1) {};
		\node [style=none] (81) at (0.5, 1) {};
		\node [style=none] (82) at (-2, 0.5) {$3$};
		\node [style=miniBlue] (83) at (0, 1) {};
		\node [style=miniU] (84) at (-1, 1) {};
		\node [style=none] (85) at (-1, 0.5) {$2$};
		\node [style=none] (86) at (0, 0.5) {$3$};
		\node [style=miniU] (87) at (-2, 1) {};
		\node [style=miniU] (88) at (0, 1) {};
		\node [style=miniBlue] (89) at (-1, 1) {};
	\end{pgfonlayer}
	\begin{pgfonlayer}{edgelayer}
		\draw (2) to (0);
		\draw (0) to (1);
		\draw (48) to (47);
		\draw (55) to (59);
		\draw (60) to (56);
		\draw (57) to (64.center);
		\draw (58) to (56);
		\draw [style=brace1] (73.center) to (72.center);
		\draw (78) to (65.center);
		\draw (83) to (81.center);
		\draw (58) to (80.center);
		\draw (84) to (78);
		\draw (84) to (83);
	\end{pgfonlayer}
\end{tikzpicture}
} 
} &   $\overline{\mathcal{O}}^{\mathfrak{so}(4n+2)}_{\text{min}}= d_{2n+1}$  & \scalebox{0.8}{
 
\begin{tabular}{cc}
      \raisebox{-.5\height}{
    \begin{tikzpicture}
	\begin{pgfonlayer}{nodelayer}
		\node [style=miniBlue] (1) at (1, -1.25) {};
		\node [style=none] (18) at (1, -1.75) {2};
		\node [style=miniU] (27) at (0, -1.25) {};
		\node [style=none] (28) at (0, -1.15) {};
		\node [style=none] (29) at (1, -1.15) {};
		\node [style=none] (30) at (0, -1.35) {};
		\node [style=none] (31) at (1, -1.35) {};
		\node [style=none] (32) at (0.175, -0.825) {};
		\node [style=none] (33) at (0.675, -1.25) {};
		\node [style=none] (34) at (0.175, -1.675) {};
		\node [style=none] (36) at (0, -1.75) {2};
		\node [style=miniBlue] (37) at (1, -1.25) {};
		\node [style=miniBlue] (38) at (0, -1.25) {};
		\node [style=miniU] (39) at (1, -1.25) {};
		\node [style=none] (42) at (0, -2) {};
		\node [style=miniU] (45) at (0, -1.25) {};
		\node [style=miniBlue] (46) at (1, -1.25) {};
		\node [style=miniBlue] (47) at (0, -1.25) {};
		\node [style=miniU] (48) at (1, -1.25) {};
		\node [style=flavourRed] (49) at (-1, -1.25) {};
		\node [style=none] (50) at (-1, -1.75) {2};
	\end{pgfonlayer}
	\begin{pgfonlayer}{edgelayer}
		\draw (28.center) to (29.center);
		\draw (30.center) to (31.center);
		\draw (32.center) to (33.center);
		\draw (33.center) to (34.center);
		\draw (49) to (47);
	\end{pgfonlayer}
\end{tikzpicture}
} & for $n=1$ 
\\
   \raisebox{-.5\height}{ \begin{tikzpicture}
	\begin{pgfonlayer}{nodelayer}
		\node [style=miniBlue] (1) at (1, -1.25) {};
		\node [style=none] (18) at (1, -1.75) {$3$};
		\node [style=miniU] (27) at (0, -1.25) {};
		\node [style=none] (28) at (0, -1.15) {};
		\node [style=none] (29) at (1, -1.15) {};
		\node [style=none] (30) at (0, -1.35) {};
		\node [style=none] (31) at (1, -1.35) {};
		\node [style=none] (32) at (0.175, -0.825) {};
		\node [style=none] (33) at (0.675, -1.25) {};
		\node [style=none] (34) at (0.175, -1.675) {};
		\node [style=none] (36) at (0, -1.75) {$2$};
		\node [style=miniBlue] (37) at (1, -1.25) {};
		\node [style=miniU] (46) at (2, -1.25) {};
		\node [style=none] (47) at (2, -1.75) {$2$};
		\node [style=none] (48) at (2.75, -1.25) {\dots};
		\node [style=none] (49) at (2.5, -1.25) {};
		\node [style=none] (50) at (3, -1.25) {};
		\node [style=miniBlue] (51) at (3.5, -1.25) {};
		\node [style=miniU] (52) at (4.5, -1.25) {};
		\node [style=none] (53) at (3.5, -1.75) {$2$};
		\node [style=none] (54) at (4.5, -1.75) {$3$};
		\node [style=miniBlue] (55) at (5.5, -1.25) {};
		\node [style=none] (56) at (5.5, -1.75) {$2$};
		\node [style=flavourRed] (57) at (5.5, -0.25) {};
		\node [style=none] (58) at (5.5, 0.25) {$1$};
		\node [style=miniU] (59) at (6.5, -1.25) {};
		\node [style=none] (60) at (6.5, -1.75) {$2$};
		\node [style=none] (61) at (0, -2) {};
		\node [style=none] (62) at (6.5, -2) {};
		\node [style=none] (63) at (3.2, -2.55) {$2n$ nodes};
		\node [style=miniBlue] (64) at (0, -1.25) {};
		\node [style=miniBlue] (65) at (2, -1.25) {};
		\node [style=miniU] (66) at (1, -1.25) {};
	\end{pgfonlayer}
	\begin{pgfonlayer}{edgelayer}
		\draw (28.center) to (29.center);
		\draw (30.center) to (31.center);
		\draw (32.center) to (33.center);
		\draw (33.center) to (34.center);
		\draw (46) to (37);
		\draw (49.center) to (46);
		\draw (51) to (50.center);
		\draw (51) to (52);
		\draw (55) to (52);
		\draw (57) to (55);
		\draw (55) to (59);
		\draw [style=brace] (62.center) to (61.center);
	\end{pgfonlayer}
\end{tikzpicture}  } & for $n>1$ 
\end{tabular}
}
&   $\overline{\mathcal{O}}^{\mathfrak{sl}(2n+1)}_{\text{min}}= a_{2n}$  \\
         \bottomrule
    \end{tabular}}
    \caption{The orthosymplectic quivers on the left have Coulomb branches that are closures of nilpotent orbits of $D_n$ algebras. Red nodes with an index $k$ denote $\mathrm{SO}(k)$ groups while blue nodes with index $2k$ denote $\mathrm{USp}(2k)$ groups. Folding these quivers along the identical legs gives the non-simply laced orthosymplectic quivers on the right. The Coulomb branches of these theories are given as well. Note that the central node for the quivers on the left are orthogonal on the first row and symplectic on the second row.   }
    \label{Dntable}
\end{table}

\begin{table}[t]
    \centering
    \hspace*{-.5cm}\scalebox{0.9}{
    \begin{tabular}{cc|cc}
    \multicolumn{2}{c}{\Large{Before folding}} &     \multicolumn{2}{c}{\Large{After folding}} \\ 
    \multicolumn{2}{c}{ } &     \multicolumn{2}{c}{ } \\  \toprule 
        Orthosymplectic quiver & Coulomb branch & Orthosymplectic quiver & Coulomb branch \\ \midrule 
         \scalebox{0.8}{  
         \raisebox{-.5\height}{\begin{tikzpicture}
	\begin{pgfonlayer}{nodelayer}
		\node [style=bluegauge] (0) at (0, 0) {};
		\node [style=redgauge] (1) at (1.25, 0) {};
		\node [style=bluegauge] (2) at (2.5, 0) {};
		\node [style=redgauge] (3) at (3.75, 0) {};
		\node [style=redgauge] (4) at (-1.25, 0) {};
		\node [style=redgauge] (5) at (-3.75, 0) {};
		\node [style=bluegauge] (6) at (-2.5, 0) {};
		\node [style=none] (8) at (0, -0.5) {4};
		\node [style=none] (9) at (1.25, -0.5) {6};
		\node [style=none] (10) at (2.5, -0.5) {6};
		\node [style=none] (11) at (3.75, -0.5) {8};
		\node [style=none] (12) at (-1.25, -0.5) {4};
		\node [style=none] (13) at (-2.5, -0.5) {2};
		\node [style=none] (14) at (-3.75, -0.5) {2};
		\node [style=bluegauge] (23) at (5, 0) {};
		\node [style=none] (25) at (5, -0.5) {6};
		\node [style=bluegauge] (27) at (10, 0) {};
		\node [style=redgauge] (28) at (11.25, 0) {};
		\node [style=redgauge] (30) at (8.75, 0) {};
		\node [style=redgauge] (31) at (6.25, 0) {};
		\node [style=bluegauge] (32) at (7.5, 0) {};
		\node [style=none] (33) at (10, -0.5) {2};
		\node [style=none] (34) at (11.25, -0.5) {2};
		\node [style=none] (36) at (8.75, -0.5) {4};
		\node [style=none] (37) at (7.5, -0.5) {4};
		\node [style=none] (38) at (6.25, -0.5) {6};
		\node [style=bluegauge] (39) at (3.75, 1.25) {};
		\node [style=none] (40) at (3.75, 1.75) {2};
	\end{pgfonlayer}
	\begin{pgfonlayer}{edgelayer}
		\draw (5) to (6);
		\draw (6) to (4);
		\draw (4) to (0);
		\draw (0) to (1);
		\draw (1) to (2);
		\draw (2) to (3);
		\draw (3) to (23);
		\draw (31) to (32);
		\draw (32) to (30);
		\draw (30) to (27);
		\draw (27) to (28);
		\draw (23) to (31);
		\draw (3) to (39);
	\end{pgfonlayer}
\end{tikzpicture}} 
} &  $\overline{\mathcal{O}}^{\mathfrak{e}_8}_{\text{min}} = e_8$ & \scalebox{0.8}{
\raisebox{-.5\height}{
\begin{tikzpicture}
	\begin{pgfonlayer}{nodelayer}
		\node [style=miniBlue] (1) at (1, -1.25) {};
		\node [style=miniU] (27) at (0, -1.25) {};
		\node [style=none] (28) at (0, -1.15) {};
		\node [style=none] (29) at (1, -1.15) {};
		\node [style=none] (30) at (0, -1.35) {};
		\node [style=none] (31) at (1, -1.35) {};
		\node [style=none] (32) at (0.175, -0.825) {};
		\node [style=none] (33) at (0.675, -1.25) {};
		\node [style=none] (34) at (0.175, -1.675) {};
		\node [style=miniBlue] (37) at (1, -1.25) {};
		\node [style=miniU] (46) at (2, -1.25) {};
		\node [style=miniBlue] (47) at (-1, -1.25) {};
		\node [style=none] (48) at (-1, -2+0.25) {2};
		\node [style=none] (49) at (0, -2+0.25) {8};
		\node [style=none] (50) at (1, -2+0.25) {6};
		\node [style=none] (51) at (2, -2+0.25) {6};
		\node [style=miniU] (52) at (4, -1.25) {};
		\node [style=miniU] (53) at (6, -1.25) {};
		\node [style=miniBlue] (54) at (5, -1.25) {};
		\node [style=miniBlue] (55) at (3, -1.25) {};
		\node [style=none] (56) at (3, -2+0.25) {4};
		\node [style=none] (57) at (4, -2+0.25) {4};
		\node [style=none] (58) at (5, -2+0.25) {2};
		\node [style=none] (59) at (6, -2+0.25) {2};
	\end{pgfonlayer}
	\begin{pgfonlayer}{edgelayer}
		\draw (28.center) to (29.center);
		\draw (30.center) to (31.center);
		\draw (32.center) to (33.center);
		\draw (33.center) to (34.center);
		\draw (46) to (37);
		\draw (47) to (27);
		\draw (46) to (55);
		\draw (55) to (52);
		\draw (52) to (54);
		\draw (54) to (53);
	\end{pgfonlayer}
\end{tikzpicture}
}}
&  $\overline{\mathcal{O}}^{\mathfrak{e}_7}_{\text{min}} = e_7$ \\
\scalebox{0.8}{
\raisebox{-.5\height}{
\begin{tikzpicture}
	\begin{pgfonlayer}{nodelayer}
		\node [style=bluegauge] (0) at (0, 0) {};
		\node [style=redgauge] (1) at (1.25, 0) {};
		\node [style=bluegauge] (2) at (2.5, 0) {};
		\node [style=redgauge] (3) at (3.75, 0) {};
		\node [style=redgauge] (4) at (-1.25, 0) {};
		\node [style=redgauge] (5) at (-3.75, 0) {};
		\node [style=bluegauge] (6) at (-2.5, 0) {};
		\node [style=none] (8) at (0, -0.5) {4};
		\node [style=none] (9) at (1.25, -0.5) {6};
		\node [style=none] (10) at (2.5, -0.5) {4};
		\node [style=none] (11) at (3.75, -0.5) {4};
		\node [style=none] (12) at (-1.25, -0.5) {4};
		\node [style=none] (13) at (-2.5, -0.5) {2};
		\node [style=none] (14) at (-3.75, -0.5) {2};
		\node [style=bluegauge] (23) at (5, 0) {};
		\node [style=redgauge] (24) at (6.25, 0) {};
		\node [style=none] (25) at (5, -0.5) {2};
		\node [style=none] (26) at (6.25, -0.5) {2};
		\node [style=bluegauge] (27) at (1.25, 1.25) {};
		\node [style=gauge3] (28) at (1.25, 2.5) {};
		\node [style=none] (29) at (2.075-0.25, 1.25) {2};
		\node [style=none] (30) at (2.075-0.25, 2.5) {1};
	\end{pgfonlayer}
	\begin{pgfonlayer}{edgelayer}
		\draw (5) to (6);
		\draw (6) to (4);
		\draw (4) to (0);
		\draw (0) to (1);
		\draw (1) to (2);
		\draw (2) to (3);
		\draw (3) to (23);
		\draw (23) to (24);
		\draw (28) to (27);
		\draw (27) to (1);
	\end{pgfonlayer}
\end{tikzpicture}
}}
& $\overline{\mathcal{O}}^{\mathfrak{e}_7}_{\text{min}} = e_7$  & \scalebox{0.8}{
\raisebox{-.5\height}{
\begin{tikzpicture}
	\begin{pgfonlayer}{nodelayer}
		\node [style=miniBlue] (1) at (1, -1.25) {};
		\node [style=miniU] (27) at (0, -1.25) {};
		\node [style=none] (28) at (0, -1.15) {};
		\node [style=none] (29) at (1, -1.15) {};
		\node [style=none] (30) at (0, -1.35) {};
		\node [style=none] (31) at (1, -1.35) {};
		\node [style=none] (32) at (0.175, -0.825) {};
		\node [style=none] (33) at (0.675, -1.25) {};
		\node [style=none] (34) at (0.175, -1.675) {};
		\node [style=miniBlue] (37) at (1, -1.25) {};
		\node [style=miniU] (46) at (2, -1.25) {};
		\node [style=miniBlue] (47) at (-1, -1.25) {};
		\node [style=none] (48) at (-1, -2+0.25) {2};
		\node [style=none] (49) at (0, -2+0.25) {6};
		\node [style=none] (50) at (1, -2+0.25) {4};
		\node [style=none] (51) at (2, -2+0.25) {4};
		\node [style=miniU] (52) at (4, -1.25) {};
		\node [style=miniBlue] (55) at (3, -1.25) {};
		\node [style=none] (56) at (3, -2+0.25) {2};
		\node [style=none] (57) at (4, -2+0.25) {2};
		\node [style=gauge3] (60) at (-2, -1.25) {};
		\node [style=none] (61) at (-2, -2+0.25) {1};
	\end{pgfonlayer}
	\begin{pgfonlayer}{edgelayer}
		\draw (28.center) to (29.center);
		\draw (30.center) to (31.center);
		\draw (32.center) to (33.center);
		\draw (33.center) to (34.center);
		\draw (46) to (37);
		\draw (47) to (27);
		\draw (46) to (55);
		\draw (55) to (52);
		\draw (47) to (60);
	\end{pgfonlayer}
\end{tikzpicture}
}}
 &  $\overline{\mathcal{O}}^{\mathfrak{e}_6}_{\text{min}} = e_6$  \\
\scalebox{0.8}{
\raisebox{-.5\height}{
\begin{tikzpicture}
	\begin{pgfonlayer}{nodelayer}
		\node [style=bluegauge] (0) at (0, 0) {};
		\node [style=redgauge] (1) at (1.25, 0) {};
		\node [style=bluegauge] (2) at (2.5, 0) {};
		\node [style=redgauge] (3) at (3.75, 0) {};
		\node [style=redgauge] (4) at (-1.25, 0) {};
		\node [style=redgauge] (5) at (-3.75, 0) {};
		\node [style=bluegauge] (6) at (-2.5, 0) {};
		\node [style=gauge3] (7) at (0, 1) {};
		\node [style=none] (8) at (0, -0.75+0.25) {4};
		\node [style=none] (9) at (1.25, -0.75+0.25) {4};
		\node [style=none] (10) at (2.5, -0.75+0.25) {2};
		\node [style=none] (11) at (3.75, -0.75+0.25) {2};
		\node [style=none] (12) at (-1.25, -0.75+0.25) {4};
		\node [style=none] (13) at (-2.5, -0.75+0.25) {2};
		\node [style=none] (14) at (-3.75, -0.75+0.25) {2};
		\node [style=none] (15) at (0, 1.5) {1};
	\end{pgfonlayer}
	\begin{pgfonlayer}{edgelayer}
		\draw (5) to (6);
		\draw (6) to (4);
		\draw (4) to (0);
		\draw (0) to (7);
		\draw (0) to (1);
		\draw (1) to (2);
		\draw (2) to (3);
	\end{pgfonlayer}
\end{tikzpicture}
}} &   $\overline{\mathcal{O}}^{\mathfrak{e}_6}_{\text{min}} = e_6$  & \scalebox{0.8}{
\raisebox{-.5\height}{
\begin{tikzpicture}
	\begin{pgfonlayer}{nodelayer}
		\node [style=miniBlue] (1) at (1, -1.25) {};
		\node [style=miniU] (27) at (0, -1.25) {};
		\node [style=none] (28) at (0, -1.15) {};
		\node [style=none] (29) at (1, -1.15) {};
		\node [style=none] (30) at (0, -1.35) {};
		\node [style=none] (31) at (1, -1.35) {};
		\node [style=none] (32) at (0.175, -0.825) {};
		\node [style=none] (33) at (0.675, -1.25) {};
		\node [style=none] (34) at (0.175, -1.675) {};
		\node [style=miniBlue] (37) at (1, -1.25) {};
		\node [style=miniU] (46) at (2, -1.25) {};
		\node [style=none] (49) at (0, -2+0.25) {4};
		\node [style=none] (50) at (1, -2+0.25) {4};
		\node [style=none] (51) at (2, -2+0.25) {2};
		\node [style=miniBlue] (55) at (3, -1.25) {};
		\node [style=none] (56) at (3, -2+0.25) {2};
		\node [style=gauge3] (60) at (-1, -1.25) {};
		\node [style=none] (61) at (-1, -2+0.25) {1};
		\node [style=miniBlue] (62) at (0, -1.25) {};
		\node [style=miniBlue] (63) at (2, -1.25) {};
		\node [style=miniU] (64) at (1, -1.25) {};
		\node [style=miniU] (65) at (3, -1.25) {};
	\end{pgfonlayer}
	\begin{pgfonlayer}{edgelayer}
		\draw (28.center) to (29.center);
		\draw (30.center) to (31.center);
		\draw (32.center) to (33.center);
		\draw (33.center) to (34.center);
		\draw (46) to (37);
		\draw (46) to (55);
		\draw (60) to (27);
	\end{pgfonlayer}
\end{tikzpicture}
}} &  $\overline{\mathcal{O}}^{\mathfrak{so}(10)}_{\text{min}} = d_5$ 
\\
\scalebox{0.8}{
\raisebox{-.5\height}{
\begin{tikzpicture}
	\begin{pgfonlayer}{nodelayer}
		\node [style=bluegauge] (0) at (0, 0) {};
		\node [style=redgauge] (1) at (1.25, 0) {};
		\node [style=bluegauge] (2) at (2.5, 0) {};
		\node [style=redgauge] (3) at (3.75, 0) {};
		\node [style=redgauge] (4) at (-1.25, 0) {};
		\node [style=none] (8) at (0, -0.75+0.25) {2};
		\node [style=none] (9) at (1.25, -0.75+0.25) {4};
		\node [style=none] (10) at (2.5, -0.75+0.25) {2};
		\node [style=none] (11) at (3.75, -0.75+0.25) {2};
		\node [style=none] (12) at (-1.25, -0.75+0.25) {2};
		\node [style=none] (29) at (1.75, 1.25) {1};
		\node [style=gauge3] (30) at (1.25, 1.25) {};
	\end{pgfonlayer}
	\begin{pgfonlayer}{edgelayer}
		\draw (4) to (0);
		\draw (0) to (1);
		\draw (1) to (2);
		\draw (2) to (3);
		\draw (30) to (1);
	\end{pgfonlayer}
\end{tikzpicture}
}} &   $ \overline{\mathcal{O}}^{\mathfrak{so}(10)}_{\text{min}} = d_5$  &\scalebox{0.8}{
\raisebox{-.5\height}{
\begin{tikzpicture}
	\begin{pgfonlayer}{nodelayer}
		\node [style=miniBlue] (1) at (1, -1.25) {};
		\node [style=miniU] (27) at (0, -1.25) {};
		\node [style=none] (28) at (0, -1.15) {};
		\node [style=none] (29) at (1, -1.15) {};
		\node [style=none] (30) at (0, -1.35) {};
		\node [style=none] (31) at (1, -1.35) {};
		\node [style=none] (32) at (0.175, -0.825) {};
		\node [style=none] (33) at (0.675, -1.25) {};
		\node [style=none] (34) at (0.175, -1.675) {};
		\node [style=miniBlue] (37) at (1, -1.25) {};
		\node [style=miniU] (46) at (2, -1.25) {};
		\node [style=none] (49) at (0, -2+0.25) {4};
		\node [style=none] (50) at (1, -2+0.25) {2};
		\node [style=none] (51) at (2, -2+0.25) {2};
		\node [style=gauge3] (60) at (-1, -1.25) {};
		\node [style=none] (61) at (-1, -2+0.25) {1};
		\node [style=miniBlue] (62) at (0, -1.25) {};
		\node [style=miniBlue] (63) at (2, -1.25) {};
		\node [style=miniU] (64) at (1, -1.25) {};
		\node [style=miniU] (66) at (0, -1.25) {};
		\node [style=miniU] (67) at (2, -1.25) {};
		\node [style=miniBlue] (68) at (1, -1.25) {};
	\end{pgfonlayer}
	\begin{pgfonlayer}{edgelayer}
		\draw (28.center) to (29.center);
		\draw (30.center) to (31.center);
		\draw (32.center) to (33.center);
		\draw (33.center) to (34.center);
		\draw (46) to (37);
		\draw (60) to (27);
	\end{pgfonlayer}
\end{tikzpicture}
}}
 &  $\overline{\mathcal{O}}^{\mathfrak{so}(8)}_{\text{min}} = d_4$  \\
 \scalebox{0.8}{
 \raisebox{-.5\height}{
 \begin{tikzpicture}
	\begin{pgfonlayer}{nodelayer}
		\node [style=redgauge] (3) at (-2, 0) {};
		\node [style=bluegauge] (9) at (-1, 0) {};
		\node [style=redgauge] (16) at (0, 0) {};
		\node [style=none] (29) at (-1, -0.75+0.25) {2};
		\node [style=none] (35) at (0, -0.75+0.25) {2};
		\node [style=none] (39) at (-2, -0.75+0.25) {2};
		\node [style=gauge3] (41) at (-1, 1) {};
 		\node [style=none] (43) at (-0.5, 1) {1};
		\node [style=none] (44) at (-0.5, 2) {$1$};
		\node [style=flavour2] (48) at (-1, 2) {};
	\end{pgfonlayer}
	\begin{pgfonlayer}{edgelayer}
		\draw (3) to (9);
		\draw (9) to (41);
		\draw (9) to (16);
		\draw [line join=round,decorate, decoration={zigzag, segment length=4,amplitude=.9,post=lineto,post length=2pt}]  (48) -- (41);
	\end{pgfonlayer}
\end{tikzpicture}
}}  &  $\overline{\mathcal{O}}^{\mathfrak{sl}(5)}_{\text{min}} = a_4$ &
\scalebox{0.8}{
\raisebox{-.5\height}{
\begin{tikzpicture}
	\begin{pgfonlayer}{nodelayer}
		\node [style=none] (28) at (0, -1.15) {};
		\node [style=none] (29) at (1, -1.15) {};
		\node [style=none] (30) at (0, -1.35) {};
		\node [style=none] (31) at (1, -1.35) {};
		\node [style=none] (32) at (0.175, -0.825) {};
		\node [style=none] (33) at (0.675, -1.25) {};
		\node [style=none] (34) at (0.175, -1.675) {};
		\node [style=none] (49) at (0, -2+0.25) {2};
		\node [style=none] (50) at (1, -2+0.25) {2};
		\node [style=flavour2] (60) at (-2.05, -1.25) {};
		\node [style=none] (61) at (-2.05, -2+0.25) {1};
		\node [style=miniU] (67) at (1, -1.25) {};
		\node [style=miniBlue] (68) at (0, -1.25) {};
		\node [style=gauge3] (69) at (-1, -1.25) {};
		\node [style=none] (70) at (-1, -2+0.25) {1};
	\end{pgfonlayer}
	\begin{pgfonlayer}{edgelayer}
		\draw (28.center) to (29.center);
		\draw (30.center) to (31.center);
		\draw (32.center) to (33.center);
		\draw (33.center) to (34.center);
		\draw (69) to (68);
		\draw [line join=round,decorate, decoration={zigzag, segment length=4,amplitude=.9,post=lineto,post length=2pt}]  (60) -- (69);
	\end{pgfonlayer}
\end{tikzpicture}
}}

 &  $\overline{\mathcal{O}}^{\mathfrak{sl}(4)}_{\text{min}} = a_3$ \\ 
         \bottomrule
    \end{tabular}}
    \caption{The orthosymplectic quivers on the left have Coulomb branches that are closures of exceptional algebras $E_n$ for $n=8,7,6,5,4$. Red nodes with an index $k$ denote $\mathrm{SO}(k)$ groups while blue nodes with index $2k$ denote $\mathrm{USp}(2k)$ groups. Folding these quivers along the identical legs gives the non-simply laced orthosymplectic quivers on the right. The Coulomb branches of these theories are given as well. In all the quivers here, there is an overall $\mathbb{Z}_2$ which is ungauged, see \cite{Bourget:2020xdz} for more details. }
    \label{Entable}
\end{table}
\end{landscape}
}

For unframed/unflavoured orthosymplectic quivers, there is a nice set of $E_n$ quivers which are studied in detail in \cite{Bourget:2020gzi,Bourget:2020xdz}. Upon folding their identical legs, one obtains the following key results:
\begin{compactitem}
\item First, folding orthosymplectic quivers, whose Coulomb branches are closures of $E_n$ minimal nilpotent orbits for $4\leq n \leq 8$, leads to non-simply laced orthosymplectic quivers, whose Coulomb branches are also closures of minimal nilpotent orbits. Folding the $E_8$, $E_7$, $E_6$, $E_5\cong D_5$, $E_4\cong A_4$ quivers, leads to non-simply laced orthosymplectic quivers, whose Coulomb branches are closures of minimal orbits of $E_7$, $E_6$, $D_5$, $D_4$, $D_3$  respectively. This can be depicted as follows: 
\begin{equation}
    \raisebox{-.5\height}{\begin{tikzpicture}
        \node (11) at (-10,1) {$E_8$};
        \node (12) at (-8,1) {$E_7$};
        \node (13) at (-6,1) {$E_6$};
        \node (14) at (-4,1) {$D_5$};
        \node (15) at (-2,1) {$A_4$};
        \node (16) at (0,1) {$\cdots$};
        \node (21) at (-9,0) {$E_7$};
        \node (22) at (-7,0) {$E_6$};
        \node (23) at (-5,0) {$D_5$};
        \node (24) at (-3,0) {$D_4$};
        \node (25) at (-1,0) {$A_3$};
        \node (26) at (1,0) {$\cdots$};
        \draw[->] (11)--(12);
        \draw[->] (12)--(13);
        \draw[->] (13)--(14);
        \draw[->] (14)--(15);
        \draw[->] (15)--(16);
        \draw[->] (21)--(22);
        \draw[->] (22)--(23);
        \draw[->] (23)--(24);
        \draw[->] (24)--(25);
        \draw[->] (25)--(26);
        \draw[->,red,thick] (11)--(21);
        \draw[->,red,thick] (12)--(22);
        \draw[->,red,thick] (13)--(23);
        \draw[->,red,thick] (14)--(24);
        \draw[->,red,thick] (15)--(25);
    \end{tikzpicture}}
\end{equation}
The red arrows denote orthosymplectic folding. Note that the top line corresponds to the standard exceptional sequence while the bottom line corresponds to a chain of inclusions of associated affine Weyl groups studied in \cite{sakai2001rational,boalch2009quivers}. 
\item Second, each member of the $E_n$ family of orthosymplectic quivers can be generalized to an infinite sequence of quivers, as shown in \cite{Bourget:2020xdz}. These quivers are magnetic quivers for 5d $\mathcal{N}=1$ SQCD theories. Each of these families of quivers can be folded, producing infinite sequences of non-simply laced orthosymplectic quivers. Some of these families  are magnetic quivers for 4d $\mathcal{N}=2$ theories. 
\end{compactitem}

The outline of the paper is as follows:
Section \ref{monopole} provides a brief recap of magnetic quivers and the monopole formula for non-simply laced orthosymplectic quivers. The alternative method of calculating such Coulomb branches via Hall-Littlewood polynomials and their related functions is also summarised. Thereafter, orthosymplectic quivers with a known unitary quiver counterpart are considered in Section \ref{Example}. 
A comparison from folding both types of quivers, namely the orthosymplectic as well the unitary realisation, demonstrates that the non-simply laced orthosymplectic quiver produces results consistent with the expectation from folding. 
Section \ref{framed} details the folding of framed orthosymplectic quivers, i.e.\ quivers that contain flavour nodes. 
Section \ref{unframed} investigates  non-simply laced orthosymplectic quivers whose Coulomb branches are closures of $E_{5,6,7}$ minimal nilpotent orbits. Section \ref{generalunframed} presents the non-simply laced orthosymplectic quivers that are new magnetic quiver constructions for certain $4$d $\mathcal{N}=2$ theories. 
Having derived a new class of magnetic quivers, Section \ref{Hasse} details the construction of their Hasse diagrams by extending the quiver subtraction algorithm to non-simply laced orthosymplectic quivers. Section \ref{branes} provides brane realisations for flavoured non-simply laced orthosymplectic quivers. 
Lastly, Section \ref{conclusion} concludes and provides an outlook.

\section{The Coulomb branch}
\label{monopole}
The notion of a \emph{magnetic quiver} was recently introduced and studied in \cite{Cremonesi:2015lsa,Mekareeya:2017jgc,Ferlito:2017xdq,Hanany:2018vph,Hanany:2018uhm,Cabrera:2018jxt,Cabrera:2019izd,Cabrera:2019dob,Bourget:2020gzi}. 
A given hyper-K\"ahler moduli space $X$ is said to have a \emph{magnetic quiver construction} if there exist finitely many quivers $\mathsf{Q}_i$ such that 
\begin{equation}
X = \bigcup_i\mathcal{C}^{3d}(\mathsf{Q}_i )
\end{equation}
holds as equality of moduli spaces, where the intersections are lower dimensional and also admit magnetic quiver constructions. In other words, each magnetic quiver is taken as input data for a $3$d $\mathcal{N}=4$ Coulomb branch  $\mathcal{C}^{3d} (\mathsf{Q}_i)$ and each of them is a symplectic singularity \cite{beauville2000symplectic} itself; in contrast, $X$ might be a union of hyper-K\"ahler cones. 
Note that $3$d $\mathcal{N}=4$ Coulomb branches are used only as a black box to construct moduli spaces of theories that do not need to be three dimensional. 
In many physically motivated examples, $X$ is taken as a Higgs branch of a theory with $8$ supercharges in space-time dimensions $d=3,4,5,6$. 

It is important to note that the magnetic quiver construction is not unique, as there are several known examples for which different magnetic quivers describe the same space $X$. For example, in Section \ref{unframed} we consider the exceptional $E_n$ families introduced in \cite{Bourget:2020gzi} or the different representations of the minimal nilpotent orbit of $E_6$ discussed in \cite{Bourget:2020xdz}.

\subsection{The monopole formula}

Given a magnetic quiver, the associated Coulomb branch moduli space can be studied via various techniques such as abelianisation \cite{Bullimore:2015lsa,Bullimore:2016hdc}, Coulomb branch quantisation \cite{Dedushenko:2017avn,Dedushenko:2018icp}, Hilbert series \cite{Cremonesi:2013lqa}, or more mathematical approaches \cite{Nakajima:2015gxa,Nakajima:2015txa,Braverman:2016wma}. 
For this work, the central tool is the \emph{monopole formula} which allows one to evaluate the Coulomb branch Hilbert series by counting dressed monopole operators. The monopole formula for simply-laced unitary quivers and simply-laced orthosymplectic quivers was introduced in \cite{Cremonesi:2013lqa}.
To briefly review, for a given 3d $\mathcal{N}=4$ gauge theory with gauge group $G$ and matter content transforming in some representation $\mathcal{R}$ of $G$, the unrefined monopole formula takes the form
\begin{align}
\begin{aligned}
    \mathrm{HS}(t) &= \sum_{m \in \Lambda \slash \mathcal{W}} P(t;m)\, t^{2 \Delta(m)} \\
    2\Delta(m ) &=  \sum_{\rho \in \mathcal{R}} |\rho(m)| - 2\sum_{\alpha \in \Phi_+} |\alpha(m)|
    \end{aligned}
    \label{monopoleeqn}
\end{align}
with $\Phi_+$ the set of positive roots of $G$. The magnetic lattice $\Lambda$ is the weight lattice of the GNO dual group $G^\vee$ \cite{Goddard:1976qe}, which has Weyl group $\mathcal{W}$. The classical factor $P(t,m)$ originates from dressings by gauge invariant combinations of the residual massless degrees of freedom in the monopole background labelled by $m$. The reader is referred to \cite{Cremonesi:2013lqa} for details. 

For an unframed orthosymplectic quiver with gauge nodes $\{g_i\}$ a possible choice of gauge group is $G'=\prod_ig_i$. If there is a subgroup $H\subset G'$ acting trivially on the matter content, one may choose a different global form of the gauge group: $G=G'/H$. This affects the magnetic lattice in a non-trivial way \cite{Goddard:1976qe}, as discussed in detail in \cite{Bourget:2020xdz}. In this paper, the discrete subgroup  for the unframed simply-laced orthosymplectic quiver is always chosen to be $H=\mathbb{Z}_2$. The magnetic lattice $\Lambda$ of $G=G'/\mathbb{Z}_2$ can be divided into two parts, $\Lambda = \Lambda_1 + \Lambda_2$, where $\Lambda_1\cong\mathbb{Z}^r$ is the magnetic lattice of $G'$, and $\Lambda_2\cong\left(\mathbb{Z}+\frac{1}{2}\right)^r$, $r$ being the rank of $G$. The full Hilbert series is
\begin{equation}
    \mathrm{HS}_{\Lambda}(t)=\mathrm{HS}_{\Lambda_1}(t) + \mathrm{HS}_{\Lambda_2}(t)\;.
\end{equation}

\begin{figure}[t]
    \centering
\begin{tikzpicture}
	\begin{pgfonlayer}{nodelayer}
		\node [style=gauge3] (0) at (-5.8, 3) {};
		\node [style=gauge3] (1) at (-4.5, 3) {};
		\node [style=none] (2) at (-5.8, 3.125) {};
		\node [style=none] (3) at (-4.6, 3.125) {};
		\node [style=none] (4) at (-5.8, 2.875) {};
		\node [style=none] (5) at (-4.6, 2.875) {};
		\node [style=none] (6) at (-5.5, 3.375) {};
		\node [style=none] (7) at (-5, 3) {};
		\node [style=none] (8) at (-5.5, 2.625) {};
		\node [style=none] (9) at (0, 3) {$\Delta_{\mathrm{edge}}=\frac{1}{2}\sum\limits_{i=1}^{k}\sum\limits_{j=1}^{l}|b\;m_{1,i}-m_{2,j}|$};
		\node [style=none] (10) at (-5.75, 2.25) {\scriptsize$\mathrm{U}(k)$};
		\node [style=none] (11) at (-4.5, 2.25) {\scriptsize$\mathrm{U}(l)$};
		\node [style=gauge3] (12) at (-5.75, 0) {};
		\node [style=gauge3] (13) at (-4.45, 0) {};
		\node [style=none] (14) at (-5.75, 0.125) {};
		\node [style=none] (15) at (-4.55, 0.125) {};
		\node [style=none] (16) at (-5.75, -0.125) {};
		\node [style=none] (17) at (-4.55, -0.125) {};
		\node [style=none] (18) at (-5.45, 0.375) {};
		\node [style=none] (19) at (-4.95, 0) {};
		\node [style=none] (20) at (-5.45, -0.375) {};
		\node [style=none] (21) at (1.5, 0) {$\Delta_{\mathrm{edge}}=\frac{1}{2}\sum\limits_{i=1}^{k}\sum\limits_{j=1}^{l}(|b\;m_{1,i}-m_{2,j}|+|b\;m_{1,i}+m_{2,j}|)$};
		\node [style=none] (22) at (-5.7, -0.75) {\scriptsize$\mathrm{USp}(2k)$};
		\node [style=none] (23) at (-4.45, -0.75) {\scriptsize$\mathrm{SO}(2l)$};
		\node [style=miniBlue] (24) at (-5.75, 0) {};
		\node [style=miniU] (25) at (-4.45, 0) {};
		\node [style=none] (26) at (-5.25, 3.75) {$b$};
		\node [style=none] (27) at (-5.2, 0.75) {$b$};
		\node [style=gauge3] (28) at (-5.7, -3) {};
		\node [style=gauge3] (29) at (-4.4, -3) {};
		\node [style=none] (30) at (-5.7, -2.875) {};
		\node [style=none] (31) at (-4.5, -2.875) {};
		\node [style=none] (32) at (-5.7, -3.125) {};
		\node [style=none] (33) at (-4.5, -3.125) {};
		\node [style=none] (34) at (-5.4, -2.625) {};
		\node [style=none] (35) at (-4.9, -3) {};
		\node [style=none] (36) at (-5.4, -3.375) {};
		\node [style=none] (38) at (-4.4, -3.75) {\scriptsize$\mathrm{USp}(2l)$};
		\node [style=none] (39) at (-5.65, -3.75) {\scriptsize$\mathrm{SO}(2k)$};
		\node [style=miniBlue] (40) at (-4.4, -3) {};
		\node [style=miniU] (41) at (-5.7, -3) {};
		\node [style=none] (42) at (-5.15, -2.25) {$b$};
		\node [style=gauge3] (43) at (-5.65, -6) {};
		\node [style=gauge3] (44) at (-4.35, -6) {};
		\node [style=none] (45) at (-5.65, -5.875) {};
		\node [style=none] (46) at (-4.45, -5.875) {};
		\node [style=none] (47) at (-5.65, -6.125) {};
		\node [style=none] (48) at (-4.45, -6.125) {};
		\node [style=none] (49) at (-5.35, -5.625) {};
		\node [style=none] (50) at (-4.85, -6) {};
		\node [style=none] (51) at (-5.35, -6.375) {};
		\node [style=none] (53) at (-5.6, -6.75) {\scriptsize$\mathrm{USp}(2k)$};
		\node [style=none] (54) at (-4.1, -6.75) {\scriptsize$\mathrm{SO}(2l+1)$};
		\node [style=miniBlue] (55) at (-5.65, -6) {};
		\node [style=miniU] (56) at (-4.35, -6) {};
		\node [style=none] (57) at (-5.1, -5.25) {$b$};
		\node [style=gauge3] (58) at (-5.6, -9) {};
		\node [style=gauge3] (59) at (-4.3, -9) {};
		\node [style=none] (60) at (-5.6, -8.875) {};
		\node [style=none] (61) at (-4.4, -8.875) {};
		\node [style=none] (62) at (-5.6, -9.125) {};
		\node [style=none] (63) at (-4.4, -9.125) {};
		\node [style=none] (64) at (-5.3, -8.625) {};
		\node [style=none] (65) at (-4.8, -9) {};
		\node [style=none] (66) at (-5.3, -9.375) {};
		\node [style=none] (68) at (-4.3, -9.75) {\scriptsize $\mathrm{USp}(2l)$};
		\node [style=none] (69) at (-5.8, -9.75) {\scriptsize $\mathrm{SO}(2k+1)$};
		\node [style=miniBlue] (70) at (-4.3, -9) {};
		\node [style=miniU] (71) at (-5.6, -9) {};
		\node [style=none] (72) at (-5.05, -8.25) {$b$};
		\node [style=none] (73) at (1.5, -3) {$\Delta_{\mathrm{edge}}=\frac{1}{2}\sum\limits_{i=1}^{k}\sum\limits_{j=1}^{l}(|b\;m_{1,i}-m_{2,j}|+|b\;m_{1,i}+m_{2,j}|)$};
		\node [style=none] (74) at (2.7, -6) {$\Delta_{\mathrm{edge}}=\frac{1}{2}\sum\limits_{i=1}^{k}\sum\limits_{j=1}^{l}(|b\;m_{1,i}-m_{2,j}|+|b\;m_{1,i}+m_{2,j}|)+\frac{1}{2}\sum\limits_{i=1}^{k}|b\;m_{1,i}|$};
		\node [style=none] (75) at (2.7, -9) {$\Delta_{\mathrm{edge}}=\frac{1}{2}\sum\limits_{i=1}^{k}\sum\limits_{j=1}^{l}(|b\;m_{1,i}-m_{2,j}|+|b\;m_{1,i}+m_{2,j}|)+\frac{1}{2}\sum\limits_{j=1}^{l}|m_{2,j}|$};
	\end{pgfonlayer}
	\begin{pgfonlayer}{edgelayer}
		\draw (2.center) to (3.center);
		\draw (4.center) to (5.center);
		\draw (6.center) to (7.center);
		\draw (7.center) to (8.center);
		\draw (14.center) to (15.center);
		\draw (16.center) to (17.center);
		\draw (18.center) to (19.center);
		\draw (19.center) to (20.center);
		\draw (30.center) to (31.center);
		\draw (32.center) to (33.center);
		\draw (34.center) to (35.center);
		\draw (35.center) to (36.center);
		\draw (45.center) to (46.center);
		\draw (47.center) to (48.center);
		\draw (49.center) to (50.center);
		\draw (50.center) to (51.center);
		\draw (60.center) to (61.center);
		\draw (62.center) to (63.center);
		\draw (64.center) to (65.center);
		\draw (65.center) to (66.center);
	\end{pgfonlayer}
\end{tikzpicture}
    \caption{The contribution of the edges to the conformal dimension $\Delta_{\mathrm{edge}}$ is given for the two-node quivers on the left. The magnetic charges for the left nodes are denoted by $\{m_{1,i}\}$ and for the right node by $\{m_{2,j}\}$. The non-simply laced edge has multiplicity $b$, which then appears as a multiplicative factor for the $m_{1,i}$ magnetic charges. The contribution of the vector multiplets is not affected by non-simply laced edges. }
    \label{conformalD}
\end{figure}
Besides the magnetic lattice, another ingredient for the monopole formula is the conformal dimension. For non-simply laced unitary quivers, the conformal dimension is proposed in \cite{Cremonesi:2014xha}. For framed non-simply laced orthosymplectic quivers, one may propose a similar set of amendments such that the conformal dimension is modified as summarised in Figure \ref{conformalD} to accommodate for the non-simply laced edge.

For unframed non-simply laced orthosymplectic quivers, one needs to take into consideration both the changes to the magnetic lattice due to $H$ as well as the change to the conformal dimension due to the non-simply laced edge. One can divide the nodes of non-simply laced quivers long and short nodes (in the sense of the long and short nodes of Dynkin diagrams). Denote by $\Lambda_L$ the magnetic lattice of the long nodes/gauge groups and by $\Lambda_S$ the magnetic lattice of the short nodes/gauge groups. 
A vector of magnetic charges $m \in \Lambda$ is represented as a pair $m\in (m_L,m_S) \in \Lambda_L \times \Lambda_S $. Let $r_L$ denote the sum of the ranks of all long nodes and $r_S$ the sum of the ranks of all short nodes. 
If the non-simply laced edge is even (i.e.\ with double, quadruple bond etc.), then the magnetic lattice to be summed over is as follows:
\begin{equation}
\begin{aligned}
 \mathbb{Z}^{r_S+r_L} \oplus (  (\mathbb{Z}+\tfrac{1}{2})^{r_L} \times \mathbb{Z}^{r_S} ) 
    \end{aligned}
\end{equation}
In contrast, if the non-simply laced edge is odd (i.e.\ with triple, quintuple bond etc.), then the magnetic lattice is:
\begin{equation}
\begin{aligned}
\mathbb{Z}^{r_S+r_L} \oplus  (\mathbb{Z}+\tfrac{1}{2})^{r_S+r_L}  
    \end{aligned}
\end{equation}
If the non-simply laced orthosymplectic quiver is framed, then the Hilbert series sum is evaluated only over the integer-valued magnetic charges, because the discrete group $H$ is trivial, see \cite{Bourget:2020xdz}. 

In this paper, moduli spaces are identified and compared by Hilbert series computations. These can be computed for the Coulomb branches of magnetic quivers by alternative methods. The central method used in this paper is the monopole formula \eqref{monopoleeqn}. This yields Hilbert series, which can be computed exactly for small rank quivers and perturbatively to high orders for larger rank quivers. For unitary quivers, the Hilbert series can be refined, but there is no known prescription for obtaining the refined Hilbert series of an orthosymplectic quiver via the monopole formula. Nonetheless, such unrefined Hilbert series can be compared with the Hilbert series of known moduli spaces.  Furthermore, when a unitary magnetic quiver construction is known for the same geometric space, the refined Hilbert series can be given in the form of a highest weight generating function (HWG) \cite{RudolphHWG} for characters of the global symmetry group. In certain cases, these HWGs can be generalised to arbitrary rank. The monopole formula results for the orthosymplectic magnetic quivers in this paper have been tested against the unrefined Hilbert series for unitary magnetic quivers, calculated either directly, or by expanding and unrefining their HWGs.

\subsection{Hall-Littlewood computations}
The relationship between the monopole formula with background charges and Hall-Littlewood polynomials was explored in \cite{CremonesiHall}. This relationship permits an alternative method of calculating Coulomb branches that is applicable to many star shaped quivers, both unitary and/or orthosymplectic, with a central node of type $G$. The approach is to identify the Coulomb branches of the linear quiver legs as framed Slodowy slices \cite{Hanany:2019tji}, and to compose these by summing over the weight lattice of the GNO dual $G^\vee$, while incorporating symmetry factors, all as described in Appendix \ref{app:HL}. This method permits the exact calculation of refined (or in some cases, partially refined) Hilbert series and HWGs, thereby providing many consistency checks on the results herein.

\subsection{Unitary vs Orthosymplectic}
\label{Example}
In this section, the folding of simply-laced orthosymplectic quivers is demonstrated on a set of examples. These examples are chosen such that the orthosymplectic quivers have unitary counterparts. The resulting non-simply laced quivers are analysed and are found to be consistent with each other.

\subsubsection{\texorpdfstring{$D_4$ affine Dynkin diagram}{D4 affine Dynkin diagram}}

Due to the isomorphisms  $\mathfrak{so}(2)\cong \mathfrak{u}(1)$ and $\mathfrak{sl}(2)\cong \mathfrak{usp}(2)$, one can construct two quivers with equivalent Coulomb branches:
\begin{equation}
\raisebox{-.5\height}{\scalebox{.811}{
\begin{tikzpicture}
	\begin{pgfonlayer}{nodelayer}
		\node [style=miniU] (0) at (-1, 1) {};
		\node [style=miniU] (1) at (1, 1) {};
		\node [style=miniU] (2) at (-1, -1) {};
		\node [style=miniU] (3) at (1, -1) {};
		\node [style=miniBlue] (4) at (0, 0) {};
		\node [style=none] (5) at (-1, 1.5) {2};
		\node [style=none] (6) at (1, 1.5) {2};
		\node [style=none] (7) at (0.5, 0) {2};
		\node [style=none] (8) at (-1, -1.5) {2};
		\node [style=none] (9) at (1, -1.5) {2};
		\node [style=gauge3] (10) at (-6, 1) {};
		\node [style=gauge3] (11) at (-4, 1) {};
		\node [style=gauge3] (12) at (-6, -1) {};
		\node [style=gauge3] (13) at (-4, -1) {};
		\node [style=gauge3] (14) at (-5, 0) {};
		\node [style=none] (15) at (-6, 1.5) {1};
		\node [style=none] (16) at (-4, 1.5) {1};
		\node [style=none] (17) at (-4.5, 0) {2};
		\node [style=none] (18) at (-6, -1.5) {1};
		\node [style=none] (19) at (-4, -1.5) {1};
		\node [style=none] (20) at (-3.25, 0) {};
		\node [style=none] (21) at (-1.5, 0) {};
	\end{pgfonlayer}
	\begin{pgfonlayer}{edgelayer}
		\draw (4) to (0);
		\draw (4) to (1);
		\draw (4) to (3);
		\draw (4) to (2);
		\draw (10) to (14);
		\draw (14) to (11);
		\draw (14) to (13);
		\draw (14) to (12);
		\draw [style=->] (20.center) to (21.center);
		\draw [style=->] (21.center) to (20.center);
	\end{pgfonlayer}
\end{tikzpicture}}
}
\end{equation}
where white nodes with label $n$ denote $\mathrm{U}(n)$ gauge groups, red nodes with label $n$ denote $\mathrm{SO}(n)$ gauge groups, and blue nodes with label $2n$ denote $\mathrm{USp}(2n)$ gauge groups. Note that the central node has rank 2 on the left hand side and rank 1 on the right hand side. As a reminder, for unframed unitary quivers, there is always a diagonal $\mathrm{U}(1)$ that one needs to ungauge. Whereas for the unframed orthosymplectic quiver, one chooses $H=\mathbb{Z}_2\subset \mathrm{USp}(2)\times \mathrm{SO}(2)^4$ as the diagonal subgroup to be ungauged.  As shown in \cite{Bourget:2020xdz}, the Coulomb branch Hilbert series of both quivers are the same.

Since the two quivers have the same shape, one can fold identical legs and check if the non-simply laced quivers reproduce the same results. Using the monopole formula, one can verify that the following foldings reproduce the same Coulomb branch Hilbert series. 
\paragraph{Folding two identical legs.}
First one folds two of the four identical legs in each quiver such that one obtains 
\begin{equation}
\raisebox{-.5\height}{\scalebox{.811}{
\begin{tikzpicture}
	\begin{pgfonlayer}{nodelayer}
		\node [style=miniU] (0) at (-1, 0.75) {};
		\node [style=miniU] (2) at (-1, -0.75) {};
		\node [style=miniBlue] (4) at (0, 0) {};
		\node [style=none] (5) at (-1, 1.25) {2};
		\node [style=none] (8) at (-1, -1.25) {2};
		\node [style=gauge3] (11) at (-5.75, 0.75) {};
		\node [style=gauge3] (12) at (-5.75, -0.75) {};
		\node [style=gauge3] (13) at (-4, 0) {};
		\node [style=gauge3] (14) at (-5, 0) {};
		\node [style=none] (16) at (-5.75, 1.25) {1};
		\node [style=none] (17) at (-5, -0.5) {2};
		\node [style=none] (18) at (-5.75, -1.25) {1};
		\node [style=none] (19) at (-4, -0.5) {1};
		\node [style=none] (20) at (-3.25, 0) {};
		\node [style=none] (21) at (-1.5, 0) {};
		\node [style=none] (22) at (-5, 0.1) {};
		\node [style=none] (23) at (-4, 0.1) {};
		\node [style=none] (24) at (-5, -0.1) {};
		\node [style=none] (25) at (-4, -0.1) {};
		\node [style=none] (26) at (-4.825, 0.425) {};
		\node [style=none] (27) at (-4.325, 0) {};
		\node [style=none] (28) at (-4.825, -0.425) {};
		\node [style=none] (29) at (0, 0.1) {};
		\node [style=none] (30) at (1, 0.1) {};
		\node [style=none] (31) at (0, -0.1) {};
		\node [style=none] (32) at (1, -0.1) {};
		\node [style=none] (33) at (0.175, 0.425) {};
		\node [style=none] (34) at (0.675, 0) {};
		\node [style=none] (35) at (0.175, -0.425) {};
		\node [style=miniU] (36) at (1, 0) {};
		\node [style=none] (37) at (1, -0.5) {2};
		\node [style=none] (38) at (0, -0.5) {2};
	\end{pgfonlayer}
	\begin{pgfonlayer}{edgelayer}
		\draw (4) to (0);
		\draw (4) to (2);
		\draw (14) to (11);
		\draw (14) to (12);
		\draw [style=->] (20.center) to (21.center);
		\draw [style=->] (21.center) to (20.center);
		\draw (22.center) to (23.center);
		\draw (24.center) to (25.center);
		\draw (26.center) to (27.center);
		\draw (27.center) to (28.center);
		\draw (29.center) to (30.center);
		\draw (31.center) to (32.center);
		\draw (33.center) to (34.center);
		\draw (34.center) to (35.center);
	\end{pgfonlayer}
\end{tikzpicture}}
}
\end{equation}
For non-simply laced quivers, the node where one ungauges the $\mathrm{U}(1)$ is important and different nodes can yield different Coulomb branches \cite{Hanany:2020jzl}. In this paper, the choice taken is to ungauge on a long node (ungaugings on long nodes all give the same moduli space). 
The Coulomb branch of the unitary quiver is known be $\overline{\mathcal{O}}^{\mathfrak{so}(7)}_{\text{min}}$ as it is the affine Dynkin diagram of $B_3$ \cite{Hanany:2016gbz}. An explicit Coulomb branch Hilbert series computation shows that the folded orthosymplectic quiver is consistent with having the same Coulomb branch:
\begin{equation}
    \mathrm{HS}_{\mathrm{U}}(t)=\mathrm{HS}_{\mathrm{OSp}}(t)=\frac{1+13t^2+28t^4+13t^6+t^8}{(1-t^2)^8}
\end{equation}
where $\mathrm{HS}_{\mathrm{U}}(t)$ and $\mathrm{HS}_{\mathrm{OSp}}(t)$ are the Hilbert series of the unitary and orthosymplectic quivers, respectively. 

\paragraph{Folding three identical legs.}
Next, one proceeds to fold three of the identical legs:
\begin{equation}
\raisebox{-.5\height}{\scalebox{.811}{
    \begin{tikzpicture}
	\begin{pgfonlayer}{nodelayer}
		\node [style=miniU] (2) at (-1, 0) {};
		\node [style=miniBlue] (4) at (0, 0) {};
		\node [style=none] (8) at (-1, -0.5) {2};
		\node [style=gauge3] (12) at (-6, 0) {};
		\node [style=gauge3] (13) at (-4, 0) {};
		\node [style=gauge3] (14) at (-5, 0) {};
		\node [style=none] (17) at (-5, -0.5) {2};
		\node [style=none] (18) at (-6, -0.5) {1};
		\node [style=none] (19) at (-4, -0.5) {1};
		\node [style=none] (20) at (-3.25, 0) {};
		\node [style=none] (21) at (-1.5, 0) {};
		\node [style=none] (22) at (-5, 0.1) {};
		\node [style=none] (23) at (-4, 0.1) {};
		\node [style=none] (24) at (-5, -0.1) {};
		\node [style=none] (25) at (-4, -0.1) {};
		\node [style=none] (26) at (-4.825, 0.425) {};
		\node [style=none] (27) at (-4.325, 0) {};
		\node [style=none] (28) at (-4.825, -0.425) {};
		\node [style=none] (29) at (0, 0.1) {};
		\node [style=none] (30) at (1, 0.1) {};
		\node [style=none] (31) at (0, -0.1) {};
		\node [style=none] (32) at (1, -0.1) {};
		\node [style=none] (33) at (0.175, 0.425) {};
		\node [style=none] (34) at (0.675, 0) {};
		\node [style=none] (35) at (0.175, -0.425) {};
		\node [style=miniU] (36) at (1, 0) {};
		\node [style=none] (37) at (1, -0.5) {2};
		\node [style=none] (38) at (0, -0.5) {2};
	\end{pgfonlayer}
	\begin{pgfonlayer}{edgelayer}
		\draw (4) to (2);
		\draw (14) to (12);
		\draw [style=->] (20.center) to (21.center);
		\draw [style=->] (21.center) to (20.center);
		\draw (22.center) to (23.center);
		\draw (24.center) to (25.center);
		\draw (26.center) to (27.center);
		\draw (27.center) to (28.center);
		\draw (29.center) to (30.center);
		\draw (31.center) to (32.center);
		\draw (33.center) to (34.center);
		\draw (34.center) to (35.center);
		\draw (14) to (13);
		\draw (4) to (36);
	\end{pgfonlayer}
\end{tikzpicture}}
}
\end{equation}
The unitary quiver is the affine Dynkin diagram of $G_2$ and, hence, the Coulomb branch is $\overline{\mathcal{O}}^{\mathfrak{g}_2}_{\text{min}}$. An explicit computation of the Coulomb branch Hilbert series of the orthosymplectic quiver 
\begin{equation}
    \mathrm{HS}_{\mathrm{U}}(t)=\mathrm{HS}_{\mathrm{OSp}}(t)=\frac{(1+t^2)(1+7t^2+t^4)}{(1-t^2)^6}
\end{equation}
 confirms the equality of the moduli spaces in terms of Hilbert series.
\paragraph{Folding four identical legs.}
Finally, one folds all the identical legs and obtains the following quivers
\begin{equation}
\raisebox{-.5\height}{\scalebox{.811}{
\begin{tikzpicture}
	\begin{pgfonlayer}{nodelayer}
		\node [style=miniBlue] (4) at (-0.75, 0) {};
		\node [style=gauge3] (13) at (-3.8, 0) {};
		\node [style=gauge3] (14) at (-5, 0) {};
		\node [style=none] (17) at (-5, -0.5) {2};
		\node [style=none] (19) at (-4, -0.5) {1};
		\node [style=none] (20) at (-3.25, 0) {};
		\node [style=none] (21) at (-1.5, 0) {};
		\node [style=none] (22) at (-5, 0.1) {};
		\node [style=none] (24) at (-5, -0.1) {};
		\node [style=none] (26) at (-4.7, 0.425) {};
		\node [style=none] (27) at (-4.2, 0) {};
		\node [style=none] (28) at (-4.7, -0.425) {};
		\node [style=none] (29) at (-0.75, 0.1) {};
		\node [style=none] (30) at (0.25, 0.1) {};
		\node [style=none] (31) at (-0.75, -0.1) {};
		\node [style=none] (32) at (0.25, -0.1) {};
		\node [style=miniU] (36) at (0.25, 0) {};
		\node [style=none] (37) at (0.25, -0.5) {2};
		\node [style=none] (40) at (-4.85, 0.15) {};
		\node [style=none] (41) at (-4.85, 0.05) {};
		\node [style=none] (42) at (-4.85, -0.05) {};
		\node [style=none] (43) at (-4.85, -0.15) {};
		\node [style=none] (44) at (-3.85, 0.15) {};
		\node [style=none] (45) at (-3.85, 0.05) {};
		\node [style=none] (46) at (-3.85, -0.05) {};
		\node [style=none] (47) at (-3.85, -0.15) {};
		\node [style=none] (49) at (-0.6, 0.425) {};
		\node [style=none] (50) at (-0.1, 0) {};
		\node [style=none] (51) at (-0.6, -0.425) {};
		\node [style=none] (52) at (-0.75, 0.15) {};
		\node [style=none] (53) at (-0.75, 0.05) {};
		\node [style=none] (54) at (-0.75, -0.05) {};
		\node [style=none] (55) at (-0.75, -0.15) {};
		\node [style=none] (56) at (0.25, 0.15) {};
		\node [style=none] (57) at (0.25, 0.05) {};
		\node [style=none] (58) at (0.25, -0.05) {};
		\node [style=none] (59) at (0.25, -0.15) {};
		\node [style=none] (60) at (-0.75, -0.5) {2};
	\end{pgfonlayer}
	\begin{pgfonlayer}{edgelayer}
		\draw [style=->] (20.center) to (21.center);
		\draw [style=->] (21.center) to (20.center);
		\draw (26.center) to (27.center);
		\draw (27.center) to (28.center);
		\draw (40.center) to (44.center);
		\draw (45.center) to (41.center);
		\draw (42.center) to (46.center);
		\draw (47.center) to (43.center);
		\draw (49.center) to (50.center);
		\draw (50.center) to (51.center);
		\draw (52.center) to (56.center);
		\draw (57.center) to (53.center);
		\draw (54.center) to (58.center);
		\draw (59.center) to (55.center);
	\end{pgfonlayer}
\end{tikzpicture}}
}
\end{equation}
The unitary non-simply laced quiver is investigated in \cite{Bourget:2020asf} and the Coulomb branch is $\overline{\mathcal{O}}^{\mathfrak{sl}_3}_{\text{min}}$. An explicit computation of the Coulomb branch Hilbert series of the orthosymplectic quiver
\begin{equation}
    \mathrm{HS}_{\mathrm{U}}(t)=\mathrm{HS}_{\mathrm{OSp}}(t)=\frac{1+4t^2+t^4}{(1-t^2)^4}
\end{equation}
shows that both magnetic quiver constructions agree in the unrefined Hilbert series.
\subsubsection{\texorpdfstring{$T_4$ theory}{T4 theory}}
Next, let us turn our attention to a more involved example. The theory known as $T_4$ is constructed by gluing together quivers whose Coulomb branches are closures of maximal nilpotent orbits of $\mathfrak{sl}(4)$. Due to the isomorphism $\mathfrak{sl}(4)\cong \mathfrak{so}(6)$, the following quivers have equivalent moduli spaces:
\begin{equation}
\raisebox{-.5\height}{\scalebox{.811}{
    \begin{tikzpicture}[scale=0.8]
	\begin{pgfonlayer}{nodelayer}
		\node [style=gauge3] (0) at (-4, 2) {};
		\node [style=gauge3] (1) at (-4, 1) {};
		\node [style=gauge3] (2) at (-4, 0) {};
		\node [style=gauge3] (3) at (-3, 0) {};
		\node [style=gauge3] (4) at (-2, 0) {};
		\node [style=gauge3] (5) at (-1, 0) {};
		\node [style=gauge3] (6) at (-4, 3) {};
		\node [style=gauge3] (7) at (-5, 0) {};
		\node [style=gauge3] (8) at (-6, 0) {};
		\node [style=gauge3] (9) at (-7, 0) {};
		\node [style=none] (10) at (-3.5, 3) {1};
		\node [style=none] (11) at (-3.5, 2) {2};
		\node [style=none] (12) at (-3.5, 1) {3};
		\node [style=none] (13) at (-4, -0.5) {4};
		\node [style=none] (14) at (-3, -0.5) {3};
		\node [style=none] (15) at (-2, -0.5) {2};
		\node [style=none] (16) at (-1, -0.5) {1};
		\node [style=none] (17) at (-5, -0.5) {3};
		\node [style=none] (18) at (-6, -0.5) {2};
		\node [style=none] (19) at (-7, -0.5) {1};
		\node [style=none] (20) at (0, 0) {};
		\node [style=none] (21) at (2, 0) {};
		\node [style=miniU] (22) at (7, 0) {};
		\node [style=miniBlue] (23) at (6, 0) {};
		\node [style=miniBlue] (24) at (4, 0) {};
		\node [style=miniBlue] (25) at (8, 0) {};
		\node [style=miniBlue] (26) at (7, 1) {};
		\node [style=miniBlue] (27) at (7, 3) {};
		\node [style=miniBlue] (28) at (10, 0) {};
		\node [style=miniU] (29) at (9, 0) {};
		\node [style=miniU] (30) at (11, 0) {};
		\node [style=miniU] (31) at (5, 0) {};
		\node [style=miniU] (32) at (3, 0) {};
		\node [style=miniU] (33) at (7, 2) {};
		\node [style=miniU] (34) at (7, 4) {};
		\node [style=none] (35) at (7, -0.5) {6};
		\node [style=none] (36) at (6, -0.5) {4};
		\node [style=none] (37) at (8, -0.5) {4};
		\node [style=none] (38) at (9, -0.5) {4};
		\node [style=none] (39) at (10, -0.5) {2};
		\node [style=none] (40) at (11, -0.5) {2};
		\node [style=none] (41) at (5, -0.5) {4};
		\node [style=none] (42) at (4, -0.5) {2};
		\node [style=none] (43) at (3, -0.5) {2};
		\node [style=none] (44) at (7.5, 1) {4};
		\node [style=none] (45) at (7.5, 2) {4};
		\node [style=none] (46) at (7.5, 3) {2};
		\node [style=none] (47) at (7.5, 4) {2};
	\end{pgfonlayer}
	\begin{pgfonlayer}{edgelayer}
		\draw (6) to (0);
		\draw (0) to (1);
		\draw (1) to (2);
		\draw (2) to (3);
		\draw (3) to (4);
		\draw (4) to (5);
		\draw (2) to (7);
		\draw (7) to (8);
		\draw (8) to (9);
		\draw [style=->] (20.center) to (21.center);
		\draw [style=->] (21.center) to (20.center);
		\draw (34) to (27);
		\draw (27) to (33);
		\draw (33) to (26);
		\draw (26) to (22);
		\draw (22) to (25);
		\draw (25) to (29);
		\draw (29) to (28);
		\draw (28) to (30);
		\draw (22) to (23);
		\draw (23) to (31);
		\draw (31) to (24);
		\draw (32) to (24);
	\end{pgfonlayer}
\end{tikzpicture}}
}
\end{equation}
Computation of the Coulomb branch Hilbert series of the orthosymplectic quiver is given in \cite[Fig.39]{Bourget:2020xdz} and is consistent with the unitary counterpart.

As a first step, one can fold two of the quiver legs which yields:
\begin{equation}
   \scalebox{.811}{ 
   \begin{tikzpicture}[scale=0.8]
	\begin{pgfonlayer}{nodelayer}
		\node [style=gauge3] (2) at (-4, 0) {};
		\node [style=gauge3] (3) at (-3, 0) {};
		\node [style=gauge3] (4) at (-2, 0) {};
		\node [style=gauge3] (5) at (-1, 0) {};
		\node [style=gauge3] (7) at (-5, 0) {};
		\node [style=gauge3] (8) at (-6, 0) {};
		\node [style=gauge3] (9) at (-7, 0) {};
		\node [style=none] (13) at (-4, -0.5) {4};
		\node [style=none] (14) at (-3, -0.5) {3};
		\node [style=none] (15) at (-2, -0.5) {2};
		\node [style=none] (16) at (-1, -0.5) {1};
		\node [style=none] (17) at (-5, -0.5) {3};
		\node [style=none] (18) at (-6, -0.5) {2};
		\node [style=none] (19) at (-7, -0.5) {1};
		\node [style=none] (20) at (0, 0) {};
		\node [style=none] (21) at (2, 0) {};
		\node [style=miniU] (22) at (7, 0) {};
		\node [style=miniBlue] (23) at (6, 0) {};
		\node [style=miniBlue] (24) at (4, 0) {};
		\node [style=miniBlue] (25) at (8, 0) {};
		\node [style=miniBlue] (28) at (10, 0) {};
		\node [style=miniU] (29) at (9, 0) {};
		\node [style=miniU] (30) at (11, 0) {};
		\node [style=miniU] (31) at (5, 0) {};
		\node [style=miniU] (32) at (3, 0) {};
		\node [style=none] (35) at (7, -0.5) {6};
		\node [style=none] (36) at (6, -0.5) {4};
		\node [style=none] (37) at (8, -0.5) {4};
		\node [style=none] (38) at (9, -0.5) {4};
		\node [style=none] (39) at (10, -0.5) {2};
		\node [style=none] (40) at (11, -0.5) {2};
		\node [style=none] (41) at (5, -0.5) {4};
		\node [style=none] (42) at (4, -0.5) {2};
		\node [style=none] (43) at (3, -0.5) {2};
		\node [style=none] (48) at (-4, 0.1) {};
		\node [style=none] (49) at (-4, -0.15) {};
		\node [style=none] (50) at (-2.875, 0.1) {};
		\node [style=none] (51) at (-2.875, -0.15) {};
		\node [style=none] (52) at (-3.75, 0.35) {};
		\node [style=none] (53) at (-3.75, -0.4) {};
		\node [style=none] (54) at (-3.325, -0.025) {};
		\node [style=none] (57) at (7, 0.1) {};
		\node [style=none] (58) at (7, -0.15) {};
		\node [style=none] (59) at (8.125, 0.1) {};
		\node [style=none] (60) at (8.125, -0.15) {};
		\node [style=none] (61) at (7.25, 0.35) {};
		\node [style=none] (62) at (7.25, -0.4) {};
		\node [style=none] (63) at (7.675, -0.025) {};
	\end{pgfonlayer}
	\begin{pgfonlayer}{edgelayer}
		\draw (3) to (4);
		\draw (4) to (5);
		\draw (2) to (7);
		\draw (7) to (8);
		\draw (8) to (9);
		\draw [style=->] (20.center) to (21.center);
		\draw [style=->] (21.center) to (20.center);
		\draw (25) to (29);
		\draw (29) to (28);
		\draw (28) to (30);
		\draw (22) to (23);
		\draw (23) to (31);
		\draw (31) to (24);
		\draw (32) to (24);
		\draw (48.center) to (50.center);
		\draw (49.center) to (51.center);
		\draw (52.center) to (54.center);
		\draw (54.center) to (53.center);
		\draw (57.center) to (59.center);
		\draw (58.center) to (60.center);
		\draw (61.center) to (63.center);
		\draw (63.center) to (62.center);
	\end{pgfonlayer}
\end{tikzpicture} }
\end{equation}
Computation of the Coulomb branch Hilbert series of both quivers yields:
\begin{equation}
    \mathrm{HS}_{\mathrm{U}}(t)=\mathrm{HS}_{\mathrm{OSp}}(t)=\frac{\small \left(\begin{array}{c}
    1+21 t^2+68 t^3+341 t^4+1300 t^5+4936 t^6+15988 t^7 \\ 
    +50242 t^8+142812 t^9+384411 t^{10}+960772 t^{11}+2270650 t^{12} \\
    +5038840 t^{13}+10601001 t^{14}+21083004 t^{15}+39862377 t^{16} \\ 
    +71590384 t^{17}+122553812 t^{18}+199944220 t^{19}+311642452 t^{20} \\ 
    +464078612 t^{21}+661421665 t^{22}+902317920 t^{23}+1179751147 t^{24} \\
    +1478423752 t^{25}+1777451140 t^{26}+2050065624 t^{27} \\ 
    +2269933494 t^{28}+2412458048 t^{29}+2462182956   t^{30}\\ 
    +\mathrm{palindromic} +\dots +21t^{58}+t^{60} \end{array} \right)}{(1-t^2)^{9} \left(1-t^3\right)^{12} \left(1-t^4\right)^9 
    }
\end{equation}

As a next step, one folds all three identical legs which yields
\begin{equation}
\raisebox{-.5\height}{\scalebox{.811}{
    \begin{tikzpicture}
	\begin{pgfonlayer}{nodelayer}
		\node [style=gauge3] (2) at (-5, 0) {};
		\node [style=none] (13) at (-5, -0.5) {4};
		\node [style=miniU] (44) at (1.25, 0) {};
		\node [style=none] (53) at (-1.25, 0) {};
		\node [style=none] (54) at (0.5, 0) {};
		\node [style=none] (62) at (1.25, 0.1) {};
		\node [style=none] (63) at (2.25, 0.1) {};
		\node [style=none] (64) at (1.25, -0.1) {};
		\node [style=none] (65) at (2.25, -0.1) {};
		\node [style=none] (66) at (1.425, 0.425) {};
		\node [style=none] (67) at (1.925, 0) {};
		\node [style=none] (68) at (1.425, -0.425) {};
		\node [style=none] (70) at (2.25, -0.5) {4};
		\node [style=none] (71) at (1.25, -0.5) {6};
		\node [style=gauge3] (73) at (-4, 0) {};
		\node [style=gauge3] (74) at (-5, 0) {};
		\node [style=none] (76) at (-4, -0.5) {3};
		\node [style=none] (77) at (-5, 0.1) {};
		\node [style=none] (78) at (-4, 0.1) {};
		\node [style=none] (79) at (-5, -0.1) {};
		\node [style=none] (80) at (-4, -0.1) {};
		\node [style=none] (81) at (-4.825, 0.425) {};
		\node [style=none] (82) at (-4.325, 0) {};
		\node [style=gauge3] (85) at (-3, 0) {};
		\node [style=gauge3] (86) at (-2, 0) {};
		\node [style=none] (87) at (-3, -0.5) {2};
		\node [style=none] (88) at (-2, -0.5) {1};
		\node [style=miniBlue] (89) at (2.25, 0) {};
		\node [style=miniBlue] (90) at (4.25, 0) {};
		\node [style=miniU] (91) at (3.25, 0) {};
		\node [style=miniU] (92) at (5.25, 0) {};
		\node [style=none] (93) at (3.25, -0.5) {4};
		\node [style=none] (94) at (4.25, -0.5) {2};
		\node [style=none] (95) at (5.25, -0.5) {2};
		\node [style=none] (96) at (-4.3, 0) {};
		\node [style=none] (97) at (-4.8, -0.425) {};
	\end{pgfonlayer}
	\begin{pgfonlayer}{edgelayer}
		\draw [style=->] (53.center) to (54.center);
		\draw [style=->] (54.center) to (53.center);
		\draw (62.center) to (63.center);
		\draw (64.center) to (65.center);
		\draw (66.center) to (67.center);
		\draw (67.center) to (68.center);
		\draw (77.center) to (78.center);
		\draw (79.center) to (80.center);
		\draw (81.center) to (82.center);
		\draw (74) to (73);
		\draw (85) to (82.center);
		\draw (86) to (85);
		\draw (89) to (91);
		\draw (91) to (90);
		\draw (90) to (92);
		\draw (89) to (44);
		\draw (96.center) to (97.center);
	\end{pgfonlayer}
\end{tikzpicture}}
}
\label{T4folded}
\end{equation}
The unitary quiver in \eqref{T4folded} is a known member of the generalised rank 1 $4$d $\mathcal{N}=2$ sequence studied in \cite{Bourget:2020asf}. An explicit computation of the Coulomb branch Hilbert series of the both quivers in \eqref{T4folded} yields 
\begin{equation}
    \mathrm{HS}_{\mathrm{U}}(t)=\mathrm{HS}_{\mathrm{OSp}}(t)=\frac{\small \left(\begin{array}{c}1-t+10 t^2+23 t^3+67 t^4+190 t^5+525 t^6+1053 t^7\\+2292 t^8+4167 t^9+7299 t^{10}+11494 t^{11}+17114 t^{12}+23080 t^{13}\\+29925
   t^{14}+35107 t^{15}+39221 t^{16}+40320 t^{17}\\ +\mathrm{palindromic} +\dots +10t^{32}-t^{33}+t^{34}\end{array} \right)}{(1-t)(1-t^2)^{5}(1-t^3)^{7} (1-t^4)^{5} } \, . 
\end{equation}
As a reminder, in all calculations in this article that involve non-simply laced unitary quivers which lack explicit framing, the overall $\mathrm{U}(1)$ framing is applied on a long node, such as the central node, in order to obtain consistent results.

The above examples reinforce the conjecture that folding orthosymplectic quivers yields valid results, so one may proceed to fold quivers where the resulting Coulomb branches cannot easily be determined from accidental isomorphisms.

\section{Folding framed orthosymplectic quivers}\label{framed}
As a next step, examine the folding of certain families of orthosymplectic quivers treated in \cite{Cremonesi:2014uva,Cabrera:2017ucb}, whose Coulomb branches are closures of nilpotent orbits. To be concrete, the focus is placed on so-called height two orbits, which are orbits of elements $x \in \mathfrak{g}$ such that $\mathrm{ad}(x)^2 \neq 0$ and $\mathrm{ad}(x)^3 = 0$ \cite{panyushev1994complexity}. These are the lowest dimensional non-trivial nilpotent orbits and yield a few clear candidates with the necessary symmetry for folding. In particular, height two orbits of type D are considered.

The monopole formula for orthosymplectic quivers only returns unrefined Hilbert series \cite{CremonesiHall}. However, these are often sufficient to identify known moduli spaces (such as nilpotent orbit closures). For these, the encoding of refined Hilbert series into HWGs is often straightforward. Indeed, for each of the orthosymplectic quivers in the following sections, the Coulomb branches turn out to be well-known moduli spaces, for which HWGs provide a concise description.
Furthermore, as is shown below, one can find natural projection maps between the HWGs for orthosymplectic quivers before and after folding.

\subsection{Height two nilpotent orbits}

\paragraph{Even D-type.} 
For $\overline{\mathcal{O}}^{\mathfrak{so}(4n)}_{[2^{2n}]}$, the orbit is the union of two identical cones \cite{Collingwood:1993fk}. One of these cones has the magnetic quiver:
\begin{equation}
\raisebox{-.5\height}{\scalebox{.811}{
    \begin{tikzpicture}
	\begin{pgfonlayer}{nodelayer}
		\node [style=miniU] (0) at (0, 1) {};
		\node [style=miniBlue] (1) at (1, 1) {};
		\node [style=miniBlue] (2) at (-1, 1) {};
		\node [style=none] (3) at (1.75, 1) {\dots};
		\node [style=none] (4) at (-1.75, 1) {\dots};
		\node [style=miniU] (5) at (2.5, 1) {};
		\node [style=miniBlue] (6) at (3.5, 1) {};
		\node [style=miniBlue] (7) at (-3.5, 1) {};
		\node [style=miniU] (8) at (-2.5, 1) {};
		\node [style=miniU] (9) at (-4.5, 1) {};
		\node [style=flavourBlue] (10) at (0, 2) {};
		\node [style=miniU] (11) at (4.5, 1) {};
		\node [style=none] (12) at (1.5, 1) {};
		\node [style=none] (13) at (2, 1) {};
		\node [style=none] (14) at (-1.5, 1) {};
		\node [style=none] (15) at (-2, 1) {};
		\node [style=none] (16) at (0, 2.5) {2};
		\node [style=none] (17) at (0, 0.5) {$2n$};
		\node [style=none] (18) at (1, 0.5) {$2n{-}2$};
		\node [style=none] (19) at (-1, 0.5) {$2n{-}2$};
		\node [style=none] (20) at (-2.5, 0.5) {4};
		\node [style=none] (21) at (-3.5, 0.5) {2};
		\node [style=none] (22) at (-4.5, 0.5) {2};
		\node [style=none] (23) at (2.5, 0.5) {4};
		\node [style=none] (24) at (3.5, 0.5) {2};
		\node [style=none] (25) at (4.5, 0.5) {2};
	\end{pgfonlayer}
	\begin{pgfonlayer}{edgelayer}
		\draw (2) to (0);
		\draw (0) to (10);
		\draw (0) to (1);
		\draw (5) to (6);
		\draw (8) to (7);
		\draw (7) to (9);
		\draw (6) to (11);
		\draw (15.center) to (8);
		\draw (14.center) to (2);
		\draw (1) to (12.center);
		\draw (13.center) to (5);
	\end{pgfonlayer}
\end{tikzpicture}}
}
\label{evenorbit}
\end{equation}
The refined Coulomb branch Hilbert series can be encoded as the HWG
\begin{equation}
    \mathrm{HWG}\eqref{evenorbit}= \mathrm{PE}\left[\sum_{i=1}^{n-1}\rho_{2i}t^{2i}+\rho_{2n}^2t^{2n} \right] \, , 
    \label{hwgeven}
\end{equation}
where $\rho_i$ for $i=1,\dots,2n$ are the highest weight fugacities of $\mathfrak{so}(4n)$. Note, the fugacity for the $\rho_{2n}$ spinor is present in the expression \cite{Ferlito:2016grh}. For the second cone, the quiver is the same as \eqref{evenorbit}, but the other spinor $\rho_{2n-1}$ is used in the HWG. The orbit is the union of the two cones and includes both spinors (whereas the intersection contains neither). 

Folding \eqref{evenorbit} results in the following quiver:
\begin{equation}
\raisebox{-.5\height}{\scalebox{.811}{
    \begin{tikzpicture}
	\begin{pgfonlayer}{nodelayer}
		\node [style=miniBlue] (1) at (1, -1.25) {};
		\node [style=none] (3) at (1.75, -1.25) {$\dots$};
		\node [style=miniU] (5) at (2.5, -1.25) {};
		\node [style=miniBlue] (6) at (3.5, -1.25) {};
		\node [style=miniU] (11) at (4.5, -1.25) {};
		\node [style=none] (12) at (1.5, -1.25) {};
		\node [style=none] (13) at (2, -1.25) {};
		\node [style=none] (18) at (1, -1.75) {$2n{-}2$};
		\node [style=none] (23) at (2.5, -1.75) {4};
		\node [style=none] (24) at (3.5, -1.75) {2};
		\node [style=none] (25) at (4.5, -1.75) {2};
		\node [style=miniU] (27) at (0, -1.25) {};
		\node [style=none] (28) at (0, -1.15) {};
		\node [style=none] (29) at (1, -1.15) {};
		\node [style=none] (30) at (0, -1.35) {};
		\node [style=none] (31) at (1, -1.35) {};
		\node [style=none] (32) at (0.175, -0.825) {};
		\node [style=none] (33) at (0.675, -1.25) {};
		\node [style=none] (34) at (0.175, -1.675) {};
		\node [style=none] (36) at (0, -1.75) {$2n$};
		\node [style=miniBlue] (37) at (1, -1.25) {};
		\node [style=flavourBlue] (38) at (-1, -1.25) {};
		\node [style=none] (39) at (-1, -1.75) {2};
	\end{pgfonlayer}
	\begin{pgfonlayer}{edgelayer}
		\draw (5) to (6);
		\draw (6) to (11);
		\draw (1) to (12.center);
		\draw (13.center) to (5);
		\draw (28.center) to (29.center);
		\draw (30.center) to (31.center);
		\draw (32.center) to (33.center);
		\draw (33.center) to (34.center);
		\draw (38) to (27);
	\end{pgfonlayer}
\end{tikzpicture}}
} 
\label{foldedeven}
\end{equation} 
Explicit computation shows the Coulomb branch of \eqref{foldedeven} to be the moduli space $\overline{\mathcal{O}}^{\mathfrak{sl}(2n)}_{[2^{n}]}$. The unitary quiver with the same Coulomb branch is well-known and takes the form: 
\begin{equation}
\raisebox{-.5\height}{\scalebox{.811}{
    \begin{tikzpicture}
	\begin{pgfonlayer}{nodelayer}
		\node [style=gauge3] (0) at (0, 0) {};
		\node [style=gauge3] (1) at (-1, 0) {};
		\node [style=gauge3] (2) at (1, 0) {};
		\node [style=flavour2] (3) at (0, 1) {};
		\node [style=none] (4) at (0, 1.5) {2};
		\node [style=none] (5) at (0, -0.5) {$n$};
		\node [style=none] (6) at (1, -0.5) {$n{-}1$};
		\node [style=none] (7) at (-1, -0.5) {$n{-}1$};
		\node [style=none] (8) at (1.5, 0) {};
		\node [style=none] (9) at (1.75, 0) {$\dots$};
		\node [style=none] (10) at (2, 0) {};
		\node [style=none] (11) at (-1.5, 0) {};
		\node [style=none] (12) at (-1.75, 0) {$\dots$};
		\node [style=none] (13) at (-2, 0) {};
		\node [style=gauge3] (14) at (2.5, 0) {};
		\node [style=gauge3] (15) at (-2.5, 0) {};
		\node [style=gauge3] (16) at (-3.5, 0) {};
		\node [style=gauge3] (18) at (3.5, 0) {};
		\node [style=none] (19) at (-2.5, -0.5) {2};
		\node [style=none] (20) at (-3.5, -0.5) {1};
		\node [style=none] (21) at (2.5, -0.5) {2};
		\node [style=none] (22) at (3.5, -0.5) {1};
	\end{pgfonlayer}
	\begin{pgfonlayer}{edgelayer}
		\draw (0) to (3);
		\draw (0) to (2);
		\draw (0) to (1);
		\draw (13.center) to (15);
		\draw (11.center) to (1);
		\draw (2) to (8.center);
		\draw (10.center) to (14);
		\draw (15) to (16);
		\draw (14) to (18);
	\end{pgfonlayer}
\end{tikzpicture}} }
\label{evenunitary}
\end{equation}
The Coulomb branches of these quivers share the HWG:
\begin{equation}
    \mathrm{HWG}\eqref{foldedeven}=\mathrm{HWG}\eqref{evenunitary}= \mathrm{PE}\left[\sum_{i=1}^{n}\mu_{i}\mu_{2n-i}t^{2i}\right]
    \label{hwgunitary},
\end{equation}
where $\mu_i$ for $i=1,\dots ,2n-1$ are the highest weight fugacities of $\mathfrak{sl}(2n)$.

When folding quivers, one observes that the creation of a non-simply laced edge leads to a change in the global symmetry. For unitary quivers this is obvious as the Dynkin diagram changes from simply laced to non-simply laced. For these unitary quivers, the action of folding also results in a mapping of the highest weight fugacities \cite{Hanany:2020jzl,Bourget:2020bxh}. As a result, in this class of examples, the HWG of the Coulomb branch of the folded quiver can be inferred if the HWG for the original quiver is known, see also Section \ref{branes}.

By studying \eqref{hwgeven} and \eqref{hwgunitary} one observes that the global symmetry changes from $\mathrm{SO}(4n)$ to $\mathrm{SU}(2n)$. Furthermore, for this family, the action of folding results in the following mapping of highest weight \emph{monomials} for even D-type:
    \begin{subequations}
  \label{fugacitymap1}
    \begin{align}
 \mathrm{For}\; i=1,\dots,n-1 \, , \qquad  (\rho_{2i})_{\mathfrak{so}(4n)} & \rightarrow  (\mu_i\mu_{2n-i})_{\mathfrak{sl}(2n)}, \\
  (\rho_{2n-1})_{\mathfrak{so}(4n)}  , (\rho_{2n})_{\mathfrak{so}(4n)} & \rightarrow (\mu_{n})_{\mathfrak{sl}(2n)},\\
  (\rho_{2n-1}^2)_{\mathfrak{so}(4n)} , (\rho_{2n-1} \rho_{2n})_{\mathfrak{so}(4n)} ,  (\rho_{2n}^2)_{\mathfrak{so}(4n)} & \rightarrow (\mu_{n}^2)_{\mathfrak{sl}(2n)}.
    \end{align}
    \end{subequations}

One can repeat this procedure for the remaining $\mathfrak{so}(4n)$ height 2 orbits: $\overline{\mathcal{O}}^{\mathfrak{so}(4n)}_{[2^{2k},1^{4n-4k}]}$, where $n \ge k \ge 1$. These geometric spaces are given by the Coulomb branches of
\begin{equation}
	\raisebox{-.5\height}{\scalebox{.811}{\begin{tikzpicture}
\begin{pgfonlayer}{nodelayer}
		\node [style=miniU] (0) at (-4.5, 1) {};
		\node [style=miniBlue] (1) at (-3.5, 1) {};
		\node [style=miniBlue] (2) at (-5.5, 1) {};
		\node [style=none] (3) at (4.25, 1) {\dots};
		\node [style=none] (4) at (-6.25, 1) {\dots};
		\node [style=miniU] (5) at (5, 1) {};
		\node [style=miniBlue] (6) at (6, 1) {};
		\node [style=miniBlue] (7) at (-8, 1) {};
		\node [style=miniU] (8) at (-7, 1) {};
		\node [style=miniU] (9) at (-9, 1) {};
		\node [style=miniU] (11) at (7, 1) {};
		\node [style=none] (13) at (4.5, 1) {};
		\node [style=none] (14) at (-6, 1) {};
		\node [style=none] (15) at (-6.5, 1) {};
		\node [style=miniBlue] (26) at (-4.5, 1) {};
		\node [style=miniBlue] (27) at (5, 1) {};
		\node [style=miniBlue] (28) at (7, 1) {};
		\node [style=miniBlue] (29) at (-7, 1) {};
		\node [style=miniBlue] (30) at (-9, 1) {};
		\node [style=miniU] (32) at (-5.5, 1) {};
		\node [style=miniU] (33) at (-3.5, 1) {};
		\node [style=miniU] (34) at (6, 1) {};
		\node [style=miniU] (35) at (-8, 1) {};
		\node [style=miniU] (38) at (-4.5, 1) {};
		\node [style=miniBlue] (39) at (-3.5, 1) {};
		\node [style=miniBlue] (40) at (-5.5, 1) {};
		\node [style=miniU] (41) at (5, 1) {};
		\node [style=miniU] (42) at (-7, 1) {};
		\node [style=miniBlue] (43) at (6, 1) {};
		\node [style=miniBlue] (44) at (-8, 1) {};
		\node [style=miniU] (45) at (7, 1) {};
		\node [style=miniU] (46) at (-9, 1) {};
		\node [style=miniU] (47) at (2.5, 1) {};
		\node [style=miniBlue] (48) at (3.5, 1) {};
		\node [style=none] (49) at (4, 1) {};
		\node [style=none] (50) at (-9, 0.5) {2};
		\node [style=none] (51) at (-8, 0.5) {2};
		\node [style=none] (52) at (-7, 0.5) {4};
		\node [style=miniU] (53) at (-5.5, 1) {};
		\node [style=miniU] (54) at (3.5, 1) {};
		\node [style=miniBlue] (55) at (-4.5, 1) {};
		\node [style=miniBlue] (56) at (2.5, 1) {};
		\node [style=miniU] (57) at (-3.5, 1) {};
		\node [style=miniU] (58) at (1.5, 1) {};
		\node [style=flavourRed] (59) at (-4.5, 2) {};
		\node [style=flavourRed] (60) at (2.5, 2) {};
		\node [style=none] (61) at (2.5, 2.5) {1};
		\node [style=none] (62) at (-4.5, 2.5) {1};
		\node [style=none] (63) at (-2.75, 1) {\dots};
		\node [style=none] (64) at (-3, 1) {};
		\node [style=none] (65) at (-2.5, 1) {};
		\node [style=none] (66) at (-5.5, 0.5) {$2k$};
		\node [style=none] (67) at (-4.5, 0.5) {$2k$};
		\node [style=none] (68) at (2.5, 0.5) {$2k$};
		\node [style=none] (69) at (3.5, 0.5) {$2k$};
		\node [style=none] (70) at (1.5, 0.5) {$2k+1$};
		\node [style=none] (71) at (-3.5, 0.5) {$2k+1$};
		\node [style=none] (72) at (-4.5, 0.25) {};
		\node [style=none] (73) at (2.5, 0.25) {};
		\node [style=none] (74) at (-1, -0.4) {$4n-4k-1$ nodes};
		\node [style=none] (75) at (5, 0.5) {4};
		\node [style=none] (76) at (6, 0.5) {2};
		\node [style=none] (77) at (7, 0.5) {2};
		\node [style=miniBlue] (78) at (-2, 1) {};
		\node [style=none] (79) at (0.75, 1) {\dots};
		\node [style=none] (80) at (1, 1) {};
		\node [style=none] (81) at (0.5, 1) {};
		\node [style=none] (82) at (-2, 0.5) {$2k$};
		\node [style=miniBlue] (83) at (0, 1) {};
		\node [style=miniU] (84) at (-1, 1) {};
		\node [style=none] (85) at (-1, 0.5) {$2k+1$};
		\node [style=none] (86) at (0, 0.5) {$2k$};
	\end{pgfonlayer}
	\begin{pgfonlayer}{edgelayer}
		\draw (2) to (0);
		\draw (0) to (1);
		\draw (5) to (6);
		\draw (8) to (7);
		\draw (7) to (9);
		\draw (6) to (11);
		\draw (15.center) to (8);
		\draw (14.center) to (2);
		\draw (13.center) to (5);
		\draw (48) to (49.center);
		\draw (48) to (47);
		\draw (55) to (59);
		\draw (60) to (56);
		\draw (57) to (64.center);
		\draw (58) to (56);
		\draw [style=brace1] (73.center) to (72.center);
		\draw (78) to (65.center);
		\draw (83) to (81.center);
		\draw (58) to (80.center);
		\draw (84) to (78);
		\draw (84) to (83);
	\end{pgfonlayer}
\end{tikzpicture}}}
\label{smallerorbitseven}
\end{equation}
and their HWGs are given by:
\begin{equation}
    \mathrm{HWG}\eqref{smallerorbitseven}= \mathrm{PE}\left[\sum_{i=1}^{k}\rho_{2i}t^{2i}\right] \,.
    \label{hwgevensmaller}
\end{equation}
After folding of \eqref{smallerorbitseven}, one obtains the quiver
\begin{equation}
\raisebox{-.5\height}{\scalebox{.811}{
\begin{tikzpicture}
	\begin{pgfonlayer}{nodelayer}
		\node [style=miniBlue] (1) at (1, -1.25) {};
		\node [style=none] (3) at (7.25, -1.25) {\dots};
		\node [style=miniU] (5) at (8, -1.25) {};
		\node [style=miniBlue] (6) at (9, -1.25) {};
		\node [style=none] (12) at (7, -1.25) {};
		\node [style=none] (13) at (7.5, -1.25) {};
		\node [style=none] (18) at (1, -1.75) {$2k$};
		\node [style=none] (23) at (8, -1.75) {2};
		\node [style=none] (24) at (9, -1.75) {2};
		\node [style=miniU] (27) at (0, -1.25) {};
		\node [style=none] (28) at (0, -1.15) {};
		\node [style=none] (29) at (1, -1.15) {};
		\node [style=none] (30) at (0, -1.35) {};
		\node [style=none] (31) at (1, -1.35) {};
		\node [style=none] (32) at (0.175, -0.825) {};
		\node [style=none] (33) at (0.675, -1.25) {};
		\node [style=none] (34) at (0.175, -1.675) {};
		\node [style=none] (36) at (0, -1.75) {$2k+1$};
		\node [style=miniBlue] (37) at (1, -1.25) {};
		\node [style=miniU] (41) at (9, -1.25) {};
		\node [style=miniBlue] (44) at (8, -1.25) {};
		\node [style=miniU] (46) at (2, -1.25) {};
		\node [style=none] (47) at (2, -1.75) {$2k+1$};
		\node [style=none] (48) at (2.75, -1.25) {\dots};
		\node [style=none] (49) at (2.5, -1.25) {};
		\node [style=none] (50) at (3, -1.25) {};
		\node [style=miniBlue] (51) at (3.5, -1.25) {};
		\node [style=miniU] (52) at (4.5, -1.25) {};
		\node [style=none] (53) at (3.5, -1.75) {$2k$};
		\node [style=none] (54) at (4.5, -1.75) {$2k+1$};
		\node [style=miniBlue] (55) at (5.5, -1.25) {};
		\node [style=none] (56) at (5.5, -1.75) {$2k$};
		\node [style=flavourRed] (57) at (5.5, -0.25) {};
		\node [style=none] (58) at (5.5, 0.25) {1};
		\node [style=miniU] (59) at (6.5, -1.25) {};
		\node [style=none] (60) at (6.5, -1.75) {$2k$};
		\node [style=none] (61) at (0, -2) {};
		\node [style=none] (62) at (5.5, -2) {};
		\node [style=none] (63) at (2.75, -2.75) {$2n-2k$ nodes};
	\end{pgfonlayer}
	\begin{pgfonlayer}{edgelayer}
		\draw (5) to (6);
		\draw (13.center) to (5);
		\draw (28.center) to (29.center);
		\draw (30.center) to (31.center);
		\draw (32.center) to (33.center);
		\draw (33.center) to (34.center);
		\draw  (46) to (37);
		\draw  (49.center) to (46);
		\draw   (51) to (50.center);
		\draw  (51) to (52);
		\draw  (55) to (52);
		\draw  (57) to (55);
		\draw   (55) to (59);
		\draw (59) to (12.center);
		\draw [style=brace1] (62.center) to (61.center);
	\end{pgfonlayer}
\end{tikzpicture}}
}
\label{smallfoldedeven}
\end{equation}
whose Coulomb branch, after computing its Hilbert series, is found to be the closure of the $\mathfrak{sl}(2n)$ orbit  $\overline{\mathcal{O}}^{\mathfrak{sl}(2n)}_{[2^{k},1^{2n-2k}]}$. The unitary quiver with the same Coulomb branch is:
\begin{equation}
\raisebox{-.5\height}{\scalebox{.811}{
\begin{tikzpicture}
	\begin{pgfonlayer}{nodelayer}
		\node [style=gauge3] (0) at (0, 0) {};
		\node [style=gauge3] (1) at (3.5, 0) {};
		\node [style=gauge3] (2) at (4.5, 0) {};
		\node [style=gauge3] (3) at (-1, 0) {};
		\node [style=none] (4) at (5.25, 0) {\dots};
		\node [style=none] (5) at (5, 0) {};
		\node [style=none] (6) at (5.5, 0) {};
		\node [style=none] (7) at (-1.75, 0) {\dots};
		\node [style=none] (8) at (-2, 0) {};
		\node [style=none] (9) at (-1.5, 0) {};
		\node [style=gauge3] (10) at (6, 0) {};
		\node [style=gauge3] (11) at (-2.5, 0) {};
		\node [style=gauge3] (12) at (-3.5, 0) {};
		\node [style=gauge3] (13) at (7, 0) {};
		\node [style=none] (14) at (7, -0.5) {1};
		\node [style=none] (15) at (6, -0.5) {2};
		\node [style=flavour2] (16) at (0, 1) {};
		\node [style=flavour2] (17) at (3.5, 1) {};
		\node [style=none] (18) at (-3.5, -0.5) {1};
		\node [style=none] (19) at (-2.5, -0.5) {2};
		\node [style=none] (20) at (0, 1.5) {1};
		\node [style=none] (21) at (3.5, 1.5) {1};
		\node [style=gauge3] (22) at (1, 0) {};
		\node [style=gauge3] (23) at (2.5, 0) {};
		\node [style=none] (24) at (1.75, 0) {\dots};
		\node [style=none] (25) at (1.5, 0) {};
		\node [style=none] (26) at (2, 0) {};
		\node [style=none] (27) at (-1, -0.5) {$k-1$};
		\node [style=none] (28) at (0, -0.5) {$k$};
		\node [style=none] (29) at (1, -0.5) {$k$};
		\node [style=none] (30) at (2.5, -0.5) {$k$};
		\node [style=none] (31) at (3.5, -0.5) {$k$};
		\node [style=none] (32) at (4.5, -0.5) {$k-1$};
		\node [style=none] (33) at (3.5, -0.8) {};
		\node [style=none] (34) at (0, -0.8) {};
		\node [style=none] (35) at (1.75, -1.55) {$2n-2k+1$ nodes};
	\end{pgfonlayer}
	\begin{pgfonlayer}{edgelayer}
		\draw (0) to (3);
		\draw (1) to (2);
		\draw (5.center) to (2);
		\draw (6.center) to (10);
		\draw (3) to (9.center);
		\draw (8.center) to (11);
		\draw (16) to (0);
		\draw (1) to (17);
		\draw (11) to (12);
		\draw (10) to (13);
		\draw (0) to (22);
		\draw (23) to (1);
		\draw (25.center) to (22);
		\draw (26.center) to (23);
		\draw [style=brace1] (33.center) to (34.center);
	\end{pgfonlayer}
\end{tikzpicture}}
}
\label{unitarysmall}
\end{equation}
The HWG of \eqref{smallfoldedeven} is therefore:
\begin{equation}
    \mathrm{HWG}\eqref{smallfoldedeven}= \mathrm{HWG}\eqref{unitarysmall}= \mathrm{PE}\left[\sum_{i=1}^{k}\mu_{i}\mu_{2n-i}t^{2i}\right] \,.
    \label{hwgunitarysmall}
\end{equation}
With the absence of spinors in the HWG, one can get from \eqref{hwgevensmaller} to \eqref{hwgunitarysmall} by the mapping (\ref{fugacitymap1}). 

The folding of a quiver whose Coulomb branch has $\mathfrak{so}(4n)$ global symmetry into a quiver whose Coulomb branch is $\mathfrak{sl}(2n)$ global symmetry is not surprising, as in the $k=1$ case it reduces to the following simple observation. One can see this by folding the affine Dynkin quiver of $D_{2n}$ along its vertical symmetry axis
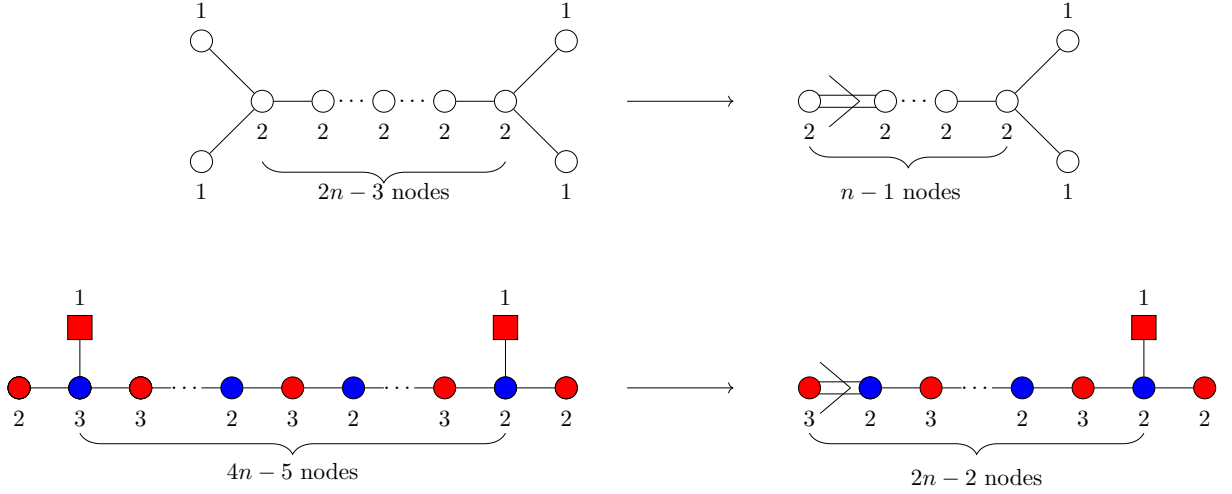
\begin{equation}
\scalebox{0.8}{\begin{tikzpicture}
	\begin{pgfonlayer}{nodelayer}
		\node [style=none] (2) at (4.25, 1.1) {};
		\node [style=none] (3) at (3, 1.1) {};
		\node [style=none] (4) at (4.25, 0.9) {};
		\node [style=none] (5) at (3, 0.9) {};
		\node [style=none] (6) at (3.325, 1.425) {};
		\node [style=none] (7) at (3.825, 1) {};
		\node [style=none] (8) at (3.325, 0.575) {};
		\node [style=none] (14) at (4.25, 0.5) {2};
		\node [style=none] (17) at (3, 0.5) {2};
		\node [style=gauge3] (23) at (3, 1) {};
		\node [style=gauge3] (24) at (4.25, 1) {};
		\node [style=gauge3] (25) at (-2, 1) {};
		\node [style=gauge3] (26) at (-1, 2) {};
		\node [style=gauge3] (27) at (-1, 0) {};
		\node [style=gauge3] (28) at (-3, 1) {};
		\node [style=gauge3] (29) at (-5, 1) {};
		\node [style=none] (30) at (-4.5, 1) {\dots};
		\node [style=gauge3] (31) at (-6, 1) {};
		\node [style=gauge3] (32) at (-7, 2) {};
		\node [style=gauge3] (33) at (-7, 0) {};
		\node [style=none] (34) at (-7, 2.5) {1};
		\node [style=none] (35) at (-7, -0.5) {1};
		\node [style=none] (36) at (-6, 0.5) {2};
		\node [style=none] (37) at (-5, 0.5) {2};
		\node [style=none] (38) at (-3, 0.5) {2};
		\node [style=none] (39) at (-2, 0.5) {2};
		\node [style=none] (40) at (-1, 2.5) {1};
		\node [style=none] (41) at (-1, -0.5) {1};
		\node [style=none] (42) at (0, 1) {};
		\node [style=none] (43) at (1.75, 1) {};
		\node [style=gauge3] (44) at (-4, 1) {};
		\node [style=none] (45) at (-3.5, 1) {\dots};
		\node [style=none] (46) at (4.75, 1) {\dots};
		\node [style=gauge3] (47) at (6.25, 1) {};
		\node [style=gauge3] (48) at (7.25, 2) {};
		\node [style=gauge3] (49) at (7.25, 0) {};
		\node [style=none] (50) at (7.25, 2.5) {1};
		\node [style=none] (51) at (7.25, -0.5) {1};
		\node [style=none] (52) at (6.25, 0.5) {2};
		\node [style=gauge3] (54) at (5.25, 1) {};
		\node [style=none] (55) at (5.25, 0.5) {2};
		\node [style=none] (56) at (-6, 0) {};
		\node [style=none] (57) at (-2, 0) {};
		\node [style=none] (58) at (3, 0.25) {};
		\node [style=none] (59) at (6.25, 0.25) {};
		\node [style=none] (60) at (-4, 0.5) {2};
		\node [style=none] (61) at (-4, -0.5) {$2n-3$ nodes};
		\node [style=none] (62) at (4.5, -0.5) {$n-1$ nodes};
		\node [style=miniU] (63) at (-9, -3.75) {};
		\node [style=miniBlue] (64) at (-8, -3.75) {};
		\node [style=miniBlue] (65) at (-10, -3.75) {};
		\node [style=miniBlue] (77) at (-9, -3.75) {};
		\node [style=miniU] (82) at (-10, -3.75) {};
		\node [style=miniU] (83) at (-8, -3.75) {};
		\node [style=miniU] (86) at (-9, -3.75) {};
		\node [style=miniBlue] (87) at (-8, -3.75) {};
		\node [style=miniBlue] (88) at (-10, -3.75) {};
		\node [style=miniU] (95) at (-2, -3.75) {};
		\node [style=miniBlue] (96) at (-1, -3.75) {};
		\node [style=miniU] (101) at (-10, -3.75) {};
		\node [style=miniU] (102) at (-1, -3.75) {};
		\node [style=miniBlue] (103) at (-9, -3.75) {};
		\node [style=miniBlue] (104) at (-2, -3.75) {};
		\node [style=miniU] (105) at (-8, -3.75) {};
		\node [style=miniU] (106) at (-3, -3.75) {};
		\node [style=flavourRed] (107) at (-9, -2.75) {};
		\node [style=flavourRed] (108) at (-2, -2.75) {};
		\node [style=none] (109) at (-2, -2.25) {1};
		\node [style=none] (110) at (-9, -2.25) {1};
		\node [style=none] (111) at (-7.25, -3.75) {\dots};
		\node [style=none] (112) at (-7.5, -3.75) {};
		\node [style=none] (113) at (-7, -3.75) {};
		\node [style=none] (114) at (-10, -4.25) {2};
		\node [style=none] (115) at (-9, -4.25) {3};
		\node [style=none] (116) at (-2, -4.25) {2};
		\node [style=none] (117) at (-1, -4.25) {2};
		\node [style=none] (118) at (-3, -4.25) {3};
		\node [style=none] (119) at (-8, -4.25) {3};
		\node [style=none] (120) at (-9, -4.5) {};
		\node [style=none] (121) at (-2, -4.5) {};
		\node [style=none] (122) at (-5.5, -5.15) {$4n-5$ nodes};
		\node [style=miniBlue] (126) at (-6.5, -3.75) {};
		\node [style=none] (127) at (-3.75, -3.75) {\dots};
		\node [style=none] (128) at (-3.5, -3.75) {};
		\node [style=none] (129) at (-4, -3.75) {};
		\node [style=none] (130) at (-6.5, -4.25) {2};
		\node [style=miniBlue] (131) at (-4.5, -3.75) {};
		\node [style=miniU] (132) at (-5.5, -3.75) {};
		\node [style=none] (133) at (-5.5, -4.25) {3};
		\node [style=none] (134) at (-4.5, -4.25) {2};
		\node [style=miniBlue] (135) at (4, -3.75) {};
		\node [style=none] (141) at (4, -4.25) {2};
		\node [style=miniU] (144) at (3, -3.75) {};
		\node [style=none] (145) at (3, -3.65) {};
		\node [style=none] (146) at (4, -3.65) {};
		\node [style=none] (147) at (3, -3.85) {};
		\node [style=none] (148) at (4, -3.85) {};
		\node [style=none] (149) at (3.175, -3.325) {};
		\node [style=none] (150) at (3.675, -3.75) {};
		\node [style=none] (151) at (3.175, -4.175) {};
		\node [style=none] (152) at (3, -4.25) {3};
		\node [style=miniBlue] (153) at (4, -3.75) {};
		\node [style=miniU] (156) at (5, -3.75) {};
		\node [style=none] (157) at (5, -4.25) {3};
		\node [style=none] (158) at (5.75, -3.75) {\dots};
		\node [style=none] (159) at (5.5, -3.75) {};
		\node [style=none] (160) at (6, -3.75) {};
		\node [style=miniBlue] (161) at (6.5, -3.75) {};
		\node [style=miniU] (162) at (7.5, -3.75) {};
		\node [style=none] (163) at (6.5, -4.25) {2};
		\node [style=none] (164) at (7.5, -4.25) {3};
		\node [style=miniBlue] (165) at (8.5, -3.75) {};
		\node [style=none] (166) at (8.5, -4.25) {2};
		\node [style=flavourRed] (167) at (8.5, -2.75) {};
		\node [style=none] (168) at (8.5, -2.25) {1};
		\node [style=miniU] (169) at (9.5, -3.75) {};
		\node [style=none] (170) at (9.5, -4.25) {2};
		\node [style=none] (171) at (3, -4.5) {};
		\node [style=none] (172) at (8.5, -4.5) {};
		\node [style=none] (173) at (5.75, -5.25) {$2n-2$ nodes};
		\node [style=none] (174) at (0, -3.75) {};
		\node [style=none] (175) at (1.75, -3.75) {};
	\end{pgfonlayer}
	\begin{pgfonlayer}{edgelayer}
		\draw (2.center) to (3.center);
		\draw (4.center) to (5.center);
		\draw (6.center) to (7.center);
		\draw (7.center) to (8.center);
		\draw (32) to (31);
		\draw (31) to (33);
		\draw (31) to (29);
		\draw (28) to (25);
		\draw (25) to (26);
		\draw (25) to (27);
		\draw [style=->] (42.center) to (43.center);
		\draw (47) to (48);
		\draw (47) to (49);
		\draw (47) to (54);
		\draw [style=brace1] (57.center) to (56.center);
		\draw [style=brace1] (59.center) to (58.center);
		\draw (65) to (63);
		\draw (63) to (64);
		\draw (96) to (95);
		\draw (103) to (107);
		\draw (108) to (104);
		\draw (105) to (112.center);
		\draw (106) to (104);
		\draw [style=brace1] (121.center) to (120.center);
		\draw (126) to (113.center);
		\draw (131) to (129.center);
		\draw (106) to (128.center);
		\draw (132) to (126);
		\draw (132) to (131);
		\draw (145.center) to (146.center);
		\draw (147.center) to (148.center);
		\draw (149.center) to (150.center);
		\draw (150.center) to (151.center);
		\draw (156) to (153);
		\draw (159.center) to (156);
		\draw (161) to (160.center);
		\draw (161) to (162);
		\draw (165) to (162);
		\draw (167) to (165);
		\draw (165) to (169);
		\draw [style=brace1] (172.center) to (171.center);
		\draw [style=->] (174.center) to (175.center);
	\end{pgfonlayer}
\end{tikzpicture}}
\end{equation}
where the diagram on the top right is the twisted affine Dynkin quiver of $A^{(2)}_{2n-1}$ \footnote{We follow the labelling of Kac \cite{Kac:1994} for twisted affine algebras, as this predicts the Coulomb branch of the balanced quiver.} whose Coulomb branch is the minimal nilpotent orbit closure of $\mathfrak{sl}(2n)$. In this case, one observes that folding either a unitary quiver or an orthosymplectic quiver, whose Coulomb branch is the closure of the minimal $\mathfrak{so}(4n)$ orbit, produces a quiver whose Coulomb branch is the minimal orbit closure of $\mathfrak{sl}(2n)$. The difficulty in reproducing this procedure for other unitary quivers whose Coulomb branch are height 2 nilpotent orbits of $\mathfrak{so}(4n)$ is that they do not have identical legs to fold.

\paragraph{Odd D-type.} 
For $\overline{\mathcal{O}}^{\mathfrak{so}(4n+2)}_{[2^{2n},1^2]}$,  there is a single cone. The moduli space is given by the Coulomb branch of:
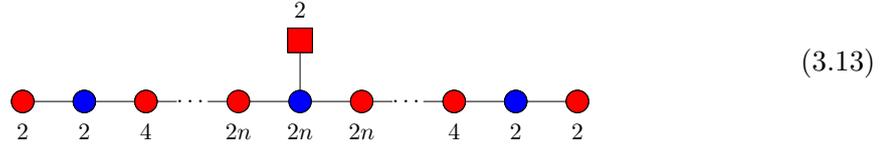
\begin{equation}
\raisebox{-.5\height}{\scalebox{.811}{
\begin{tikzpicture}
	\begin{pgfonlayer}{nodelayer}
		\node [style=miniU] (0) at (0, 1) {};
		\node [style=miniBlue] (1) at (1, 1) {};
		\node [style=miniBlue] (2) at (-1, 1) {};
		\node [style=none] (3) at (1.75, 1) {\dots};
		\node [style=none] (4) at (-1.75, 1) {\dots};
		\node [style=miniU] (5) at (2.5, 1) {};
		\node [style=miniBlue] (6) at (3.5, 1) {};
		\node [style=miniBlue] (7) at (-3.5, 1) {};
		\node [style=miniU] (8) at (-2.5, 1) {};
		\node [style=miniU] (9) at (-4.5, 1) {};
		\node [style=flavourBlue] (10) at (0, 2) {};
		\node [style=miniU] (11) at (4.5, 1) {};
		\node [style=none] (12) at (1.5, 1) {};
		\node [style=none] (13) at (2, 1) {};
		\node [style=none] (14) at (-1.5, 1) {};
		\node [style=none] (15) at (-2, 1) {};
		\node [style=none] (16) at (0, 2.5) {2};
		\node [style=none] (17) at (0, 0.5) {$2n$};
		\node [style=none] (18) at (1, 0.5) {$2n$};
		\node [style=none] (19) at (-1, 0.5) {$2n$};
		\node [style=none] (20) at (-2.5, 0.5) {4};
		\node [style=none] (21) at (-3.5, 0.5) {2};
		\node [style=none] (22) at (-4.5, 0.5) {2};
		\node [style=none] (23) at (2.5, 0.5) {4};
		\node [style=none] (24) at (3.5, 0.5) {2};
		\node [style=none] (25) at (4.5, 0.5) {2};
		\node [style=miniBlue] (26) at (0, 1) {};
		\node [style=miniBlue] (27) at (2.5, 1) {};
		\node [style=miniBlue] (28) at (4.5, 1) {};
		\node [style=miniBlue] (29) at (-2.5, 1) {};
		\node [style=miniBlue] (30) at (-4.5, 1) {};
		\node [style=miniU] (32) at (-1, 1) {};
		\node [style=miniU] (33) at (1, 1) {};
		\node [style=miniU] (34) at (3.5, 1) {};
		\node [style=miniU] (35) at (-3.5, 1) {};
		\node [style=flavourRed] (36) at (0, 2) {};
		\node [style=miniU] (37) at (-2.5, 1) {};
		\node [style=miniU] (38) at (-4.5, 1) {};
		\node [style=miniU] (39) at (4.5, 1) {};
		\node [style=miniU] (40) at (2.5, 1) {};
		\node [style=miniBlue] (41) at (3.5, 1) {};
		\node [style=miniBlue] (42) at (-3.5, 1) {};
	\end{pgfonlayer}
	\begin{pgfonlayer}{edgelayer}
		\draw (2) to (0);
		\draw (0) to (10);
		\draw (0) to (1);
		\draw (5) to (6);
		\draw (8) to (7);
		\draw (7) to (9);
		\draw (6) to (11);
		\draw (15.center) to (8);
		\draw (14.center) to (2);
		\draw (1) to (12.center);
		\draw (13.center) to (5);
	\end{pgfonlayer}
\end{tikzpicture}}
}
\label{beforefoldodd}
\end{equation}
with the following HWG:
\begin{equation}
    \mathrm{HWG}\eqref{beforefoldodd}=\mathrm{PE}\left[\sum_{i=1}^{n-1} \rho_{2i} t^{2i} + \rho_{2n}\rho_{2n+1}t^{2n} \right] \,.
    \label{hwgbefore}
\end{equation}
Folding the quiver \eqref{beforefoldodd} gives:
\begin{equation}
\raisebox{-.5\height}{\scalebox{.811}{
\begin{tikzpicture}
	\begin{pgfonlayer}{nodelayer}
		\node [style=miniBlue] (1) at (1, -1.25) {};
		\node [style=none] (3) at (1.75, -1.25) {\dots};
		\node [style=miniU] (5) at (2.5, -1.25) {};
		\node [style=miniBlue] (6) at (3.5, -1.25) {};
		\node [style=none] (12) at (1.5, -1.25) {};
		\node [style=none] (13) at (2, -1.25) {};
		\node [style=none] (18) at (1, -1.75) {$2n$};
		\node [style=none] (23) at (2.5, -1.75) {2};
		\node [style=none] (24) at (3.5, -1.75) {2};
		\node [style=miniU] (27) at (0, -1.25) {};
		\node [style=none] (28) at (0, -1.15) {};
		\node [style=none] (29) at (1, -1.15) {};
		\node [style=none] (30) at (0, -1.35) {};
		\node [style=none] (31) at (1, -1.35) {};
		\node [style=none] (32) at (0.175, -0.825) {};
		\node [style=none] (33) at (0.675, -1.25) {};
		\node [style=none] (34) at (0.175, -1.675) {};
		\node [style=none] (36) at (0, -1.75) {$2n$};
		\node [style=miniBlue] (37) at (1, -1.25) {};
		\node [style=none] (39) at (-1, -1.75) {2};
		\node [style=flavourRed] (40) at (-1, -1.25) {};
		\node [style=miniU] (41) at (3.5, -1.25) {};
		\node [style=miniU] (42) at (1, -1.25) {};
		\node [style=miniBlue] (43) at (0, -1.25) {};
		\node [style=miniBlue] (44) at (2.5, -1.25) {};
	\end{pgfonlayer}
	\begin{pgfonlayer}{edgelayer}
		\draw (5) to (6);
		\draw (1) to (12.center);
		\draw (13.center) to (5);
		\draw (28.center) to (29.center);
		\draw (30.center) to (31.center);
		\draw (32.center) to (33.center);
		\draw (33.center) to (34.center);
		\draw  (43) to (40);
	\end{pgfonlayer}
\end{tikzpicture} }
}
\label{foldedodd}
\end{equation} 
The Coulomb branch of \eqref{foldedodd} turns out to be $\overline{\mathcal{O}}^{\mathfrak{sl}(2n+1)}_{[2^{n},1]}$. The unitary quiver counterpart with the same Coulomb branch takes the form:
\begin{equation}
\raisebox{-.5\height}{\scalebox{.811}{
 \begin{tikzpicture}
	\begin{pgfonlayer}{nodelayer}
		\node [style=gauge3] (0) at (0, 0) {};
		\node [style=gauge3] (1) at (1, 0) {};
		\node [style=gauge3] (2) at (2, 0) {};
		\node [style=gauge3] (3) at (-1, 0) {};
		\node [style=none] (4) at (2.75, 0) {\dots};
		\node [style=none] (5) at (2.5, 0) {};
		\node [style=none] (6) at (3, 0) {};
		\node [style=none] (7) at (-1.75, 0) {\dots};
		\node [style=none] (8) at (-2, 0) {};
		\node [style=none] (9) at (-1.5, 0) {};
		\node [style=gauge3] (10) at (3.5, 0) {};
		\node [style=gauge3] (11) at (-2.5, 0) {};
		\node [style=gauge3] (12) at (-3.5, 0) {};
		\node [style=gauge3] (13) at (4.5, 0) {};
		\node [style=none] (14) at (4.5, -0.5) {1};
		\node [style=none] (15) at (3.5, -0.5) {2};
		\node [style=flavour2] (16) at (0, 1) {};
		\node [style=flavour2] (17) at (1, 1) {};
		\node [style=none] (18) at (-3.5, -0.5) {1};
		\node [style=none] (19) at (-2.5, -0.5) {2};
		\node [style=none] (20) at (0, 1.5) {1};
		\node [style=none] (21) at (1, 1.5) {1};
		\node [style=none] (22) at (0, -0.5) {$n$};
		\node [style=none] (23) at (1, -0.5) {$n$};
		\node [style=none] (24) at (2, -0.5) {$n-1$};
		\node [style=none] (25) at (-1, -0.5) {$n-1$};
	\end{pgfonlayer}
	\begin{pgfonlayer}{edgelayer}
		\draw (0) to (3);
		\draw (1) to (0);
		\draw (1) to (2);
		\draw (5.center) to (2);
		\draw (6.center) to (10);
		\draw (3) to (9.center);
		\draw (8.center) to (11);
		\draw (16) to (0);
		\draw (1) to (17);
		\draw (11) to (12);
		\draw (10) to (13);
	\end{pgfonlayer}
\end{tikzpicture} }
\label{unitary6}
}
\end{equation}
The HWG of the folded quiver is:
\begin{equation}
    \mathrm{HWG}\eqref{foldedodd}=\mathrm{HWG}\eqref{unitary6}=\mathrm{PE}\left[\sum_{i=1}^{n}\mu_{i}\mu_{2n+1-i}t^{2i} \right] \,.
    \label{hwgafter}
\end{equation}
By comparing \eqref{hwgbefore} and \eqref{hwgafter}, one observes that there is again a mapping of the highest weight monomials of $\mathfrak{so}(4n+2)$ to $\mathfrak{sl}(2n+1)$:
\begin{subequations}
\label{fugacitymap2}
   \begin{align}
  (\rho_{2i})_{\mathfrak{so}(4n+2)} &\rightarrow  (\mu_i\mu_{2n+1-i})_{\mathfrak{sl}(2n+1)} \; \; ,\mathrm{for}\; i=1,\dots,n-1 \\
  (\rho_{2n})_{\mathfrak{so}(4n+2)} &\rightarrow (\mu_{n})_{\mathfrak{sl}(2n+1)} \\
  (\rho_{2n+1})_{\mathfrak{so}(4n+2)} &\rightarrow (\mu_{n+1})_{\mathfrak{sl}(2n+1)} \\
  (\rho_{2n} \rho_{2n+1})_{\mathfrak{so}(4n+2)} &\rightarrow (\mu_{n} \mu_{n+1})_{\mathfrak{sl}(2n+1)} 
    \end{align}
    \end{subequations}
for odd D-type.

As with the even D-type case, one can repeat the same folding procedure for the remaining height 2 orbits  $\overline{\mathcal{O}}^{\mathfrak{so}(4n+2)}_{[2^{2k},1^{4n-4k+2}]}$, for $n \ge k \ge 1$. This moduli space is given by the Coulomb branch of
\begin{equation}
\raisebox{-.5\height}{\scalebox{0.811}{
\begin{tikzpicture}
	\begin{pgfonlayer}{nodelayer}
		\node [style=miniU] (0) at (-4.5, 1) {};
		\node [style=miniBlue] (1) at (-3.5, 1) {};
		\node [style=miniBlue] (2) at (-5.5, 1) {};
		\node [style=none] (3) at (4.25, 1) {\dots};
		\node [style=none] (4) at (-6.25, 1) {\dots};
		\node [style=miniU] (5) at (5, 1) {};
		\node [style=miniBlue] (6) at (6, 1) {};
		\node [style=miniBlue] (7) at (-8, 1) {};
		\node [style=miniU] (8) at (-7, 1) {};
		\node [style=miniU] (9) at (-9, 1) {};
		\node [style=miniU] (11) at (7, 1) {};
		\node [style=none] (13) at (4.5, 1) {};
		\node [style=none] (14) at (-6, 1) {};
		\node [style=none] (15) at (-6.5, 1) {};
		\node [style=miniBlue] (26) at (-4.5, 1) {};
		\node [style=miniBlue] (27) at (5, 1) {};
		\node [style=miniBlue] (28) at (7, 1) {};
		\node [style=miniBlue] (29) at (-7, 1) {};
		\node [style=miniBlue] (30) at (-9, 1) {};
		\node [style=miniU] (32) at (-5.5, 1) {};
		\node [style=miniU] (33) at (-3.5, 1) {};
		\node [style=miniU] (34) at (6, 1) {};
		\node [style=miniU] (35) at (-8, 1) {};
		\node [style=miniU] (38) at (-4.5, 1) {};
		\node [style=miniBlue] (39) at (-3.5, 1) {};
		\node [style=miniBlue] (40) at (-5.5, 1) {};
		\node [style=miniU] (41) at (5, 1) {};
		\node [style=miniU] (42) at (-7, 1) {};
		\node [style=miniBlue] (43) at (6, 1) {};
		\node [style=miniBlue] (44) at (-8, 1) {};
		\node [style=miniU] (45) at (7, 1) {};
		\node [style=miniU] (46) at (-9, 1) {};
		\node [style=miniU] (47) at (2.5, 1) {};
		\node [style=miniBlue] (48) at (3.5, 1) {};
		\node [style=none] (49) at (4, 1) {};
		\node [style=none] (50) at (-9, 0.5) {2};
		\node [style=none] (51) at (-8, 0.5) {2};
		\node [style=none] (52) at (-7, 0.5) {4};
		\node [style=miniU] (53) at (-5.5, 1) {};
		\node [style=miniU] (54) at (3.5, 1) {};
		\node [style=miniBlue] (55) at (-4.5, 1) {};
		\node [style=miniBlue] (56) at (2.5, 1) {};
		\node [style=miniU] (57) at (-3.5, 1) {};
		\node [style=miniU] (58) at (1.5, 1) {};
		\node [style=flavourRed] (59) at (-4.5, 2) {};
		\node [style=flavourRed] (60) at (2.5, 2) {};
		\node [style=none] (61) at (2.5, 2.5) {1};
		\node [style=none] (62) at (-4.5, 2.5) {1};
		\node [style=none] (63) at (-2.75, 1) {\dots};
		\node [style=none] (64) at (-3, 1) {};
		\node [style=none] (65) at (-2.5, 1) {};
		\node [style=none] (66) at (-5.5, 0.5) {$2k$};
		\node [style=none] (67) at (-4.5, 0.5) {$2k$};
		\node [style=none] (68) at (2.5, 0.5) {$2k$};
		\node [style=none] (69) at (3.5, 0.5) {$2k$};
		\node [style=none] (70) at (1.5, 0.5) {$2k+1$};
		\node [style=none] (71) at (-3.5, 0.5) {$2k+1$};
		\node [style=none] (72) at (-4.5, 0.25) {};
		\node [style=none] (73) at (2.5, 0.25) {};
		\node [style=none] (74) at (-1, -0.4) {$4n+1-4k$ nodes};
		\node [style=none] (75) at (5, 0.5) {4};
		\node [style=none] (76) at (6, 0.5) {2};
		\node [style=none] (77) at (7, 0.5) {2};
		\node [style=miniBlue] (78) at (-2, 1) {};
		\node [style=none] (79) at (0.75, 1) {\dots};
		\node [style=none] (80) at (1, 1) {};
		\node [style=none] (81) at (0.5, 1) {};
		\node [style=none] (82) at (-2, 0.5) {$2k+1$};
		\node [style=miniBlue] (83) at (0, 1) {};
		\node [style=miniU] (84) at (-1, 1) {};
		\node [style=none] (85) at (-1, 0.5) {$2k$};
		\node [style=none] (86) at (0, 0.5) {$2k+1$};
		\node [style=miniU] (87) at (-2, 1) {};
		\node [style=miniU] (88) at (0, 1) {};
		\node [style=miniBlue] (89) at (-1, 1) {};
	\end{pgfonlayer}
	\begin{pgfonlayer}{edgelayer}
		\draw (2) to (0);
		\draw (0) to (1);
		\draw (5) to (6);
		\draw (8) to (7);
		\draw (7) to (9);
		\draw (6) to (11);
		\draw (15.center) to (8);
		\draw (14.center) to (2);
		\draw (13.center) to (5);
		\draw (48) to (49.center);
		\draw (48) to (47);
		\draw (55) to (59);
		\draw (60) to (56);
		\draw (57) to (64.center);
		\draw (58) to (56);
		\draw [style=brace1] (73.center) to (72.center);
		\draw (78) to (65.center);
		\draw (83) to (81.center);
		\draw (58) to (80.center);
		\draw (84) to (78);
		\draw (84) to (83);
	\end{pgfonlayer}
\end{tikzpicture}}}
\label{smallerorbitsodd}
\end{equation}
with the HWG being the same as \eqref{hwgevensmaller}. After folding of \eqref{smallerorbitsodd}, one obtains
\begin{equation}
\raisebox{-.5\height}{\scalebox{.811}{
\begin{tikzpicture}
	\begin{pgfonlayer}{nodelayer}
		\node [style=miniBlue] (1) at (1, -1.25) {};
		\node [style=none] (3) at (7.25, -1.25) {\dots};
		\node [style=miniU] (5) at (8, -1.25) {};
		\node [style=miniBlue] (6) at (9, -1.25) {};
		\node [style=none] (12) at (7, -1.25) {};
		\node [style=none] (13) at (7.5, -1.25) {};
		\node [style=none] (18) at (1, -1.75) {$2k+1$};
		\node [style=none] (23) at (8, -1.75) {2};
		\node [style=none] (24) at (9, -1.75) {2};
		\node [style=miniU] (27) at (0, -1.25) {};
		\node [style=none] (28) at (0, -1.15) {};
		\node [style=none] (29) at (1, -1.15) {};
		\node [style=none] (30) at (0, -1.35) {};
		\node [style=none] (31) at (1, -1.35) {};
		\node [style=none] (32) at (0.175, -0.825) {};
		\node [style=none] (33) at (0.675, -1.25) {};
		\node [style=none] (34) at (0.175, -1.675) {};
		\node [style=none] (36) at (0, -1.75) {$2k$};
		\node [style=miniBlue] (37) at (1, -1.25) {};
		\node [style=miniU] (41) at (9, -1.25) {};
		\node [style=miniBlue] (44) at (8, -1.25) {};
		\node [style=miniU] (46) at (2, -1.25) {};
		\node [style=none] (47) at (2, -1.75) {$2k$};
		\node [style=none] (48) at (2.75, -1.25) {\dots};
		\node [style=none] (49) at (2.5, -1.25) {};
		\node [style=none] (50) at (3, -1.25) {};
		\node [style=miniBlue] (51) at (3.5, -1.25) {};
		\node [style=miniU] (52) at (4.5, -1.25) {};
		\node [style=none] (53) at (3.5, -1.75) {$2k$};
		\node [style=none] (54) at (4.5, -1.75) {$2k+1$};
		\node [style=miniBlue] (55) at (5.5, -1.25) {};
		\node [style=none] (56) at (5.5, -1.75) {$2k$};
		\node [style=flavourRed] (57) at (5.5, -0.25) {};
		\node [style=none] (58) at (5.5, 0.25) {1};
		\node [style=miniU] (59) at (6.5, -1.25) {};
		\node [style=none] (60) at (6.5, -1.75) {$2k$};
		\node [style=none] (61) at (0, -2) {};
		\node [style=none] (62) at (5.5, -2) {};
		\node [style=none] (63) at (2.75, -2.55) {$2n+1-2k$ nodes};
		\node [style=miniBlue] (64) at (0, -1.25) {};
		\node [style=miniBlue] (65) at (2, -1.25) {};
		\node [style=miniU] (66) at (1, -1.25) {};
	\end{pgfonlayer}
	\begin{pgfonlayer}{edgelayer}
		\draw (5) to (6);
		\draw (13.center) to (5);
		\draw (28.center) to (29.center);
		\draw (30.center) to (31.center);
		\draw (32.center) to (33.center);
		\draw (33.center) to (34.center);
		\draw  (46) to (37);
		\draw  (49.center) to (46);
		\draw   (51) to (50.center);
		\draw  (51) to (52);
		\draw  (55) to (52);
		\draw   (57) to (55);
		\draw   (55) to (59);
		\draw   (59) to (12.center);
		\draw  [style=brace] (62.center) to (61.center);
	\end{pgfonlayer}
\end{tikzpicture}}
}
\label{foldedoddd}
\end{equation}
whose Coulomb branch is the moduli space  $\overline{\mathcal{O}}^{\mathfrak{sl}(2n+1)}_{[2^{k},1^{2n+1-2k}]}$. The unitary quiver with the same Coulomb branch is therefore:
\begin{equation}
\raisebox{-.5\height}{\scalebox{0.811}{
\begin{tikzpicture}
	\begin{pgfonlayer}{nodelayer}
		\node [style=gauge3] (0) at (0, 0) {};
		\node [style=gauge3] (1) at (3.5, 0) {};
		\node [style=gauge3] (2) at (4.5, 0) {};
		\node [style=gauge3] (3) at (-1, 0) {};
		\node [style=none] (4) at (5.25, 0) {\dots};
		\node [style=none] (5) at (5, 0) {};
		\node [style=none] (6) at (5.5, 0) {};
		\node [style=none] (7) at (-1.75, 0) {\dots};
		\node [style=none] (8) at (-2, 0) {};
		\node [style=none] (9) at (-1.5, 0) {};
		\node [style=gauge3] (10) at (6, 0) {};
		\node [style=gauge3] (11) at (-2.5, 0) {};
		\node [style=gauge3] (12) at (-3.5, 0) {};
		\node [style=gauge3] (13) at (7, 0) {};
		\node [style=none] (14) at (7, -0.5) {1};
		\node [style=none] (15) at (6, -0.5) {2};
		\node [style=flavour2] (16) at (0, 1) {};
		\node [style=flavour2] (17) at (3.5, 1) {};
		\node [style=none] (18) at (-3.5, -0.5) {1};
		\node [style=none] (19) at (-2.5, -0.5) {2};
		\node [style=none] (20) at (0, 1.5) {1};
		\node [style=none] (21) at (3.5, 1.5) {1};
		\node [style=gauge3] (22) at (1, 0) {};
		\node [style=gauge3] (23) at (2.5, 0) {};
		\node [style=none] (24) at (1.75, 0) {\dots};
		\node [style=none] (25) at (1.5, 0) {};
		\node [style=none] (26) at (2, 0) {};
		\node [style=none] (27) at (-1, -0.5) {$k-1$};
		\node [style=none] (28) at (0, -0.5) {$k$};
		\node [style=none] (29) at (1, -0.5) {$k$};
		\node [style=none] (30) at (2.5, -0.5) {$k$};
		\node [style=none] (31) at (3.5, -0.5) {$k$};
		\node [style=none] (32) at (4.5, -0.5) {$k-1$};
		\node [style=none] (33) at (3.5, -0.8) {};
		\node [style=none] (34) at (0, -0.8) {};
		\node [style=none] (35) at (1.75, -1.55) {$2n-2k+2$ nodes};
	\end{pgfonlayer}
	\begin{pgfonlayer}{edgelayer}
		\draw (0) to (3);
		\draw (1) to (2);
		\draw (5.center) to (2);
		\draw (6.center) to (10);
		\draw (3) to (9.center);
		\draw (8.center) to (11);
		\draw (16) to (0);
		\draw (1) to (17);
		\draw (11) to (12);
		\draw (10) to (13);
		\draw (0) to (22);
		\draw (23) to (1);
		\draw (25.center) to (22);
		\draw (26.center) to (23);
		\draw [style=brace1] (33.center) to (34.center);
	\end{pgfonlayer}
\end{tikzpicture}
}}
\label{unitarysmall2}
\end{equation}
with the HWG 
\begin{equation}
    \mathrm{HWG}\eqref{foldedoddd}= \mathrm{HWG}\eqref{unitarysmall2}= \mathrm{PE}\left[\sum_{i=1}^{k}\mu_{i}\mu_{2n+1-i}t^{2i}\right] \,.
    \label{hwgunitarysmall1}
\end{equation}

\paragraph{Comment.}
For orthosymplectic quivers whose Coulomb branch are closures of nilpotent orbits of classical algebras, only the height 2 orbits of $\mathfrak{so}(2n)$ have symmetric quivers that can be folded. For orthosymplectic quivers whose Coulomb branch are closures of height 2  orbits of $\mathfrak{so}(2n{+}1)$, the quivers  have \emph{orthogonal} gauge group(s) on one of the legs, making the quiver asymmetric. For closures of height 2 orbits of $\mathfrak{usp}(2n)$, the quiver legs contain `bad' nodes in the sense of \cite{Gaiotto:2008ak} that cause the monopole formula to diverge. Therefore, this section provides an exhaustive list of orthosymplectic quivers that are closures of nilpotent orbits, which can both be folded and have their Hilbert series computed with the monopole formula.

\subsection{Coulomb branch global symmetry}
For quivers composed of unitary gauge groups, it is easy to read off (in most cases \footnote{One does observe, however, that more complicated quivers such as moduli space of $k$-instantons \cite{Cremonesi:2014xha} and some non-simply laced unitary quivers \cite{Bourget:2020mez} have factors in $\mathfrak{g}_\mathrm{global}$ which cannot be read off from the balance of gauge groups. In such cases, the best way to obtain the global symmetry group is an explicit computation of the Hilbert series to order $t^2$ which reveals the dimension of the global symmetry group.}) the algebra $\mathfrak{g}_\mathrm{global}$ of the global symmetry group by studying the balance of the gauge groups. A $\mathrm{U}(k)$ gauge group is balanced if the flavour from the neighboring nodes is $N_f=2k$. The balanced nodes form the Dynkin diagram of $\mathfrak{h}$ which is a subalgebra of $\mathfrak{g}_\mathrm{global}$. In most cases where all the gauge nodes are balanced with $n$ nodes overbalanced, one finds that $\mathfrak{g}_\mathrm{global}=\prod_i \mathfrak{h}_i \times \mathfrak{u}(1)^n$ which considers all balanced subset of nodes that are connected and form Dynkin diagrams $\mathfrak{h}_i$. If the unitary quiver is unframed, then an overall $\mathfrak{u}(1)$ factor needs to be removed from the global symmetry. 

The same idea can be carried on for orthosymplectic quivers. In \cite{Gaiotto:2008ak}, the balance conditions for (special) orthogonal and symplectic gauge groups with $N_f$ fundamental hypermultiplets (i.e.\ $2N_f$ half-hypermultiplets) are as follows: 
\begin{equation}
\begin{split}
    \mathrm{SO}(2k): N_f & = 2k - 1, \\ 
    \mathrm{SO}(2k+1): N_f & = 2k , \\ 
    \mathrm{USp}(2k) : N_f & = 2k + 1. \\
\end{split}
\end{equation}
It has been shown in \cite{Gaiotto:2008ak} that a linear chain of $n$ balanced orthosymplectic gauge nodes gives a global symmetry of $\mathfrak{so}(n+1)$. To read this full global symmetry, it may be necessary to add balanced $\mathrm{USp}(0)$ nodes, but these are omitted in the drawings as they do not contribute to the Coulomb branch computations. The above is true regardless of the gauge groups being O or SO, noting that $\mathrm{USp}(0)$ nodes should not be attached to $\mathrm{O}(2)$ nodes.

Building on the investigation of non-simply laced orthosymplectic quivers, the following subsets of balanced nodes:
\begin{equation}
\raisebox{-.5\height}{\scalebox{0.811}{
    \begin{tikzpicture}
	\begin{pgfonlayer}{nodelayer}
		\node [style=miniBlue] (1) at (1, -1.25) {};
		\node [style=none] (3) at (1.75, -1.25) {\dots};
		\node [style=miniU] (5) at (2.5, -1.25) {};
		\node [style=miniBlue] (6) at (4.5, -1.25) {};
		\node [style=none] (12) at (1.5, -1.25) {};
		\node [style=none] (13) at (2, -1.25) {};
		\node [style=miniU] (27) at (0, -1.25) {};
		\node [style=none] (28) at (0, -1.15) {};
		\node [style=none] (29) at (1, -1.15) {};
		\node [style=none] (30) at (0, -1.35) {};
		\node [style=none] (31) at (1, -1.35) {};
		\node [style=none] (32) at (0.175, -0.825) {};
		\node [style=none] (33) at (0.675, -1.25) {};
		\node [style=none] (34) at (0.175, -1.675) {};
		\node [style=miniBlue] (37) at (1, -1.25) {};
		\node [style=miniU] (41) at (4.5, -1.25) {};
		\node [style=miniU] (42) at (1, -1.25) {};
		\node [style=miniBlue] (43) at (0, -1.25) {};
		\node [style=miniBlue] (44) at (2.5, -1.25) {};
		\node [style=none] (45) at (-0.5, -1.25) {};
		\node [style=none] (46) at (-1, -1.25) {\dots};
		\node [style=none] (47) at (5.5, -1.25) {\dots};
		\node [style=none] (48) at (5, -1.25) {};
		\node [style=none] (49) at (0, -1.75) {};
		\node [style=none] (50) at (4.5, -1.75) {};
		\node [style=none] (51) at (2.25, -2.5) {$n-1$ nodes};
		\node [style=miniBlue] (53) at (1, -1.25) {};
		\node [style=redgauge] (54) at (2.5, -1.25) {};
		\node [style=redgauge] (55) at (0, -1.25) {};
		\node [style=miniBlue] (56) at (3.5, -1.25) {};
	\end{pgfonlayer}
	\begin{pgfonlayer}{edgelayer}
		\draw (5) to (6);
		\draw (1) to (12.center);
		\draw (13.center) to (5);
		\draw (28.center) to (29.center);
		\draw (30.center) to (31.center);
		\draw (32.center) to (33.center);
		\draw (33.center) to (34.center);
		\draw (45.center) to (43);
		\draw (48.center) to (41);
		\draw [style=brace] (50.center) to (49.center);
	\end{pgfonlayer}
\end{tikzpicture}
}}
\end{equation}
\begin{equation}
\raisebox{-.5\height}{\scalebox{0.811}{
\begin{tikzpicture}
	\begin{pgfonlayer}{nodelayer}
		\node [style=miniBlue] (1) at (1, -1.25) {};
		\node [style=none] (3) at (1.75, -1.25) {\dots};
		\node [style=miniU] (5) at (2.5, -1.25) {};
		\node [style=miniBlue] (6) at (3.5, -1.25) {};
		\node [style=none] (12) at (1.5, -1.25) {};
		\node [style=none] (13) at (2, -1.25) {};
		\node [style=miniU] (27) at (0, -1.25) {};
		\node [style=none] (28) at (0, -1.15) {};
		\node [style=none] (29) at (1, -1.15) {};
		\node [style=none] (30) at (0, -1.35) {};
		\node [style=none] (31) at (1, -1.35) {};
		\node [style=none] (32) at (0.175, -0.825) {};
		\node [style=none] (33) at (0.675, -1.25) {};
		\node [style=none] (34) at (0.175, -1.675) {};
		\node [style=miniBlue] (37) at (1, -1.25) {};
		\node [style=miniU] (41) at (3.5, -1.25) {};
		\node [style=miniU] (42) at (1, -1.25) {};
		\node [style=miniBlue] (43) at (0, -1.25) {};
		\node [style=miniBlue] (44) at (2.5, -1.25) {};
		\node [style=none] (45) at (-0.5, -1.25) {};
		\node [style=none] (46) at (-1, -1.25) {\dots};
		\node [style=none] (47) at (4.5, -1.25) {\dots};
		\node [style=none] (48) at (4, -1.25) {};
		\node [style=none] (49) at (0, -1.75) {};
		\node [style=none] (50) at (3.5, -1.75) {};
		\node [style=none] (51) at (1.75, -2.25) {$n-1$ nodes};
	\end{pgfonlayer}
	\begin{pgfonlayer}{edgelayer}
		\draw (5) to (6);
		\draw (1) to (12.center);
		\draw (13.center) to (5);
		\draw (28.center) to (29.center);
		\draw (30.center) to (31.center);
		\draw (32.center) to (33.center);
		\draw (33.center) to (34.center);
		\draw (45.center) to (43);
		\draw (48.center) to (41);
		\draw [style=brace] (50.center) to (49.center);
	\end{pgfonlayer}
\end{tikzpicture}
}}
\end{equation}
 both contribute an $\mathfrak{sl}(n)$ factor to the global symmetry. Here, red nodes are $\mathrm{SO}$ and blue nodes are $\mathrm{USp}$. 
 
 \paragraph{Presence of $\mathrm{SO}(2)$.}
 As highlighted in \cite{Gaiotto:2008ak}, when an $\mathrm{SO}(2)$ gauge group is present, the global symmetry from a chain of $n$ balanced orthosymplectic gauge nodes is enhanced. This is due to the accidental isomorphism $\mathrm{SO}(2)\cong \mathrm{U}(1)$. However, this can be understood from an alternative point of view using D3-D5-NS5 brane configurations with O3 planes. As shown in \cite{Cabrera:2017njm}, whenever a balanced $\mathrm{SO}(2)$ gauge node is present, it is implied that a $\mathrm{USp}(0)$ gauge group is connected to it. Therefore, the following two quivers are identical: 
 \begin{equation}
\raisebox{-.5\height}{\scalebox{0.811}{
     \begin{tikzpicture}
	\begin{pgfonlayer}{nodelayer}
		\node [style=miniBlue] (1) at (1, -1.25) {};
		\node [style=none] (3) at (1.75, -1.25) {\dots};
		\node [style=miniU] (5) at (2.5, -1.25) {};
		\node [style=miniBlue] (6) at (4.5, -1.25) {};
		\node [style=none] (12) at (1.5, -1.25) {};
		\node [style=none] (13) at (2, -1.25) {};
		\node [style=miniU] (27) at (0, -1.25) {};
		\node [style=none] (28) at (0, -1.15) {};
		\node [style=none] (29) at (1, -1.15) {};
		\node [style=none] (30) at (0, -1.35) {};
		\node [style=none] (31) at (1, -1.35) {};
		\node [style=none] (32) at (0.175, -0.825) {};
		\node [style=none] (33) at (0.675, -1.25) {};
		\node [style=none] (34) at (0.175, -1.675) {};
		\node [style=miniBlue] (37) at (1, -1.25) {};
		\node [style=miniU] (41) at (4.5, -1.25) {};
		\node [style=miniU] (42) at (1, -1.25) {};
		\node [style=miniBlue] (43) at (0, -1.25) {};
		\node [style=miniBlue] (44) at (2.5, -1.25) {};
		\node [style=none] (45) at (-0.5, -1.25) {};
		\node [style=none] (46) at (-1, -1.25) {\dots};
		\node [style=none] (49) at (0, -1.75) {};
		\node [style=none] (50) at (4.5, -1.75) {};
		\node [style=none] (51) at (2.25, -2.5) {$n-1$ nodes};
		\node [style=miniBlue] (53) at (1, -1.25) {};
		\node [style=redgauge] (54) at (2.5, -1.25) {};
		\node [style=redgauge] (55) at (0, -1.25) {};
		\node [style=miniBlue] (56) at (3.5, -1.25) {};
		\node [style=none] (57) at (4.5, -0.75) {2};
		\node [style=miniBlue] (58) at (8.5, -1.25) {};
		\node [style=none] (59) at (9.25, -1.25) {\dots};
		\node [style=miniU] (60) at (10, -1.25) {};
		\node [style=miniBlue] (61) at (12, -1.25) {};
		\node [style=none] (62) at (9, -1.25) {};
		\node [style=none] (63) at (9.5, -1.25) {};
		\node [style=miniU] (64) at (7.5, -1.25) {};
		\node [style=none] (65) at (7.5, -1.15) {};
		\node [style=none] (66) at (8.5, -1.15) {};
		\node [style=none] (67) at (7.5, -1.35) {};
		\node [style=none] (68) at (8.5, -1.35) {};
		\node [style=none] (69) at (7.675, -0.825) {};
		\node [style=none] (70) at (8.175, -1.25) {};
		\node [style=none] (71) at (7.675, -1.675) {};
		\node [style=miniBlue] (72) at (8.5, -1.25) {};
		\node [style=miniU] (73) at (12, -1.25) {};
		\node [style=miniU] (74) at (8.5, -1.25) {};
		\node [style=miniBlue] (75) at (7.5, -1.25) {};
		\node [style=miniBlue] (76) at (10, -1.25) {};
		\node [style=none] (77) at (7, -1.25) {};
		\node [style=none] (78) at (6.5, -1.25) {\dots};
		\node [style=none] (79) at (7.5, -1.75) {};
		\node [style=none] (80) at (13, -1.75) {};
		\node [style=none] (81) at (10.25, -2.5) {$n$ nodes};
		\node [style=miniBlue] (82) at (8.5, -1.25) {};
		\node [style=redgauge] (83) at (10, -1.25) {};
		\node [style=redgauge] (84) at (7.5, -1.25) {};
		\node [style=miniBlue] (85) at (11, -1.25) {};
		\node [style=none] (86) at (12, -0.75) {2};
		\node [style=none] (87) at (5.5, -1.25) {=};
		\node [style=miniBlue] (89) at (13, -1.25) {};
		\node [style=none] (90) at (13, -0.75) {0};
	\end{pgfonlayer}
	\begin{pgfonlayer}{edgelayer}
		\draw (5) to (6);
		\draw (1) to (12.center);
		\draw (13.center) to (5);
		\draw (28.center) to (29.center);
		\draw (30.center) to (31.center);
		\draw (32.center) to (33.center);
		\draw (33.center) to (34.center);
		\draw (45.center) to (43);
		\draw [style=brace] (50.center) to (49.center);
		\draw (60) to (61);
		\draw (58) to (62.center);
		\draw (63.center) to (60);
		\draw (65.center) to (66.center);
		\draw (67.center) to (68.center);
		\draw (69.center) to (70.center);
		\draw (70.center) to (71.center);
		\draw (77.center) to (75);
		\draw [style=brace] (80.center) to (79.center);
		\draw (73) to (89);
	\end{pgfonlayer}
\end{tikzpicture}}
}
 \end{equation}
Both quivers contribute an $\mathfrak{sl}(n)$ factor to the global symmetry group. The quiver on the right hand side is in a more convenient form as it allows us to apply the same rule of reading off the global symmetry based on the number of balanced nodes.
Throughout the paper, it is implicit that whenever there is a balanced $\mathrm{SO}(2)$ gauge group, there is always a balanced $\mathrm{USp}(0)$ gauge node connected to it.
 
These rules for reading off the global symmetry based on balanced of gauge groups work for all framed non-simply laced orthosymplectic quiver. However, there can be rare cases among unframed orthosymplectic quivers where the global symmetry is enhanced. This is already observed in some cases for unframed simply-laced orthosymplectic in \cite{Bourget:2020xdz}. In the next section it is shown how the global symmetry of unframed non-simply laced orthosymplectic quivers can become enhanced to exceptional $\mathfrak{e}_n$ algebras.

\section{Folding unframed orthosymplectic quivers}\label{unframed}
In this section, \emph{unframed} quivers are considered, i.e.\ quivers without flavour nodes. Unframed orthosymplectic quivers have been investigated recently in \cite{Bourget:2020gzi,Bourget:2020xdz}. The simplest of these theories are magnetic quivers of  $5$d $\mathcal{N}=1$ SQCD theories. To begin with, the magnetic quivers corresponding to rank 1 $5$d SQCD theories are considered; their Coulomb branches are closures of the minimal orbits of exceptional algebras $E_n$. Thereafter, one focuses on the generalisation of these families of \cite{Bourget:2020gzi} and folds them into new families of non-simply laced unframed orthosymplectic quivers. Some of these fall into the category of star shaped quivers whose Coulomb branches can also be evaluated using the Slodowy slice approach of Appendix \ref{app:HL}.

\subsection{\texorpdfstring{$E_n$ orbits}{En orbits}}
To begin with, consider the folding of orthosymplectic quivers whose Coulomb branches are closures of $E_n$ minimal nilpotent orbits: $\overline{\mathcal{O}}^{\mathfrak{e}_n}_{\text{min}}$ for $n=4,5,6,7,8$. Since the quivers are all unframed, there is a non-trivial choice of the discrete group $H\subset \mathbb{Z}_2$ that one can ungauge. For all the quivers in this section, the Coulomb branches are defined by the choice $H=\mathbb{Z}_2$, see \cite{Bourget:2020xdz} for more details. The results are summarised in Table \ref{Entable} along with the identification of the Coulomb branch. Below, some observations for the individual cases are discussed in turn and how they are compared with folding their unitary quiver counterparts. 
\paragraph{$E_8$ orbit.}
The unitary quiver whose Coulomb branch is the closure of the minimal $E_8$ orbit takes the form of the affine Dynkin diagram of $E_8$. The unitary quiver does not have any identical legs and, therefore, cannot be folded. In contrast, the orthosymplectic quiver with the same Coulomb branch is given in the first row of Table \ref{Entable} and has two identical legs that one can fold. Folding these identical legs gives a non-simply laced quiver, see Table \ref{Entable}, whose Coulomb branch is the closure of the minimal $E_7$ orbit   $\overline{\mathcal{O}}^{\mathfrak{e}_7}_{\text{min}}$.

\paragraph{$E_7$ orbit.}
The unitary quiver whose Coulomb branch is the closure of the minimal $E_7$ orbit takes the form of the affine Dynkin diagram of $E_7$ and, hence, has two identical legs one can fold. Folding them yields the non-simply laced unitary quiver whose Coulomb branch  is  $\overline{\mathcal{O}}^{\mathfrak{e}_6}_{\text{min}}$. The orthosymplectic quiver of $E_7$ is provided in the second row of Table \ref{Entable} and has two identical legs. Folding these two legs also gives the non-simply laced orthosymplectic quiver whose Coulomb branch is $\overline{\mathcal{O}}^{\mathfrak{e}_6}_{\text{min}}$. 
\paragraph{$E_6$ orbit.}
The unitary quiver is the affine $E_6$ Dynkin diagram which has three identical legs. When two of the legs are folded, the resulting Coulomb branch is $\overline{\mathcal{O}}^{\mathfrak{f}_4}_{\text{min}}$ \cite{Hanany:2020jzl}.\footnote{Folding all three identical legs gives the minimal nilpotent orbit of $\mathfrak{so}(8)$ \cite{Bourget:2020asf}.} The  orthosymplectic quiver of $E_6$ is listed in the third row of Table \ref{Entable} and has only two identical legs. Folding them results in the  non-simply laced quiver whose Coulomb branch is  $\overline{\mathcal{O}}^{\mathfrak{e}_5}_{\text{min}} \cong \overline{\mathcal{O}}^{\mathfrak{so}(10)}_{\text{min}}$. The discrepancy is not necessarily a surprise since there are several different embeddings of $\mathbb{Z}_2$ in $E_6$. 

To summarise, one reaches the surprising statement that folding orthosymplectic quivers whose Coulomb branch are closures of the minimal  $E_8, E_7, E_6$ nilpotent orbits gives non-simply laced quivers whose Coulomb branches are closures of the $E_7, E_6, E_5 \cong D_5$ orbits respectively.

\paragraph{$E_5$ orbit.} 
The unitary quiver is the affine $D_5$ Dynkin diagram. Folding the pairs of identical nodes on the two sides of the diagram produces a  quiver with two non-simply laced edges
\begin{equation}
\raisebox{-.5\height}{\scalebox{0.811}{
\begin{tikzpicture}
	\begin{pgfonlayer}{nodelayer}
		\node [style=none] (28) at (0, -1.15) {};
		\node [style=none] (29) at (1, -1.15) {};
		\node [style=none] (30) at (0, -1.35) {};
		\node [style=none] (31) at (1, -1.35) {};
		\node [style=none] (32) at (0.175, -0.825) {};
		\node [style=none] (33) at (0.675, -1.25) {};
		\node [style=none] (34) at (0.175, -1.675) {};
		\node [style=none] (35) at (-2, -1.15) {};
		\node [style=none] (36) at (-1, -1.15) {};
		\node [style=none] (37) at (-2, -1.35) {};
		\node [style=none] (38) at (-1, -1.35) {};
		\node [style=none] (39) at (-1.075, -0.825) {};
		\node [style=none] (40) at (-1.575, -1.25) {};
		\node [style=none] (41) at (-1.075, -1.675) {};
		\node [style=gauge3] (42) at (-1, -1.25) {};
		\node [style=gauge3] (43) at (0, -1.25) {};
		\node [style=gauge3] (44) at (1, -1.25) {};
		\node [style=gauge3] (45) at (-2, -1.25) {};
		\node [style=none] (46) at (-4.75, -1.25) {};
		\node [style=none] (47) at (-3.25, -1.25) {};
		\node [style=none] (48) at (-3.25, -1.25) {};
		\node [style=gauge3] (49) at (-7, -1.25) {};
		\node [style=gauge3] (50) at (-8, -1.25) {};
		\node [style=gauge3] (51) at (-6.25, -0.5) {};
		\node [style=gauge3] (52) at (-6.25, -2) {};
		\node [style=gauge3] (53) at (-8.75, -0.5) {};
		\node [style=gauge3] (54) at (-8.75, -2) {};
		\node [style=none] (55) at (-9.25, -0.5) {1};
		\node [style=none] (56) at (-5.75, -0.5) {1};
		\node [style=none] (57) at (-5.75, -2) {1};
		\node [style=none] (58) at (-9.25, -2) {1};
		\node [style=none] (59) at (-8, -1.75) {2};
		\node [style=none] (60) at (-7, -1.75) {2};
		\node [style=none] (61) at (-2, -1.75) {1};
		\node [style=none] (62) at (-1, -1.75) {2};
		\node [style=none] (63) at (0, -1.75) {2};
		\node [style=none] (64) at (1, -1.75) {1};
	\end{pgfonlayer}
	\begin{pgfonlayer}{edgelayer}
		\draw (28.center) to (29.center);
		\draw (30.center) to (31.center);
		\draw (32.center) to (33.center);
		\draw (33.center) to (34.center);
		\draw (35.center) to (36.center);
		\draw (37.center) to (38.center);
		\draw (39.center) to (40.center);
		\draw (40.center) to (41.center);
		\draw (43) to (42);
		\draw [style=->] (46.center) to (48.center);
		\draw (53) to (50);
		\draw (50) to (54);
		\draw (50) to (49);
		\draw (49) to (51);
		\draw (49) to (52);
	\end{pgfonlayer}
\end{tikzpicture}
}}
\label{dtype}
\end{equation}
The Coulomb branches of the quivers on the right are the minimal orbits of $\overline{\mathcal{O}}^{\mathfrak{so}(8)}_{\text{min}}$ \cite{Bourget:2020asf}. The orthosymplectic quiver of $D_5$ is listed in the fourth row of Table \ref{Entable} (which is reproduced here):
\begin{equation}
  \raisebox{-.5\height}{\scalebox{0.811}{
  \begin{tikzpicture}
	\begin{pgfonlayer}{nodelayer}
		\node [style=redgauge] (65) at (-7.5, -5) {};
		\node [style=bluegauge] (66) at (-6.5, -5) {};
		\node [style=bluegauge] (67) at (-8.5, -5) {};
		\node [style=redgauge] (68) at (-5.5, -5) {};
		\node [style=redgauge] (69) at (-9.5, -5) {};
		\node [style=gauge3] (70) at (-7.5, -4) {};
		\node [style=none] (71) at (-7.5, -3.5) {1};
		\node [style=none] (72) at (-7.5, -5.5) {4};
		\node [style=none] (73) at (-6.5, -5.5) {2};
		\node [style=none] (74) at (-5.5, -5.5) {2};
		\node [style=none] (75) at (-8.5, -5.5) {2};
		\node [style=none] (76) at (-9.5, -5.5) {2};
		\node [style=none] (89) at (-4.75, -5) {};
		\node [style=none] (90) at (-3.25, -5) {};
		\node [style=none] (91) at (-3.25, -5) {};
		\node [style=none] (92) at (-1, -4.9) {};
		\node [style=none] (93) at (0, -4.9) {};
		\node [style=none] (94) at (-1, -5.1) {};
		\node [style=none] (95) at (0, -5.1) {};
		\node [style=none] (96) at (-0.825, -4.575) {};
		\node [style=none] (97) at (-0.325, -5) {};
		\node [style=none] (98) at (-0.825, -5.425) {};
		\node [style=redgauge] (103) at (-1, -5) {};
		\node [style=bluegauge] (104) at (0, -5) {};
		\node [style=redgauge] (105) at (1, -5) {};
		\node [style=none] (106) at (-1, -5.5) {4};
		\node [style=none] (107) at (0, -5.5) {2};
		\node [style=none] (108) at (1, -5.5) {2};
		\node [style=gauge3] (109) at (-2, -5) {};
		\node [style=none] (110) at (-2, -5.5) {1};
	\end{pgfonlayer}
	\begin{pgfonlayer}{edgelayer}
		\draw (69) to (68);
		\draw (70) to (65);
		\draw [style=->] (89.center) to (91.center);
		\draw (92.center) to (93.center);
		\draw (94.center) to (95.center);
		\draw (96.center) to (97.center);
		\draw (97.center) to (98.center);
		\draw (104) to (105);
		\draw (109) to (103);
	\end{pgfonlayer}
\end{tikzpicture}}}
\end{equation}
One can verify that the Coulomb branch of the folded orthosymplectic quiver is also $\overline{\mathcal{O}}^{\mathfrak{so}(8)}_{\text{min}}$. 

As a comment, \eqref{dtype} has the $D_5$ Dynkin diagram on the left and the twisted affine $D^{(2)}_4$ Dynkin diagram on the right \cite{fuchs2003symmetries}. This pattern generalises to any $n$, meaning that the affine $D_n$ Dynkin quiver, whose Coulomb branch is $\overline{\mathcal{O}}^{\mathfrak{so}(2n)}_{\text{min}}$, can be folded to the twisted affine $D^{(2)}_{n-1}$ Dynkin quiver, whose Coulomb branch is  $\overline{\mathcal{O}}^{\mathfrak{so}(2n-2)}_{\text{min}}$.

\paragraph{$E_4$ orbit.}
The unitary quiver is the affine $A_4$ Dynkin diagram, which after framing does not have identical legs attached to a pivot node, and hence cannot be folded in the common way. The orthosymplectic quiver with the same Coulomb branch is listed in the fifth row of Table \ref{Entable}, which does have two identical legs that one can fold. The wiggly line denotes a charge 2 hypermultiplet, see \cite{Bourget:2020gzi} for more details. The Coulomb branch of the folded non-simply laced orthosymplectic quiver  is $\overline{\mathcal{O}}^{\mathfrak{sl}(4)}_{\text{min}}$. 
\\ 
\\
A feature
of orthosymplectic quivers whose Coulomb branches are closures of exceptional algebras is that they always have two identical legs one can fold. This reflection symmetry is not always present in the unitary quiver counterparts.

\subsection{\texorpdfstring{$\mathbb{Z}_2$ projection on representations}{Z2 projection on representations}}
The results can be explained using representation theory. In \cite{Zafrir:2016wkk}, 5d $\mathcal{N}=1$ theories are compactified on a circle with $\mathbb{Z}_2$ twist. First, one seeks to find a subgroup $H_{\mathrm{5d}}$ of the global symmetry group $G_{\mathrm{5d}}$ of the 5d SCFT such that $H_{\mathrm{5d}}\cong H_1\times H_1 \times H_2$. In other words, $H_{\mathrm{5d}}$ must contain two identical groups. Next, consider the $\mathbb{Z}_2$ invariant part of  $H_1 \times H_1$. This way, during the compactification, the $\mathbb{Z}_2$ acts diagonally and only representations invariant under this action remain. 

Consider the $E_8$ quiver. 
$E_8$ contains the following subgroup:
\begin{equation}
    E_8 \supset \mathrm{SO}(8)_A\times \mathrm{SO}(8)_B  \,.
\end{equation}
The adjoint representation of $E_8$ decomposes as:
\begin{align}
\begin{aligned}
    (\mu_8)_{E_8} \rightarrow &(\mu_1)_{\mathrm{SO}(8)_A}(\mu_1)_{\mathrm{SO}(8)_B} + (\mu_2)_{\mathrm{SO}(8)_A} + (\mu_2)_{\mathrm{SO}(8)_B}  \\
    &+ (\mu_3)_{\mathrm{SO}(8)_A}(\mu_4)_{\mathrm{SO}(8)_B} + (\mu_4)_{\mathrm{SO}(8)_A}(\mu_3)_{\mathrm{SO}(8)_B} 
    \end{aligned}
\end{align}
where $(\mu_i)_{E_8}$, $(\mu_i)_{A}$, $(\mu_i)_{B}\;  $ are the highest weight fugacities of $E_8$, $\mathrm{SO}(8)_A$, and $\mathrm{SO}(8)_B$ respectively. 
The $\mathbb{Z}_2$ group acts on the adjoint representation as follows:
\begin{equation}
\begin{aligned}
   (\mu_2)_{\mathrm{SO}(8)_A} + (\mu_2)_{\mathrm{SO}(8)_B} & \xrightarrow[]{} (\mu_2)_{\mathrm{SO}(8)_{\mathrm{diag}}}  \\ 
   (\mu_1)_{\mathrm{SO}(8)_A}(\mu_1)_{\mathrm{SO}(8)_B} & \xrightarrow[]{} (\mu_1^2)_{\mathrm{SO}(8)_{\mathrm{diag}}}\\
    (\mu_3)_{\mathrm{SO}(8)_A}(\mu_4)_{\mathrm{SO}(8)_B}  + (\mu_4)_{\mathrm{SO}(8)_A}(\mu_3)_{\mathrm{SO}(8)_B}  & \xrightarrow[]{} (\mu_3^2)_{\mathrm{SO}(8)_{\mathrm{diag}}} +(\mu_4^2)_{\mathrm{SO}(8)_{\mathrm{diag}}} \\
   \end{aligned}
\end{equation}
Since $\mathrm{SO}(8)_{\mathrm{diag}}\subset \mathrm{SU}(8) \subset E_7$, the irreducible representation after the projection precisely gives the adjoint representation for $E_7$:
\begin{equation}
    (\rho_1)_{E_7} = (\kappa_1 \kappa_7)_{\mathrm{SU}(8)} + (\kappa_4)_{\mathrm{SU}(8)} = (\mu_2)_{\mathrm{SO}(8)} + (\mu_1^2)_{\mathrm{SO}(8)} + (\mu_3^2)_{\mathrm{SO}(8)}+ (\mu_4^2)_{\mathrm{SO}(8)}
\end{equation}
where $\rho_i$, $\kappa_i$ are highest weight fugacities of $E_7$ and $\mathrm{SU}(8)$ respectively. 

\afterpage{
\begin{landscape}
\begin{table}[p]
   \hspace*{-1cm}\scalebox{0.9}{
    \vspace*{-1.5cm}
\begin{tabular}{c|ccc|cc} 
        \multicolumn{1}{c}{ } & \multicolumn{3}{c}{\Large{Before folding}} & \multicolumn{2}{c}{\Large{After folding}} \\ 
            \multicolumn{1}{c}{ } & \multicolumn{3}{c}{ } & \multicolumn{2}{c}{ }  \\ 
    \toprule 
       Family & Magnetic quiver (infinite coupling) & 5d quiver  & Global Symmetry & Magnetic quiver & Global Symmetry \\
\bottomrule
    \end{tabular}}
    \caption{Generalised families of $E_n$ orthosymplectic quivers of those in Table \ref{Entable}. The orthosymplectic quivers before folding are magnetic quivers of certain $5$d $\mathcal{N}=1$ SQCD theories at infinite gauge coupling. The subscript next to the gauge group is the Chern-Simons level. For the $E_{3-2l}$ family, the Coulomb branch of the magnetic quiver is only one of the two cones in the Higgs branch of the 5d theory. The global symmetry is given for $k>1$ and $k>l+1$, it enhances for $k=1$ or $k=l+1$ as shown in Table \ref{Entable}. }
    \label{generalfamily}
\end{table}
\end{landscape}}

\afterpage{
\begin{landscape}
\begin{table}[t]
    \centering
    \hspace*{-1cm}
\\ 
         \bottomrule
    \end{tabular}
    \caption{The non-simply laced orthosymplectic quiver families and the unitary quiver family which have the same Coulomb branches. The highest weight generating function (HWG) is presented in the form of a plethystic logarithm (PL). The fugacities correspond to the global symmetry given in the last column of Table \ref{generalfamily}, with $q$ denoting a $\mathfrak{u}(1)$ factor and $\nu$ denoting an $\mathfrak{su}(2)$ factor when present.  }
    \label{generalhwg}
\end{table}
\end{landscape}
}

This process is beautifully encoded in the folding procedure. When given a quiver, the balance of the gauge nodes determines the global symmetry group\footnote{To be more precise, it gives the algebra of the global symmetry group.}. If one singles out a subset of balanced nodes, then this subquiver gives a subgroup of the global symmetry. For the $E_8$ quiver, one natural branching to subgroups is to identify the identical legs:
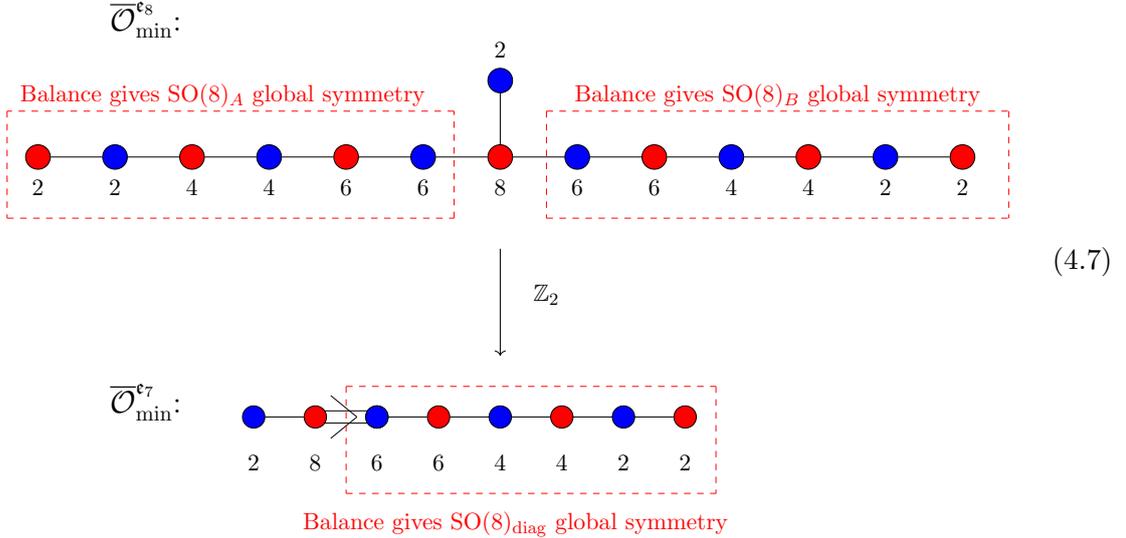
\begin{equation}
\raisebox{-.5\height}{\scalebox{0.811}{
\begin{tikzpicture}
	\begin{pgfonlayer}{nodelayer}
		\node [style=bluegauge] (0) at (0, 0) {};
		\node [style=redgauge] (1) at (1.25, 0) {};
		\node [style=bluegauge] (2) at (2.5, 0) {};
		\node [style=redgauge] (3) at (3.75, 0) {};
		\node [style=redgauge] (4) at (-1.25, 0) {};
		\node [style=redgauge] (5) at (-3.75, 0) {};
		\node [style=bluegauge] (6) at (-2.5, 0) {};
		\node [style=none] (8) at (0, -0.5) {4};
		\node [style=none] (9) at (1.25, -0.5) {6};
		\node [style=none] (10) at (2.5, -0.5) {6};
		\node [style=none] (11) at (3.75, -0.5) {8};
		\node [style=none] (12) at (-1.25, -0.5) {4};
		\node [style=none] (13) at (-2.5, -0.5) {2};
		\node [style=none] (14) at (-3.75, -0.5) {2};
		\node [style=bluegauge] (23) at (5, 0) {};
		\node [style=none] (25) at (5, -0.5) {6};
		\node [style=bluegauge] (27) at (10, 0) {};
		\node [style=redgauge] (28) at (11.25, 0) {};
		\node [style=redgauge] (30) at (8.75, 0) {};
		\node [style=redgauge] (31) at (6.25, 0) {};
		\node [style=bluegauge] (32) at (7.5, 0) {};
		\node [style=none] (33) at (10, -0.5) {2};
		\node [style=none] (34) at (11.25, -0.5) {2};
		\node [style=none] (36) at (8.75, -0.5) {4};
		\node [style=none] (37) at (7.5, -0.5) {4};
		\node [style=none] (38) at (6.25, -0.5) {6};
		\node [style=bluegauge] (39) at (3.75, 1.25) {};
		\node [style=none] (40) at (3.75, 1.75) {2};
		\node [style=none] (41) at (-4.25, 0.75) {};
		\node [style=none] (42) at (-4.25, -1) {};
		\node [style=none] (43) at (3, 0.75) {};
		\node [style=none] (44) at (3, -1) {};
		\node [style=none] (45) at (4.5, 0.75) {};
		\node [style=none] (46) at (4.5, -1) {};
		\node [style=none] (47) at (12, -1) {};
		\node [style=none] (48) at (12, 0.75) {};
		\node [style=none] (54) at (8.25, 1) {\color{red}{Balance gives $\mathrm{SO}(8)_B$ global symmetry}};
		\node [style=miniBlue] (55) at (1.75, -4.25) {};
		\node [style=miniU] (56) at (0.75, -4.25) {};
		\node [style=none] (57) at (0.75, -4.15) {};
		\node [style=none] (58) at (1.75, -4.15) {};
		\node [style=none] (59) at (0.75, -4.35) {};
		\node [style=none] (60) at (1.75, -4.35) {};
		\node [style=none] (61) at (1, -3.9) {};
		\node [style=none] (62) at (1.425, -4.25) {};
		\node [style=none] (63) at (1, -4.6) {};
		\node [style=miniBlue] (64) at (1.75, -4.25) {};
		\node [style=miniU] (65) at (2.75, -4.25) {};
		\node [style=miniBlue] (66) at (-0.25, -4.25) {};
		\node [style=none] (67) at (3.75, -1.5) {};
		\node [style=none] (68) at (3.75, -3.25) {};
		\node [style=none] (69) at (4.5, -2.25) {$\mathbb{Z}_2$};
		\node [style=miniU] (71) at (4.75, -4.25) {};
		\node [style=miniBlue] (72) at (3.75, -4.25) {};
		\node [style=miniU] (73) at (6.75, -4.25) {};
		\node [style=miniBlue] (74) at (5.75, -4.25) {};
		\node [style=none] (75) at (-0.25, -5) {2};
		\node [style=none] (76) at (0.75, -5) {8};
		\node [style=none] (77) at (1.75, -5) {6};
		\node [style=none] (78) at (2.75, -5) {6};
		\node [style=none] (79) at (3.75, -5) {4};
		\node [style=none] (80) at (4.75, -5) {4};
		\node [style=none] (81) at (5.75, -5) {2};
		\node [style=none] (82) at (6.75, -5) {2};
		\node [style=none] (83) at (1.25, -3.75) {};
		\node [style=none] (84) at (1.25, -5.5) {};
		\node [style=none] (85) at (7.25, -5.5) {};
		\node [style=none] (86) at (7.25, -3.75) {};
		\node [style=none] (89) at (4, -6) {\color{red}{Balance gives $\mathrm{SO}(8)_{\mathrm{diag}}$ global symmetry }};
		\node [style=none] (90) at (-2, -4) {\Large $\overline{\mathcal{O}}^{\mathfrak{e}_7}_{\text{min}}$:};
		\node [style=none] (91) at (-2, 2.25) {\Large $\overline{\mathcal{O}}^{\mathfrak{e}_8}_{\text{min}}$:};
		\node [style=none] (92) at (-0.75, 1) {\color{red}{Balance gives $\mathrm{SO}(8)_A$ global symmetry}};
	\end{pgfonlayer}
	\begin{pgfonlayer}{edgelayer}
		\draw (5) to (6);
		\draw (6) to (4);
		\draw (4) to (0);
		\draw (0) to (1);
		\draw (1) to (2);
		\draw (2) to (3);
		\draw (3) to (23);
		\draw (31) to (32);
		\draw (32) to (30);
		\draw (30) to (27);
		\draw (27) to (28);
		\draw (23) to (31);
		\draw (3) to (39);
		\draw [style=Reddotted] (41.center) to (42.center);
		\draw [style=Reddotted] (42.center) to (44.center);
		\draw [style=Reddotted] (44.center) to (43.center);
		\draw [style=Reddotted] (43.center) to (41.center);
		\draw [style=Reddotted] (45.center) to (46.center);
		\draw [style=Reddotted] (46.center) to (47.center);
		\draw [style=Reddotted] (45.center) to (48.center);
		\draw [style=Reddotted] (48.center) to (47.center);
		\draw (57.center) to (58.center);
		\draw (59.center) to (60.center);
		\draw (61.center) to (62.center);
		\draw (62.center) to (63.center);
		\draw (65) to (64);
		\draw (66) to (56);
		\draw [style=->] (67.center) to (68.center);
		\draw (65) to (73);
		\draw [style=Reddotted] (83.center) to (84.center);
		\draw [style=Reddotted] (84.center) to (85.center);
		\draw [style=Reddotted] (83.center) to (86.center);
		\draw [style=Reddotted] (86.center) to (85.center);
	\end{pgfonlayer}
\end{tikzpicture}
}}
\end{equation}
For a unitary magnetic quiver of $E_8$, the quiver takes the form of the $E_8$ affine Dynkin diagram which does not have a natural $\mathrm{SO}(8)\times \mathrm{SO}(8)$ subgroup one can identify and fold. This is an advantage of the $E_n$ orthosymplectic quivers in general which always has a natural $\mathbb{Z}_2$ symmetry one can fold. 

One can repeat this procedure for the remaining $E_n$ families and the result reproduce the global symmetry of the folded quivers.

\subsection{\texorpdfstring{Magnetic quivers of 4d $\mathcal{N}=2$}{Magnetic quivers of 4d N=2}}
\label{generalunframed}
In \cite{Bourget:2020asf}, a class of unitary magnetic quivers of $5$d $\mathcal{N}=1$ SQCD has been folded to produce general sequences whose limiting cases are $4$d $\mathcal{N}=2$ rank 1 theories. In cases where the folding involves two identical legs, this procedure produces the Higgs branches of $5$d theories compactified on a circle with a $\mathbb{Z}_2$ twist \cite{Zafrir:2016wkk}. However, note that folding magnetic quivers of 5d $\mathcal{N}=1$ theories does not always give rise to magnetic quivers of 4d $\mathcal{N}=2$ theories.
As seen in the above subsection, some unitary magnetic quivers, which do not have identical legs, have orthosymplectic counterparts that do have identical legs. The  orthosymplectic quivers studied here are examples like that where the unitary counterparts (tabulated in \cite[Tab.\ 1]{Bourget:2020gzi}) lack this symmetry. In this section, the generalised families of orthosymplectic quivers are considered and folded. The results are summarised in Table \ref{generalfamily}.

Like the unitary quivers, one conjectures that some of the folded orthosymplectic quivers are magnetic quivers of known 4d $\mathcal{N}=2$ theories. 
In other words, the Coulomb branch of these folded orthosymplectic quivers are the Higgs branch of 4d $\mathcal{N}=2$ theories. To be concrete, focus on the rank 1 cases in Table \ref{Entable}. After folding the $E_8$, $E_7$, $D_5$ orthosymplectic quivers, the resulting Coulomb branches are minimal nilpotent orbit closures of $E_7$, $E_6$, $D_4$ respectively. These are Higgs branches of known 4d $\mathcal{N}=2$ rank 1 theories. On the other hand, folding the $E_6$, $A_4$ orthosymplectic quivers give Coulomb branches that are minimal nilpotent orbit closures of $D_5$, $A_3$ respectively which are not the Higgs branches of known rank 1 4d theories \cite{Argyres:2016xua}. It has been shown in \cite{Shimizu:2017kzs}, via anomaly matching on the Higgs branch, that $D_5$, $A_3$ minimal nilpotent orbit closures (or equivalently, one-instanton moduli spaces) are excluded as Higgs branches of rank 1 4d $\mathcal{N}=2$ theories. This shows only a subset of the folded orthosymplectic quivers are actually magnetic quivers for 4d $\mathcal{N}=2$ theories.

Following this argument, one can generalise each $E_n$ non-simply laced orthosymplectic quiver to infinite families as in Table \ref{generalfamily}. The families obtained from folding the $E_8$, $E_7$ and $E_5\cong D_5$ families give rise to known 4d $\mathcal{N}=2$ theories. These are all class S theories. For $E_8$ and $E_7$ folded families, these are Sicilian theories with $A$-type punctures (A-type $6d$ $\mathcal{N}=(2,0)$ theories  compactified on a  sphere with 3 punctures)  as studied in \cite{DistlerA}. Using the parameterisation given in Table \ref{generalhwg}, the folded $E_8$ family gives the $[k+3],[k+3],[2^2,1^{k-1}]$ Sicilian theory where punctures are labeled by their partition data. The folded $E_7$ family gives $[k+2],[k+2],[3,1^{k-1}]$ Sicilian theory. Finally, the $E_5 \cong D_5$ folded family is the magnetic quiver for  the 4d $\mathcal{N}=2$ SCFT of $\mathrm{SU}(k+1)$ with $2k+2$ flavours. 

For the remaining three families in Table \ref{generalhwg}, the theories do not resemble magnetic quivers of known 4d $\mathcal{N}=2$ theories. Nevertheless, they are magnetic quivers for 5d $\mathcal{N}=1$ theories. For the $E_6$ folded family, the corresponding 5d theory is $\mathrm{SU}(k+1)_{\pm1}$ with $2k+2$ flavours at infinite gauge coupling. The $E_{4-2l}$ folded family is a magnetic quiver of one of the two cones of the Higgs branch of the 5d $\mathrm{SU}(k+1)_{\pm\frac{1}{2}}$ with $2k-2l+1$ flavours at infinite gauge coupling. The $E_{3-2l}$ folded family is a magnetic quiver of one of the two cones of the Higgs branch of the 5d $\mathrm{SU}(k+1)_0$ with $2k-2l+1$ flavours at infinite gauge coupling.  

The HWGs in Table \ref{generalhwg} can be obtained by taking the Coulomb branch HWG of the magnetic quiver before folding, see \cite{Bourget:2020xdz},  and applying the projection \eqref{fugacitymap1}/\eqref{fugacitymap2}, for the global symmetries  $\mathfrak{so}(4n)$/$\mathfrak{so}(4n+2)$ respectively. The HWGs of the folded quivers are already computed in \cite{Benvenuti:2010pq, Sicilian2,Ferlito:2017xdq,SQCD}.

\section{Hasse diagrams}
\label{Hasse}
Hasse diagrams are useful tools for understanding the geometry of Higgs branches of various supersymmetric gauge theories. In particular, for theories with 8 supercharges in $3,4,5,6$ dimensions, this is explored in detail in recent works \cite{Cabrera:2016vvv,Cabrera:2017njm,Cabrera:2019dob,Bourget:2019aer,Bourget:2020asf,Grimminger:2020dmg,Bourget:2020gzi}. Instead of studying the Higgs branch of an electric theory directly, one can equivalently study the 3d $\mathcal{N}=4$ Coulomb branch of the corresponding magnetic quiver(s). This allows one to obtain the Hasse diagram using \emph{quiver subtraction} \cite{Cabrera:2018ann}. Starting with a magnetic quiver, one can systematically subtract quivers to find the different transverse slices and symplectic leaves of the moduli space. The Coulomb branches of quivers one can subtract are elementary slices. The most up to date list of such quivers can be found in \cite[Table 1]{Bourget:2021siw} for unitary quivers and in \cite{Bourget:2020gzi} for simply-laced orthosymplectic quivers. Here, the list is extended by non-simply laced orthosymplectic quivers whose Coulomb branches are elementary slices. These are summarised in Table \ref{Dntable}, for framed quivers giving closures of nilpotent orbits of type $A$, and Table \ref{Entable}, for unframed quivers. 

\subsection{Maximal height 2 orbits}
\paragraph{Quiver subtraction for orthosymplectic quivers.}
The rules for quiver subtraction between unitary quivers are given in \cite{Cabrera:2018ann}. The general idea is to align two quivers and subtract the respective gauge nodes $G_i-G'_i\rightarrow G''_i$ such that $\mathrm{rank}(G_i)-\mathrm{rank}(G'_i) = \mathrm{rank}(G''_i)$. The resulting gauge groups need to be rebalanced by adding flavours. For orthosymplectic quivers, rules are given in \cite{Hanany:2019tji}.

\begin{figure}[p]
\centering
\scalebox{0.7}{
\begin{tikzpicture}
	\begin{pgfonlayer}{nodelayer}
		\node [style=miniBlue] (1) at (0.5, 1) {};
		\node [style=none] (18) at (0.5, 0.5) {6};
		\node [style=miniU] (27) at (-0.5, 1) {};
		\node [style=none] (28) at (-0.5, 1.1) {};
		\node [style=none] (29) at (0.5, 1.1) {};
		\node [style=none] (30) at (-0.5, 0.9) {};
		\node [style=none] (31) at (0.5, 0.9) {};
		\node [style=none] (32) at (-0.325, 1.425) {};
		\node [style=none] (33) at (0.175, 1) {};
		\node [style=none] (34) at (-0.325, 0.575) {};
		\node [style=none] (36) at (-0.5, 0.5) {8};
		\node [style=miniBlue] (37) at (0.5, 1) {};
		\node [style=miniBlue] (38) at (-0.5, 1) {};
		\node [style=miniU] (39) at (0.5, 1) {};
		\node [style=miniU] (45) at (-0.5, 1) {};
		\node [style=miniBlue] (46) at (0.5, 1) {};
		\node [style=none] (50) at (-1.5, 0.5) {2};
		\node [style=bd] (51) at (-3, 1) {};
		\node [style=bd] (52) at (-3, -2) {};
		\node [style=bd] (53) at (-3, -5) {};
		\node [style=bd] (54) at (-3, -8) {};
		\node [style=bd] (55) at (-3, -11) {};
		\node [style=miniU] (57) at (1.5, 1) {};
		\node [style=miniBlue] (58) at (2.5, 1) {};
		\node [style=miniU] (59) at (3.5, 1) {};
		\node [style=miniU] (60) at (5.5, 1) {};
		\node [style=miniBlue] (61) at (4.5, 1) {};
		\node [style=flavourBlue] (62) at (-1.5, 1) {};
		\node [style=none] (63) at (1.5, 0.5) {6};
		\node [style=none] (64) at (2.5, 0.5) {4};
		\node [style=none] (65) at (3.5, 0.5) {4};
		\node [style=none] (66) at (4.5, 0.5) {2};
		\node [style=none] (67) at (5.5, 0.5) {2};
		\node [style=miniBlue] (68) at (0.5, -2) {};
		\node [style=none] (69) at (0.5, -2.5) {6};
		\node [style=miniU] (70) at (-0.5, -2) {};
		\node [style=none] (71) at (-0.5, -1.9) {};
		\node [style=none] (72) at (0.5, -1.9) {};
		\node [style=none] (73) at (-0.5, -2.1) {};
		\node [style=none] (74) at (0.5, -2.1) {};
		\node [style=none] (75) at (-0.325, -1.575) {};
		\node [style=none] (76) at (0.175, -2) {};
		\node [style=none] (77) at (-0.325, -2.425) {};
		\node [style=none] (78) at (-0.5, -2.5) {7};
		\node [style=miniBlue] (79) at (0.5, -2) {};
		\node [style=miniBlue] (80) at (-0.5, -2) {};
		\node [style=miniU] (81) at (0.5, -2) {};
		\node [style=miniU] (82) at (-0.5, -2) {};
		\node [style=miniBlue] (83) at (0.5, -2) {};
		\node [style=miniU] (85) at (1.5, -2) {};
		\node [style=miniBlue] (86) at (2.5, -2) {};
		\node [style=miniU] (87) at (3.5, -2) {};
		\node [style=miniU] (88) at (5.5, -2) {};
		\node [style=miniBlue] (89) at (4.5, -2) {};
		\node [style=none] (91) at (1.5, -2.5) {6};
		\node [style=none] (92) at (2.5, -2.5) {4};
		\node [style=none] (93) at (3.5, -2.5) {4};
		\node [style=none] (94) at (4.5, -2.5) {2};
		\node [style=none] (95) at (5.5, -2.5) {2};
		\node [style=flavourRed] (96) at (0.5, -1) {};
		\node [style=none] (97) at (1, -1) {1};
		\node [style=miniBlue] (98) at (0.5, -5) {};
		\node [style=none] (99) at (0.5, -5.5) {4};
		\node [style=miniU] (100) at (-0.5, -5) {};
		\node [style=none] (101) at (-0.5, -4.9) {};
		\node [style=none] (102) at (0.5, -4.9) {};
		\node [style=none] (103) at (-0.5, -5.1) {};
		\node [style=none] (104) at (0.5, -5.1) {};
		\node [style=none] (105) at (-0.325, -4.575) {};
		\node [style=none] (106) at (0.175, -5) {};
		\node [style=none] (107) at (-0.325, -5.425) {};
		\node [style=none] (108) at (-0.5, -5.5) {5};
		\node [style=miniBlue] (109) at (0.5, -5) {};
		\node [style=miniBlue] (110) at (-0.5, -5) {};
		\node [style=miniU] (111) at (0.5, -5) {};
		\node [style=miniU] (112) at (-0.5, -5) {};
		\node [style=miniBlue] (113) at (0.5, -5) {};
		\node [style=miniU] (114) at (1.5, -5) {};
		\node [style=miniBlue] (115) at (2.5, -5) {};
		\node [style=miniU] (116) at (3.5, -5) {};
		\node [style=miniU] (117) at (5.5, -5) {};
		\node [style=miniBlue] (118) at (4.5, -5) {};
		\node [style=none] (119) at (1.5, -5.5) {5};
		\node [style=none] (120) at (2.5, -5.5) {4};
		\node [style=none] (121) at (3.5, -5.5) {4};
		\node [style=none] (122) at (4.5, -5.5) {2};
		\node [style=none] (123) at (5.5, -5.5) {2};
		\node [style=flavourRed] (124) at (2.5, -4) {};
		\node [style=none] (125) at (3, -4) {1};
		\node [style=miniBlue] (126) at (0.5, -8) {};
		\node [style=none] (127) at (0.5, -8.5) {2};
		\node [style=miniU] (128) at (-0.5, -8) {};
		\node [style=none] (129) at (-0.5, -7.9) {};
		\node [style=none] (130) at (0.5, -7.9) {};
		\node [style=none] (131) at (-0.5, -8.1) {};
		\node [style=none] (132) at (0.5, -8.1) {};
		\node [style=none] (133) at (-0.325, -7.575) {};
		\node [style=none] (134) at (0.175, -8) {};
		\node [style=none] (135) at (-0.325, -8.425) {};
		\node [style=none] (136) at (-0.5, -8.5) {3};
		\node [style=miniBlue] (137) at (0.5, -8) {};
		\node [style=miniBlue] (138) at (-0.5, -8) {};
		\node [style=miniU] (139) at (0.5, -8) {};
		\node [style=miniU] (140) at (-0.5, -8) {};
		\node [style=miniBlue] (141) at (0.5, -8) {};
		\node [style=miniU] (142) at (1.5, -8) {};
		\node [style=miniBlue] (143) at (2.5, -8) {};
		\node [style=miniU] (144) at (3.5, -8) {};
		\node [style=miniU] (145) at (5.5, -8) {};
		\node [style=miniBlue] (146) at (4.5, -8) {};
		\node [style=none] (147) at (1.5, -8.5) {3};
		\node [style=none] (148) at (2.5, -8.5) {2};
		\node [style=none] (149) at (3.5, -8.5) {3};
		\node [style=none] (150) at (4.5, -8.5) {2};
		\node [style=none] (151) at (5.5, -8.5) {2};
		\node [style=flavourRed] (152) at (4.5, -7) {};
		\node [style=none] (153) at (5, -7) {1};
		\node [style=miniBlue] (154) at (-9.75, -9.75) {};
		\node [style=none] (155) at (-9.75, -10.25) {2};
		\node [style=miniU] (156) at (-10.75, -9.75) {};
		\node [style=none] (157) at (-10.75, -9.65) {};
		\node [style=none] (158) at (-9.75, -9.65) {};
		\node [style=none] (159) at (-10.75, -9.85) {};
		\node [style=none] (160) at (-9.75, -9.85) {};
		\node [style=none] (161) at (-10.575, -9.325) {};
		\node [style=none] (162) at (-10.075, -9.75) {};
		\node [style=none] (163) at (-10.575, -10.175) {};
		\node [style=none] (164) at (-10.75, -10.25) {3};
		\node [style=miniBlue] (165) at (-9.75, -9.75) {};
		\node [style=miniBlue] (166) at (-10.75, -9.75) {};
		\node [style=miniU] (167) at (-9.75, -9.75) {};
		\node [style=miniU] (168) at (-10.75, -9.75) {};
		\node [style=miniBlue] (169) at (-9.75, -9.75) {};
		\node [style=miniU] (170) at (-8.75, -9.75) {};
		\node [style=miniBlue] (171) at (-7.75, -9.75) {};
		\node [style=miniU] (172) at (-6.75, -9.75) {};
		\node [style=miniU] (173) at (-4.75, -9.75) {};
		\node [style=miniBlue] (174) at (-5.75, -9.75) {};
		\node [style=none] (175) at (-8.75, -10.25) {3};
		\node [style=none] (176) at (-7.75, -10.25) {2};
		\node [style=none] (177) at (-6.75, -10.25) {3};
		\node [style=none] (178) at (-5.75, -10.25) {2};
		\node [style=none] (179) at (-4.75, -10.25) {2};
		\node [style=flavourRed] (180) at (-5.75, -8.75) {};
		\node [style=none] (181) at (-5.25, -8.75) {1};
		\node [style=miniU] (182) at (-9.5, -0.5) {};
		\node [style=flavourBlue] (183) at (-10.5, -0.5) {};
		\node [style=none] (184) at (-10.5, -1) {2};
		\node [style=none] (185) at (-9.5, -1) {2};
		\node [style=miniBlue] (186) at (-9.75, -4) {};
		\node [style=none] (187) at (-9.75, -4.5) {2};
		\node [style=miniU] (188) at (-10.75, -4) {};
		\node [style=none] (189) at (-10.75, -3.9) {};
		\node [style=none] (190) at (-9.75, -3.9) {};
		\node [style=none] (191) at (-10.75, -4.1) {};
		\node [style=none] (192) at (-9.75, -4.1) {};
		\node [style=none] (193) at (-10.575, -3.575) {};
		\node [style=none] (194) at (-10.075, -4) {};
		\node [style=none] (195) at (-10.575, -4.425) {};
		\node [style=none] (196) at (-10.75, -4.5) {3};
		\node [style=miniBlue] (197) at (-9.75, -4) {};
		\node [style=miniBlue] (198) at (-10.75, -4) {};
		\node [style=miniU] (199) at (-9.75, -4) {};
		\node [style=miniU] (200) at (-10.75, -4) {};
		\node [style=miniBlue] (201) at (-9.75, -4) {};
		\node [style=flavourRed] (202) at (-9.75, -3) {};
		\node [style=none] (203) at (-9.25, -3) {1};
		\node [style=miniBlue] (204) at (-9.75, -7) {};
		\node [style=none] (205) at (-9.75, -7.5) {2};
		\node [style=miniU] (206) at (-10.75, -7) {};
		\node [style=none] (207) at (-10.75, -6.9) {};
		\node [style=none] (208) at (-9.75, -6.9) {};
		\node [style=none] (209) at (-10.75, -7.1) {};
		\node [style=none] (210) at (-9.75, -7.1) {};
		\node [style=none] (211) at (-10.575, -6.575) {};
		\node [style=none] (212) at (-10.075, -7) {};
		\node [style=none] (213) at (-10.575, -7.425) {};
		\node [style=none] (214) at (-10.75, -7.5) {3};
		\node [style=miniBlue] (215) at (-9.75, -7) {};
		\node [style=miniBlue] (216) at (-10.75, -7) {};
		\node [style=miniU] (217) at (-9.75, -7) {};
		\node [style=miniU] (218) at (-10.75, -7) {};
		\node [style=miniBlue] (219) at (-9.75, -7) {};
		\node [style=miniU] (220) at (-8.75, -7) {};
		\node [style=miniBlue] (221) at (-7.75, -7) {};
		\node [style=none] (222) at (-8.75, -7.5) {3};
		\node [style=none] (223) at (-7.75, -7.5) {2};
		\node [style=flavourRed] (224) at (-7.75, -6) {};
		\node [style=none] (225) at (-7.25, -6) {1};
		\node [style=miniU] (226) at (-8.75, -4) {};
		\node [style=miniU] (227) at (-6.75, -7) {};
		\node [style=none] (228) at (-8.75, -4.5) {2};
		\node [style=none] (229) at (-6.75, -7.5) {2};
		\node [style=none] (230) at (-0.5, -11) {$\{1\}$};
		\node [style=none] (231) at (-3.75, -0.5) {$a_1$};
		\node [style=none] (232) at (-3.75, -4) {$a_3$};
		\node [style=none] (233) at (-3.75, -7) {$a_5$};
		\node [style=none] (234) at (-3.75, -9.75) {$a_7$};
	\end{pgfonlayer}
	\begin{pgfonlayer}{edgelayer}
		\draw (28.center) to (29.center);
		\draw (30.center) to (31.center);
		\draw (32.center) to (33.center);
		\draw (33.center) to (34.center);
		\draw (71.center) to (72.center);
		\draw (73.center) to (74.center);
		\draw (75.center) to (76.center);
		\draw (76.center) to (77.center);
		\draw (101.center) to (102.center);
		\draw (103.center) to (104.center);
		\draw (105.center) to (106.center);
		\draw (106.center) to (107.center);
		\draw (129.center) to (130.center);
		\draw (131.center) to (132.center);
		\draw (133.center) to (134.center);
		\draw (134.center) to (135.center);
		\draw (157.center) to (158.center);
		\draw (159.center) to (160.center);
		\draw (161.center) to (162.center);
		\draw (162.center) to (163.center);
		\draw (189.center) to (190.center);
		\draw (191.center) to (192.center);
		\draw (193.center) to (194.center);
		\draw (194.center) to (195.center);
		\draw (207.center) to (208.center);
		\draw (209.center) to (210.center);
		\draw (211.center) to (212.center);
		\draw (212.center) to (213.center);
		\draw (46) to (60);
		\draw (88) to (83);
		\draw (83) to (96);
		\draw (113) to (117);
		\draw (124) to (115);
		\draw (152) to (146);
		\draw (145) to (141);
		\draw (182) to (183);
		\draw (202) to (201);
		\draw (201) to (226);
		\draw (227) to (219);
		\draw (224) to (221);
		\draw (180) to (174);
		\draw (173) to (169);
		\draw (51) to (55);
		\draw (62) to (45);
	\end{pgfonlayer}
\end{tikzpicture}
}
\caption{Hasse diagram of $\overline{\mathcal{O}}^{\mathfrak{sl}(8)}_{(2^4)}$. }
\label{fig:2222}
\end{figure}
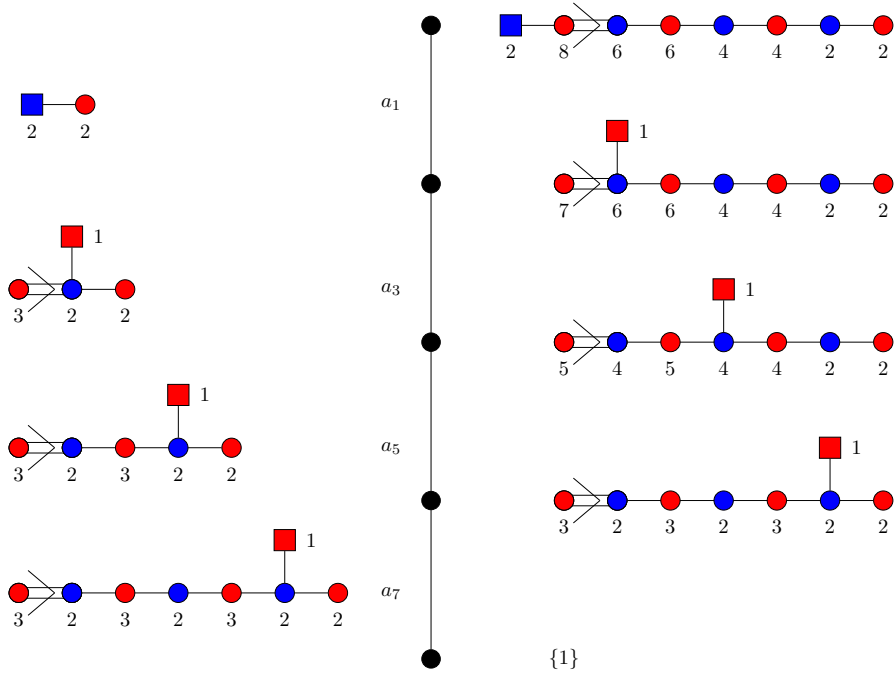

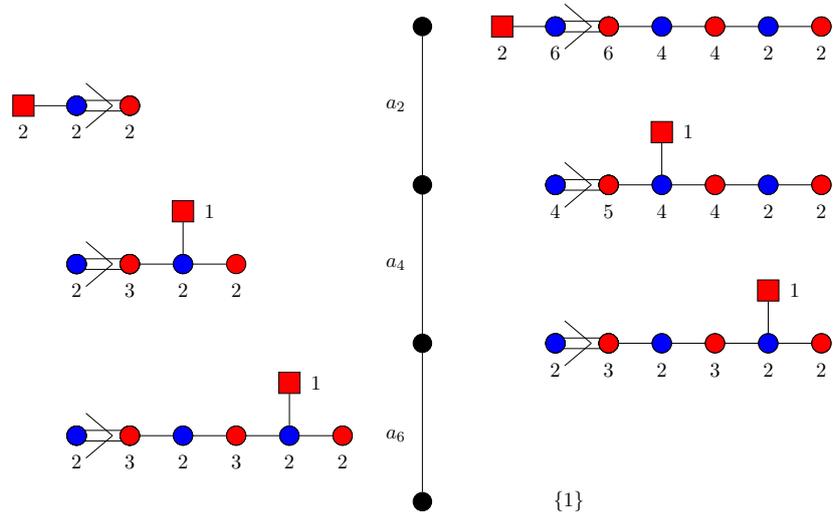
\begin{figure}[p]
\centering
\scalebox{0.7}{
\begin{tikzpicture}
	\begin{pgfonlayer}{nodelayer}
		\node [style=miniBlue] (0) at (11.5, 1) {};
		\node [style=none] (1) at (11.5, 0.5) {6};
		\node [style=miniU] (2) at (10.5, 1) {};
		\node [style=none] (3) at (10.5, 1.1) {};
		\node [style=none] (4) at (11.5, 1.1) {};
		\node [style=none] (5) at (10.5, 0.9) {};
		\node [style=none] (6) at (11.5, 0.9) {};
		\node [style=none] (7) at (10.675, 1.425) {};
		\node [style=none] (8) at (11.175, 1) {};
		\node [style=none] (9) at (10.675, 0.575) {};
		\node [style=none] (10) at (10.5, 0.5) {6};
		\node [style=miniBlue] (11) at (11.5, 1) {};
		\node [style=miniBlue] (12) at (10.5, 1) {};
		\node [style=miniU] (13) at (11.5, 1) {};
		\node [style=miniU] (14) at (10.5, 1) {};
		\node [style=miniBlue] (15) at (11.5, 1) {};
		\node [style=none] (16) at (9.5, 0.5) {2};
		\node [style=bd] (17) at (8, 1) {};
		\node [style=bd] (18) at (8, -2) {};
		\node [style=bd] (19) at (8, -5) {};
		\node [style=bd] (20) at (8, -8) {};
		\node [style=miniU] (22) at (12.5, 1) {};
		\node [style=miniBlue] (23) at (13.5, 1) {};
		\node [style=miniU] (24) at (14.5, 1) {};
		\node [style=miniBlue] (26) at (15.5, 1) {};
		\node [style=none] (28) at (12.5, 0.5) {4};
		\node [style=none] (29) at (13.5, 0.5) {4};
		\node [style=none] (30) at (14.5, 0.5) {2};
		\node [style=none] (31) at (15.5, 0.5) {2};
		\node [style=miniBlue] (33) at (11.5, -2) {};
		\node [style=miniU] (35) at (10.5, -2) {};
		\node [style=none] (36) at (10.5, -1.9) {};
		\node [style=none] (37) at (11.5, -1.9) {};
		\node [style=none] (38) at (10.5, -2.1) {};
		\node [style=none] (39) at (11.5, -2.1) {};
		\node [style=none] (40) at (10.675, -1.575) {};
		\node [style=none] (41) at (11.175, -2) {};
		\node [style=none] (42) at (10.675, -2.425) {};
		\node [style=none] (43) at (10.5, -2.5) {4};
		\node [style=miniBlue] (44) at (11.5, -2) {};
		\node [style=miniBlue] (45) at (10.5, -2) {};
		\node [style=miniU] (46) at (11.5, -2) {};
		\node [style=miniU] (47) at (10.5, -2) {};
		\node [style=miniBlue] (48) at (11.5, -2) {};
		\node [style=miniU] (49) at (12.5, -2) {};
		\node [style=miniBlue] (50) at (13.5, -2) {};
		\node [style=miniU] (51) at (14.5, -2) {};
		\node [style=miniBlue] (53) at (15.5, -2) {};
		\node [style=miniBlue] (61) at (11.5, -5) {};
		\node [style=none] (62) at (11.5, -5.5) {3};
		\node [style=miniU] (63) at (10.5, -5) {};
		\node [style=none] (64) at (10.5, -4.9) {};
		\node [style=none] (65) at (11.5, -4.9) {};
		\node [style=none] (66) at (10.5, -5.1) {};
		\node [style=none] (67) at (11.5, -5.1) {};
		\node [style=none] (68) at (10.675, -4.575) {};
		\node [style=none] (69) at (11.175, -5) {};
		\node [style=none] (70) at (10.675, -5.425) {};
		\node [style=none] (71) at (10.5, -5.5) {2};
		\node [style=miniBlue] (72) at (11.5, -5) {};
		\node [style=miniBlue] (73) at (10.5, -5) {};
		\node [style=miniU] (74) at (11.5, -5) {};
		\node [style=miniU] (75) at (10.5, -5) {};
		\node [style=miniBlue] (76) at (11.5, -5) {};
		\node [style=miniU] (77) at (12.5, -5) {};
		\node [style=miniBlue] (78) at (13.5, -5) {};
		\node [style=miniU] (79) at (14.5, -5) {};
		\node [style=miniBlue] (81) at (15.5, -5) {};
		\node [style=none] (82) at (12.5, -5.5) {2};
		\node [style=none] (83) at (13.5, -5.5) {3};
		\node [style=none] (84) at (14.5, -5.5) {2};
		\node [style=none] (85) at (15.5, -5.5) {2};
		\node [style=none] (193) at (10.75, -8) {$\{1\}$};
		\node [style=miniBlue] (194) at (10.5, 1) {};
		\node [style=miniBlue] (195) at (12.5, 1) {};
		\node [style=miniBlue] (196) at (14.5, 1) {};
		\node [style=miniBlue] (197) at (10.5, -2) {};
		\node [style=miniBlue] (198) at (12.5, -2) {};
		\node [style=miniBlue] (199) at (14.5, -2) {};
		\node [style=miniBlue] (200) at (14.5, -5) {};
		\node [style=miniBlue] (201) at (12.5, -5) {};
		\node [style=miniBlue] (202) at (10.5, -5) {};
		\node [style=miniU] (206) at (11.5, 1) {};
		\node [style=miniU] (207) at (13.5, 1) {};
		\node [style=miniU] (208) at (15.5, 1) {};
		\node [style=miniU] (209) at (15.5, -2) {};
		\node [style=miniU] (210) at (13.5, -2) {};
		\node [style=miniU] (211) at (11.5, -2) {};
		\node [style=miniU] (212) at (15.5, -5) {};
		\node [style=miniU] (213) at (13.5, -5) {};
		\node [style=miniU] (214) at (11.5, -5) {};
		\node [style=none] (218) at (11.5, -2.5) {5};
		\node [style=none] (219) at (12.5, -2.5) {4};
		\node [style=none] (220) at (13.5, -2.5) {4};
		\node [style=none] (221) at (14.5, -2.5) {2};
		\node [style=none] (222) at (15.5, -2.5) {2};
		\node [style=miniBlue] (223) at (2.5, -0.5) {};
		\node [style=none] (224) at (2.5, -1) {2};
		\node [style=miniU] (225) at (1.5, -0.5) {};
		\node [style=none] (226) at (1.5, -0.4) {};
		\node [style=none] (227) at (2.5, -0.4) {};
		\node [style=none] (228) at (1.5, -0.6) {};
		\node [style=none] (229) at (2.5, -0.6) {};
		\node [style=none] (230) at (1.675, -0.075) {};
		\node [style=none] (231) at (2.175, -0.5) {};
		\node [style=none] (232) at (1.675, -0.925) {};
		\node [style=none] (233) at (1.5, -1) {2};
		\node [style=miniBlue] (234) at (2.5, -0.5) {};
		\node [style=miniBlue] (235) at (1.5, -0.5) {};
		\node [style=miniU] (236) at (2.5, -0.5) {};
		\node [style=miniU] (237) at (1.5, -0.5) {};
		\node [style=miniBlue] (238) at (2.5, -0.5) {};
		\node [style=miniBlue] (243) at (1.5, -0.5) {};
		\node [style=miniU] (244) at (2.5, -0.5) {};
		\node [style=flavourRed] (245) at (0.5, -0.5) {};
		\node [style=none] (246) at (0.5, -1) {2};
		\node [style=flavourRed] (247) at (9.5, 1) {};
		\node [style=flavourRed] (248) at (12.5, -1) {};
		\node [style=none] (249) at (13, -1) {1};
		\node [style=miniBlue] (250) at (2.5, -3.5) {};
		\node [style=miniU] (251) at (1.5, -3.5) {};
		\node [style=none] (252) at (1.5, -3.4) {};
		\node [style=none] (253) at (2.5, -3.4) {};
		\node [style=none] (254) at (1.5, -3.6) {};
		\node [style=none] (255) at (2.5, -3.6) {};
		\node [style=none] (256) at (1.675, -3.075) {};
		\node [style=none] (257) at (2.175, -3.5) {};
		\node [style=none] (258) at (1.675, -3.925) {};
		\node [style=none] (259) at (1.5, -4) {2};
		\node [style=miniBlue] (260) at (2.5, -3.5) {};
		\node [style=miniBlue] (261) at (1.5, -3.5) {};
		\node [style=miniU] (262) at (2.5, -3.5) {};
		\node [style=miniU] (263) at (1.5, -3.5) {};
		\node [style=miniBlue] (264) at (2.5, -3.5) {};
		\node [style=miniU] (265) at (3.5, -3.5) {};
		\node [style=miniBlue] (266) at (1.5, -3.5) {};
		\node [style=miniBlue] (267) at (3.5, -3.5) {};
		\node [style=miniU] (268) at (2.5, -3.5) {};
		\node [style=none] (269) at (2.5, -4) {3};
		\node [style=none] (270) at (3.5, -4) {2};
		\node [style=flavourRed] (271) at (3.5, -2.5) {};
		\node [style=none] (272) at (4, -2.5) {1};
		\node [style=miniU] (273) at (4.5, -3.5) {};
		\node [style=none] (274) at (4.5, -4) {2};
		\node [style=flavourRed] (275) at (14.5, -4) {};
		\node [style=none] (276) at (15, -4) {1};
		\node [style=miniBlue] (278) at (2.5, -6.75) {};
		\node [style=none] (279) at (2.5, -7.25) {3};
		\node [style=miniU] (280) at (1.5, -6.75) {};
		\node [style=none] (281) at (1.5, -6.65) {};
		\node [style=none] (282) at (2.5, -6.65) {};
		\node [style=none] (283) at (1.5, -6.85) {};
		\node [style=none] (284) at (2.5, -6.85) {};
		\node [style=none] (285) at (1.675, -6.325) {};
		\node [style=none] (286) at (2.175, -6.75) {};
		\node [style=none] (287) at (1.675, -7.175) {};
		\node [style=none] (288) at (1.5, -7.25) {2};
		\node [style=miniBlue] (289) at (2.5, -6.75) {};
		\node [style=miniBlue] (290) at (1.5, -6.75) {};
		\node [style=miniU] (291) at (2.5, -6.75) {};
		\node [style=miniU] (292) at (1.5, -6.75) {};
		\node [style=miniBlue] (293) at (2.5, -6.75) {};
		\node [style=miniU] (294) at (3.5, -6.75) {};
		\node [style=miniBlue] (295) at (4.5, -6.75) {};
		\node [style=miniU] (296) at (5.5, -6.75) {};
		\node [style=miniBlue] (297) at (6.5, -6.75) {};
		\node [style=none] (298) at (3.5, -7.25) {2};
		\node [style=none] (299) at (4.5, -7.25) {3};
		\node [style=none] (300) at (5.5, -7.25) {2};
		\node [style=none] (301) at (6.5, -7.25) {2};
		\node [style=miniBlue] (302) at (5.5, -6.75) {};
		\node [style=miniBlue] (303) at (3.5, -6.75) {};
		\node [style=miniBlue] (304) at (1.5, -6.75) {};
		\node [style=miniU] (305) at (6.5, -6.75) {};
		\node [style=miniU] (306) at (4.5, -6.75) {};
		\node [style=miniU] (307) at (2.5, -6.75) {};
		\node [style=flavourRed] (308) at (5.5, -5.75) {};
		\node [style=none] (309) at (6, -5.75) {1};
		\node [style=none] (310) at (7.5, -0.5) {$a_2$};
		\node [style=none] (311) at (7.5, -3.5) {$a_4$};
		\node [style=none] (312) at (7.5, -6.75) {$a_6$};
	\end{pgfonlayer}
	\begin{pgfonlayer}{edgelayer}
		\draw (3.center) to (4.center);
		\draw (5.center) to (6.center);
		\draw (7.center) to (8.center);
		\draw (8.center) to (9.center);
		\draw (36.center) to (37.center);
		\draw (38.center) to (39.center);
		\draw (40.center) to (41.center);
		\draw (41.center) to (42.center);
		\draw (64.center) to (65.center);
		\draw (66.center) to (67.center);
		\draw (68.center) to (69.center);
		\draw (69.center) to (70.center);
		\draw (226.center) to (227.center);
		\draw (228.center) to (229.center);
		\draw (230.center) to (231.center);
		\draw (231.center) to (232.center);
		\draw (243) to (245);
		\draw (206) to (208);
		\draw (247) to (194);
		\draw (211) to (209);
		\draw (198) to (248);
		\draw (252.center) to (253.center);
		\draw (254.center) to (255.center);
		\draw (256.center) to (257.center);
		\draw (257.center) to (258.center);
		\draw (267) to (271);
		\draw (267) to (268);
		\draw (273) to (267);
		\draw (214) to (212);
		\draw (275) to (200);
		\draw (281.center) to (282.center);
		\draw (283.center) to (284.center);
		\draw (285.center) to (286.center);
		\draw (286.center) to (287.center);
		\draw (307) to (305);
		\draw (308) to (302);
		\draw (17) to (20);
	\end{pgfonlayer}
\end{tikzpicture}}
\caption{Hasse diagram for $\overline{\mathcal{O}}^{\mathfrak{sl}(7)}_{(2^3,1)}$}
\label{fig:22211}
\end{figure}

From a brane perspective, subtracting orthosymplectic quivers for $\mathfrak{so}(2n)$ orbits corresponds to Kraft-Procesi transitions \cite{Cabrera:2017njm}.\footnote{For general orthosymplectic quivers, while the rules for electric quiver subtractions are straightforward, there are many complications when dealing with magnetic quivers, such as the choice of $\mathrm{SO/O}$ gauge groups, and shifts between SO(odd) vs SO(even) gauge nodes to obtain nodes of the desired dimensions and rank, all of which make the recipe for subtraction quite involved.} It turns out that all one needs for a Kraft-Procesi decomposition of the families of orthosymplectic quivers treated in this paper are the following subtraction guides \cite{Cabrera:2019dob}, which apply to subtraction of \emph{special minimal} nilpotent orbits:
\begin{subequations}
\label{rules}
\begin{align}
    \mathrm{SO}(2k)-\mathrm{SO}(2r) &\rightarrow \mathrm{SO}(2k-2r+1), \\
    \mathrm{SO}(2k+1)-\mathrm{SO}(2r+1) &\rightarrow \mathrm{SO}(2k-2r+1), \\
   \mathrm{USp}(2k) - \mathrm{USp}(2r) & \rightarrow \mathrm{USp}(2k-2r).
\end{align} 
\end{subequations}
This means that one can subtract minimal nilpotent orbit closures of type $a_n$, $d_n$, $e_{6,7,8}$. 

To begin with, focus on a framed orthosymplectic quiver explored in Section \ref{framed}. Taking \eqref{foldedeven} with $n=4$, one obtains the Hasse diagram shown in Figure \ref{fig:2222}. The quivers on the right have Coulomb branches that are closures of the symplectic leaves denoted by \scalebox{0.8}{
}. The quivers on the left are the subtracted quivers, where the labeling $a_{n-1}$ means the Coulomb branch is the closure of the minimal nilpotent orbit of $\mathfrak{sl}(n)$. The result is consistent with the Hasse diagram of the $\overline{\mathcal{O}}^{\mathfrak{sl}(8)}_{(2^4)}$ \cite{Kraft1982}. Considering \eqref{foldedodd}, instead, for the case $n=3$, the Hasse diagram is shown in Figure \ref{fig:22211}. 
This is the Hasse diagram for $\overline{\mathcal{O}}^{\mathfrak{sl}(7)}_{(2^3,1)}$ \cite{Kraft1982}.

\begin{figure}[t]
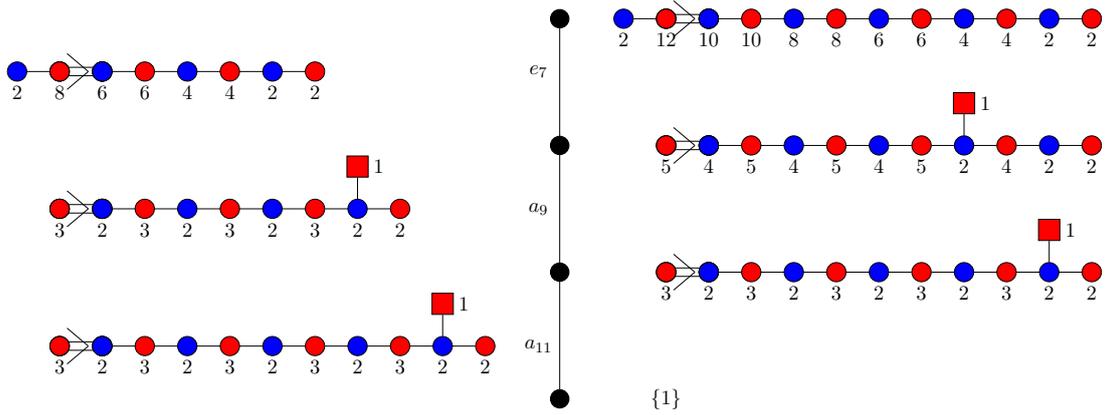

\centering
\scalebox{0.7}{
    %
}
\caption{Hasse diagram of the folded $E_8$ family for $k=3$.}
\label{fig:Hasse_E_8}
\end{figure}
\begin{figure}[t]
\centering
   \scalebox{0.7}{ 
   %
}
\caption{Hasse diagram of the folded $E_7$ family for $k=3$.}
\label{fig:Hasse_E_7}
\end{figure}
\begin{figure}[t]
\centering
    \scalebox{0.7}{
    %
}
\caption{Hasse diagram of the folded $E_6$ family for $k=3$.}
\label{fig:Hasse_E_6}
\end{figure}

\begin{figure}[t]
\centering
  \scalebox{0.7}{  
  %
}   
\caption{Hasse diagram of the folded $E_{4-2l}$ family for $k=3$ and $l=1$.}
\label{fig:Hasse_E_4-2l}
\end{figure}
\begin{figure}[t]
\centering
 \scalebox{0.7}{
 %
} 
\caption{Hasse diagram of the folded $E_{3-2l}$ family for $k=3$ and $l=1$.}
\label{fig:Hasse_E_3-2l}
\end{figure}

\subsection{General families}
Now, one performs quiver subtraction to obtain Hasse diagrams for the general families tabulated in Table \ref{generalfamily}. The only difference here is that the orthosymplectic quiver is unframed. The first non-simply laced quiver one subtracts are those in Table \ref{Entable} whose Coulomb branches are also closures of minimal nilpotent orbits. The subtraction rules for unframed orthosymplectic magnetic quivers are incomplete. However, the unframed orthosymplectic quivers (before folding) are studied in \cite{Hanany:2018uhm,Cabrera:2019dob,Bourget:2020gzi,Bourget:2020xdz} where the Hasse diagrams are obtained through transitions on brane webs. By observing those transitions, one realises that the rules for subtracting the nodes for these families are the same as \eqref{rules}. The resulting Hasse diagrams are again what one expects through quiver subtraction on their unitary quiver counterparts. Below, the Hasse diagram is drawn for one example from each of the families as the general Hasse diagram is already known from the unitary quivers and is given in \cite{Bourget:2019aer}. 

The examples for $k=3$ are shown in Figures \ref{fig:Hasse_E_8}, \ref{fig:Hasse_E_7}, \ref{fig:Hasse_E_6}, \ref{fig:Hasse_E_5}, \ref{fig:Hasse_E_4-2l}, \ref{fig:Hasse_E_3-2l}. 
In fact, for the entire $E_{4-2l}$ and $E_{3-2l}$ families ($l \geq 1$), the Coulomb branch of the first quiver to subtract is always a $A_l = \mathbb{C}^2/\mathbb{Z}_{l+1}$ Kleinian singularity given by a $\mathrm{U}(1)$ with $l+1$ charge 2 hypermultiplets. This identification follows from the rules of quiver subtraction established in \cite{Cabrera:2018ann,Bourget:2019aer,Hanany:2019tji}.

We remark that the moduli spaces for the folded $E_6$ and $E_5$ families (third and fourth rows in Table \ref{generalhwg}) coincide with the classical Higgs branches of SQCD theories, specifically $\mathrm{SU}(k+1)$ gauge theories with respectively $2k+3$ and $2k+2$ fundamental hypermultiplets. Therefore these can be realized as hyper-K\"ahler quotients.

\clearpage 

\section{Brane configurations}
\label{branes}
Inspired by the results above, one can study the respective brane configurations that give rise to the non-simply laced orthosymplectic quivers. This section focuses only on the framed orthosymplectic quivers of Section \ref{framed}. The quiver gauge theories are 3d $\mathcal{N}=4$ effective field theories living on D3 branes that are stretched between 5-branes in the presence of orientifold planes.

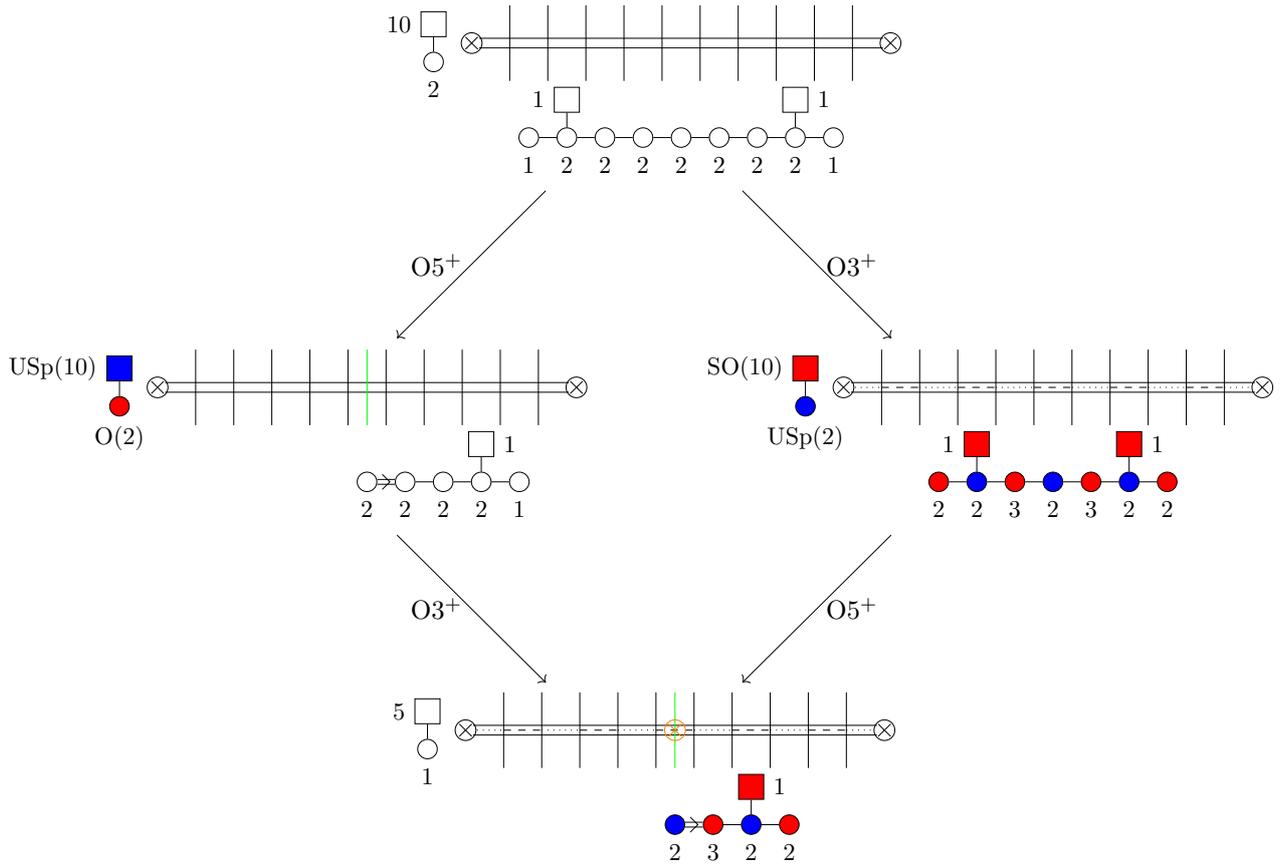
\begin{figure}[t]
    \makebox[\textwidth][c]{
\scalebox{0.911}{\begin{tikzpicture}
\node (1) at (0,0) {
\begin{tikzpicture}[scale=0.55]
\draw (0,-1)--(0,1) (1,-1)--(1,1) (2,-1)--(2,1) (3,-1)--(3,1) (4,-1)--(4,1) (5,-1)--(5,1) (6,-1)--(6,1) (7,-1)--(7,1) (8,-1)--(8,1) (9,-1)--(9,1);
\node at (-1,0) {\NS};
\node at (10,0) {\NS};
\draw[transform canvas={yshift=-2pt}] (-0.8,0)--(9.8,0);
\draw[transform canvas={yshift=2pt}] (-0.8,0)--(9.8,0);
\node[gaw,label=below:{\small 1}] (a) at (0.5,-2.5) {};
\node[gaw,label=below:{\small 2}] (b) at (1.5,-2.5) {};
\node[gaw,label=below:{\small 2}] (c) at (2.5,-2.5) {};
\node[gaw,label=below:{\small 2}] (d) at (3.5,-2.5) {};
\node[gaw,label=below:{\small 2}] (e) at (4.5,-2.5) {};
\node[gaw,label=below:{\small 2}] (f) at (5.5,-2.5) {};
\node[gaw,label=below:{\small 2}] (g) at (6.5,-2.5) {};
\node[gaw,label=below:{\small 2}] (h) at (7.5,-2.5) {};
\node[gaw,label=below:{\small 1}] (i) at (8.5,-2.5) {};
\node[flaw,label=left:{\small 1}] (b1) at (1.5,-1.5) {};
\node[flaw,label=right:{\small 1}] (h1) at (7.5,-1.5) {};
\draw (a)--(b)--(c)--(d)--(e)--(f)--(g)--(h)--(i) (b)--(b1) (h)--(h1);
\node[gaw, label=below:{\small 2}] (ea) at (-2,-0.5) {};
\node[flaw,label=left:{\small 10}] (ea1) at (-2,0.5) {};
\draw (ea)--(ea1);
\end{tikzpicture}};
\node (2) at (-5,-5) {\begin{tikzpicture}[scale=0.55]
\draw (0,-1)--(0,1) (1,-1)--(1,1) (2,-1)--(2,1) (3,-1)--(3,1) (4,-1)--(4,1) (5,-1)--(5,1) (6,-1)--(6,1) (7,-1)--(7,1) (8,-1)--(8,1) (9,-1)--(9,1);
\node at (-1,0) {\NS};
\node at (10,0) {\NS};
\draw[transform canvas={yshift=-2pt}] (-0.8,0)--(9.8,0);
\draw[transform canvas={yshift=2pt}] (-0.8,0)--(9.8,0);
\draw[green] (4.5,-1)--(4.5,1);
\node[gaw,label=below:{\small 2}] (e) at (4.5,-2.5) {};
\node[gaw,label=below:{\small 2}] (f) at (5.5,-2.5) {};
\node[gaw,label=below:{\small 2}] (g) at (6.5,-2.5) {};
\node[gaw,label=below:{\small 2}] (h) at (7.5,-2.5) {};
\node[gaw,label=below:{\small 1}] (i) at (8.5,-2.5) {};
\node[flaw,label=right:{\small 1}] (h1) at (7.5,-1.5) {};
\draw (f)--(g)--(h)--(i) (h)--(h1);
\draw[transform canvas={yshift=-1pt}] (e)--(f);
\draw[transform canvas={yshift=1pt}] (e)--(f);
\draw (4.9,-2.3)--(5.1,-2.5)--(4.9,-2.7);
\node[gar, label=below:{\small O(2)}] (ea) at (-2,-0.5) {};
\node[flab,label=left:{\small $\mathrm{USp}(10)$}] (ea1) at (-2,0.5) {};
\draw (ea)--(ea1);
\end{tikzpicture}};
\node (3) at (5,-5) {\begin{tikzpicture}[scale=0.55]
\draw (0,-1)--(0,1) (1,-1)--(1,1) (2,-1)--(2,1) (3,-1)--(3,1) (4,-1)--(4,1) (5,-1)--(5,1) (6,-1)--(6,1) (7,-1)--(7,1) (8,-1)--(8,1) (9,-1)--(9,1);
\node at (-1,0) {\NS};
\node at (10,0) {\NS};
\draw[transform canvas={yshift=-2pt}] (-0.8,0)--(9.8,0);
\draw[transform canvas={yshift=2pt}] (-0.8,0)--(9.8,0);
\draw[dotted] (-0.7,0)--(0,0) (1,0)--(2,0) (3,0)--(4,0) (5,0)--(6,0) (7,0)--(8,0) (9,0)--(9.7,0);
\draw[dashed] (0,0)--(1,0) (2,0)--(3,0) (4,0)--(5,0) (6,0)--(7,0) (8,0)--(9,0);
\node[gar,label=below:{\small 2}] (b) at (1.5,-2.5) {};
\node[gab,label=below:{\small 2}] (c) at (2.5,-2.5) {};
\node[gar,label=below:{\small 3}] (d) at (3.5,-2.5) {};
\node[gab,label=below:{\small 2}] (e) at (4.5,-2.5) {};
\node[gar,label=below:{\small 3}] (f) at (5.5,-2.5) {};
\node[gab,label=below:{\small 2}] (g) at (6.5,-2.5) {};
\node[gar,label=below:{\small 2}] (h) at (7.5,-2.5) {};
\node[flar,label=left:{\small 1}] (c1) at (2.5,-1.5) {};
\node[flar,label=right:{\small 1}] (g1) at (6.5,-1.5) {};
\draw (b)--(c)--(d)--(e)--(f)--(g)--(h) (c)--(c1) (g)--(g1);
\node[gab, label=below:{\small $\mathrm{USp}(2)$}] (ea) at (-2,-0.5) {};
\node[flar,label=left:{\small SO(10)}] (ea1) at (-2,0.5) {};
\draw (ea)--(ea1);
\end{tikzpicture}};
\node (4) at (0,-10) {\begin{tikzpicture}[scale=0.55]
\draw (0,-1)--(0,1) (1,-1)--(1,1) (2,-1)--(2,1) (3,-1)--(3,1) (4,-1)--(4,1) (5,-1)--(5,1) (6,-1)--(6,1) (7,-1)--(7,1) (8,-1)--(8,1) (9,-1)--(9,1);
\node at (-1,0) {\NS};
\node at (10,0) {\NS};
\draw[transform canvas={yshift=-2pt}] (-0.8,0)--(9.8,0);
\draw[transform canvas={yshift=2pt}] (-0.8,0)--(9.8,0);
\draw[green] (4.5,-1)--(4.5,1);
\draw[dotted] (-0.7,0)--(0,0) (1,0)--(2,0) (3,0)--(4,0) (5,0)--(6,0) (7,0)--(8,0) (9,0)--(9.7,0);
\draw[dashed] (0,0)--(1,0) (2,0)--(3,0) (4,0)--(5,0) (6,0)--(7,0) (8,0)--(9,0);
\node at (4.5,0) {\ONp};
\node[gab,label=below:{\small 2}] (e) at (4.5,-2.5) {};
\node[gar,label=below:{\small 3}] (f) at (5.5,-2.5) {};
\node[gab,label=below:{\small 2}] (g) at (6.5,-2.5) {};
\node[gar,label=below:{\small 2}] (h) at (7.5,-2.5) {};
\node[flar,label=right:{\small 1}] (g1) at (6.5,-1.5) {};
\draw (f)--(g)--(h) (g)--(g1);
\draw[transform canvas={yshift=-1pt}] (e)--(f);
\draw[transform canvas={yshift=1pt}] (e)--(f);
\draw (4.9,-2.3)--(5.1,-2.5)--(4.9,-2.7);
\node[gaw, label=below:{\small 1}] (ea) at (-2,-0.5) {};
\node[flaw,label=left:{\small 5}] (ea1) at (-2,0.5) {};
\draw (ea)--(ea1);
\end{tikzpicture}};
\draw[->] (1)--(2);
\draw[->] (2)--(4);
\draw[->] (1)--(3);
\draw[->] (3)--(4);
\node at (-3,-2.5) {O$5^+$};
\node at (3,-7.5) {O$5^+$};
\node at (3,-2.5) {O$3^+$};
\node at (-3,-7.5) {O$3^+$};
\end{tikzpicture}}}
\caption{The commutative diamond represents the construction of the non-simply laced orthosymplectic quiver. Vertical lines depict D$5$ branes, circles with crosses NS$5$ branes, and horizontal lines D$3$ branes. The vertical green line depicts an O$5^+$ plane, the horizontal dotted line an O$3^+$ plane, the horizontal dashed line an $\widetilde{\text{O}}3{}^+$ plane, and the orange circle with a cross an ON${}^+$ plane. Electric (left) and magnetic (below) quivers for the brane systems are provided. The magnetic quivers are most conveniently read after suitable Hanany-Witten transitions. The O$5
^+$ insertion on the left is the brane realisation of what is called folding in \cite{Bourget:2020bxh}.}
\label{fig:diamond}
\end{figure}

To begin with, consider the magnetic quiver in Figure \ref{fig:diamond}, whose Coulomb branch is the next-to-minimal nilpotent orbit of $\mathfrak{sl}(2n)$. The brane configuration is presented on the right. By inserting O3 planes, one produces an orthosymplectic quiver that is the minimal nilpotent orbit of $\mathfrak{so}(2n)$. One can insert an O5${}^+$ plane in the centre in combination with the presence of the O3 planes\footnote{One can also insert O5${}^-$ plane which will give a B-type non-simply laced orthosymplectic quiver. Such quivers will be studied in detail somewhere else.}. An ON${}^+$ plane then sits at the intersection of the two orientifolds \cite{Kutasov:1995te}. The proposal is that the resulting brane configuration is the folded orthosymplectic quiver whose Coulomb branch is the minimal nilpotent orbit of $\mathfrak{sl}(n)$, which is expected as the gauge theory on 2k D3 branes between two NS5 branes in the presence of this intersection of orientifolds is U(k) \cite{Hanany:1997gh,Gulotta:2012yd}. This shows the insertion of O3 and O5 branes is a commutative process.

Similarly, one can start with the $\mathfrak{sl}(2n)$ unitary quiver and insert an O5${}^+$ plane in the middle. This creates a non-simply laced unitary quiver \cite{Cremonesi:2014xha} whose Coulomb branch is the next-to-minimal orbit of $\mathfrak{usp}(2n)$. Inserting O3 planes after that reproduces the same folded orthosymplectic quiver as above.

\subsection{Generalisation of height two quivers}
The above procedure can be generalised by starting with a unitary quiver whose Coulomb branch is the closure of a general height 2 $\mathfrak{sl}(2n)$ nilpotent orbit.

\paragraph{$n$ even.}
For the maximal height 2 orbit of $\mathfrak{sl}(2n)$ the partition is $(2^n)$. Adding O3 planes to the brane configuration produces the orthosymplectic quiver in \eqref{evenorbit} whose Coulomb branch is (one of the two identical cones) of the  $(2^n)$ orbit of $\mathfrak{so}(2n)$. Adding O5 planes then yield the folded orthosymplectic quiver \eqref{foldedeven} whose Coulomb branch is the $(2^{\frac{n}{2}})$ orbit of $\mathfrak{sl}(n)$. This process is demonstrated in Figure \ref{height2max}.

The remaining height 2 orbits of $\mathfrak{sl}(2n)$ are identified by the partition $(2^{k+1},1^{2n-4k-2})$ with $n>2k$. Adding an orientifold produces the orthosymplectic quiver in \eqref{smallerorbitseven} whose Coulomb branch is the closure of $(2^{2k},{1^{2n-4k}})$ orbit of $\mathfrak{so}(2n)$. Adding an O5${}^+$ plane reproduces \eqref{smallfoldedeven} whose Coulomb branch is the closure of $(2^k,{1^{n-2k}})$ orbit of $\mathfrak{sl}(n)$. This process is demonstrated in Figure \ref{evencase}.

\paragraph{$n$ odd.}
For $\mathfrak{sl}(2n)$ and $n$ odd,  height 2 orbits are identified by the partition $(2^{2k+1},1^{2n-4k-2})$ for $n\geq 2k+1$. Putting O3 planes reproduces \eqref{beforefoldodd} for $n=2k+1$ and \eqref{smallerorbitsodd} for $n>2k+1$ whose Coulomb branches are closures of $(2^{n-1},1^2)$ and $(2^{2k-1},1^{2n-4k+2})$ nilpotent orbits of $\mathfrak{so}(2n)$ respectively. Adding an O5${}^+$ plane gives the folded orthosymplectic quiver in \eqref{foldedodd} and \eqref{foldedoddd} whose Coulomb branches are the closure of the $(2^{\frac{n-1}{2}},1)$ and $(2^k,1^{n-2k})$ orbit of $\mathfrak{sl}(n)$ respectively. This process is demonstrated in Figure \ref{oddcase}.

\begin{figure}[t]
    \centering
\scalebox{0.811}{\begin{tikzpicture}
	\begin{pgfonlayer}{nodelayer}
		\node [style=gauge3] (0) at (0, 2.5) {};
		\node [style=gauge3] (1) at (-1, 2.5) {};
		\node [style=gauge3] (2) at (1, 2.5) {};
		\node [style=flavour2] (3) at (0, 3.5) {};
		\node [style=none] (4) at (0, 4) {2};
		\node [style=none] (8) at (1.5, 2.5) {};
		\node [style=none] (9) at (1.75, 2.5) {\dots};
		\node [style=none] (10) at (2, 2.5) {};
		\node [style=none] (11) at (-1.5, 2.5) {};
		\node [style=none] (12) at (-1.75, 2.5) {$\dots$};
		\node [style=none] (13) at (-2, 2.5) {};
		\node [style=gauge3] (14) at (2.5, 2.5) {};
		\node [style=gauge3] (15) at (-2.5, 2.5) {};
		\node [style=gauge3] (16) at (-3.5, 2.5) {};
		\node [style=gauge3] (18) at (3.5, 2.5) {};
		\node [style=none] (19) at (-2.5, 2) {2};
		\node [style=none] (20) at (-3.5, 2) {1};
		\node [style=none] (21) at (2.5, 2) {2};
		\node [style=none] (22) at (3.5, 2) {1};
		\node [style=miniU] (25) at (0, -2.5) {};
		\node [style=miniBlue] (26) at (1, -2.5) {};
		\node [style=miniBlue] (27) at (-1, -2.5) {};
		\node [style=none] (28) at (1.75, -2.5) {\dots};
		\node [style=none] (29) at (-1.75, -2.5) {\dots};
		\node [style=miniU] (30) at (2.5, -2.5) {};
		\node [style=miniBlue] (31) at (3.5, -2.5) {};
		\node [style=miniBlue] (32) at (-3.5, -2.5) {};
		\node [style=miniU] (33) at (-2.5, -2.5) {};
		\node [style=miniU] (34) at (-4.5, -2.5) {};
		\node [style=flavourBlue] (35) at (0, -1.5) {};
		\node [style=miniU] (36) at (4.5, -2.5) {};
		\node [style=none] (37) at (1.5, -2.5) {};
		\node [style=none] (38) at (2, -2.5) {};
		\node [style=none] (39) at (-1.5, -2.5) {};
		\node [style=none] (40) at (-2, -2.5) {};
		\node [style=none] (41) at (0, -1) {\small{$2$}};
		\node [style=none] (42) at (0, -3) {\small{$n$}};
		\node [style=none] (43) at (1, -3) {\small{$n{-}2$}};
		\node [style=none] (44) at (-1, -3) {\small{$n{-}2$}};
		\node [style=none] (45) at (-2.5, -3) {\small{$4$}};
		\node [style=none] (46) at (-3.5, -3) {\small{$2$}};
		\node [style=none] (47) at (-4.5, -3) {\small{$2$}};
		\node [style=none] (48) at (2.5, -3) {\small{$4$}};
		\node [style=none] (49) at (3.5, -3) {\small{$2$}};
		\node [style=none] (50) at (4.5, -3) {\small{$2$}};
		\node [style=none] (53) at (0.7, 0.75) {O$3^+$};
		\node [style=none] (54) at (0.75, -4.75) {O$5^+$};
		\node [style=miniBlue] (55) at (-0.75, -6.75) {};
		\node [style=none] (56) at (0, -6.75) {\dots};
		\node [style=miniU] (57) at (0.75, -6.75) {};
		\node [style=miniBlue] (58) at (1.75, -6.75) {};
		\node [style=miniU] (59) at (2.75, -6.75) {};
		\node [style=none] (60) at (-0.25, -6.75) {};
		\node [style=none] (61) at (0.25, -6.75) {};
		\node [style=none] (62) at (-0.75, -7.25) {\small{$n{-}2$}};
		\node [style=none] (63) at (0.75, -7.25) {\small{$4$}};
		\node [style=none] (64) at (1.75, -7.25) {\small{$2$}};
		\node [style=none] (65) at (2.75, -7.25) {\small{$2$}};
		\node [style=miniU] (66) at (-1.75, -6.75) {};
		\node [style=none] (67) at (-1.75, -6.65) {};
		\node [style=none] (68) at (-0.75, -6.65) {};
		\node [style=none] (69) at (-1.75, -6.85) {};
		\node [style=none] (70) at (-0.75, -6.85) {};
		\node [style=none] (71) at (-1.575, -6.325) {};
		\node [style=none] (72) at (-1.075, -6.75) {};
		\node [style=none] (73) at (-1.575, -7.175) {};
		\node [style=none] (74) at (-1.75, -7.25) {\small{$n$}};
		\node [style=miniBlue] (75) at (-0.75, -6.75) {};
		\node [style=flavourBlue] (76) at (-2.75, -6.75) {};
		\node [style=none] (77) at (-2.75, -7.25) {\small{$2$}};
		\node [style=gauge3] (82) at (8.5, -6.75) {};
		\node [style=gauge3] (83) at (7.5, -6.75) {};
		\node [style=gauge3] (84) at (9.5, -6.75) {};
		\node [style=flavour2] (85) at (8.5, -5.75) {};
		\node [style=none] (86) at (9, -5.75) {\small{$2$}};
		\node [style=none] (87) at (8.5, -7.25) {\small{$\frac{n}{2}$}};
		\node [style=none] (88) at (9.5, -7.25) {\small{$\frac{n}{2}{-}1$}};
		\node [style=none] (89) at (7.5, -7.25) {\small{$\frac{n}{2}{-}1$}};
		\node [style=none] (90) at (10, -6.75) {};
		\node [style=none] (91) at (10.25, -6.75) {\dots};
		\node [style=none] (92) at (10.5, -6.75) {};
		\node [style=none] (93) at (7, -6.75) {};
		\node [style=none] (94) at (6.75, -6.75) {$\dots$};
		\node [style=none] (95) at (6.5, -6.75) {};
		\node [style=gauge3] (96) at (11, -6.75) {};
		\node [style=gauge3] (97) at (6, -6.75) {};
		\node [style=gauge3] (98) at (5, -6.75) {};
		\node [style=gauge3] (99) at (12, -6.75) {};
		\node [style=none] (100) at (6, -7.25) {\small{$2$}};
		\node [style=none] (101) at (5, -7.25) {\small{$1$}};
		\node [style=none] (102) at (11, -7.25) {\small{$2$}};
		\node [style=none] (103) at (12, -7.25) {\small{$1$}};
		\node [style=none] (104) at (-1, 2) {\small{$n{-}1$}};
		\node [style=none] (105) at (0, 2) {\small{$n$}};
		\node [style=none] (106) at (1, 2) {\small{$n{-}1$}};
		\node [style=none] (107) at (4, -6.75) {=};
		\node [style=none] (108) at (0, 1.25) {};
		\node [style=none] (109) at (0, -0.25) {};
		\node [style=none] (110) at (0, -4) {};
		\node [style=none] (111) at (0, -5.5) {};
	\end{pgfonlayer}
	\begin{pgfonlayer}{edgelayer}
		\draw (0) to (3);
		\draw (0) to (2);
		\draw (0) to (1);
		\draw (13.center) to (15);
		\draw (11.center) to (1);
		\draw (2) to (8.center);
		\draw (10.center) to (14);
		\draw (15) to (16);
		\draw (14) to (18);
		\draw (27) to (25);
		\draw (25) to (35);
		\draw (25) to (26);
		\draw (30) to (31);
		\draw (33) to (32);
		\draw (32) to (34);
		\draw (31) to (36);
		\draw (40.center) to (33);
		\draw (39.center) to (27);
		\draw (26) to (37.center);
		\draw (38.center) to (30);
		\draw (57) to (58);
		\draw (58) to (59);
		\draw (55) to (60.center);
		\draw (61.center) to (57);
		\draw (67.center) to (68.center);
		\draw (69.center) to (70.center);
		\draw (71.center) to (72.center);
		\draw (72.center) to (73.center);
		\draw (76) to (66);
		\draw (82) to (85);
		\draw (82) to (84);
		\draw (82) to (83);
		\draw (95.center) to (97);
		\draw (93.center) to (83);
		\draw (84) to (90.center);
		\draw (92.center) to (96);
		\draw (97) to (98);
		\draw (96) to (99);
		\draw [style=->] (108.center) to (109.center);
		\draw [style=->] (110.center) to (111.center);
	\end{pgfonlayer}
\end{tikzpicture}}
\caption{The even $n$ case. The Coulomb branch of the top quiver is the $\mathfrak{sl}(2n)$ nilpotent orbit closure with partition $(2^n)$. With the addition of $\mathrm{O3}^{+}$ planes, the following orthosymplectic quiver is the $\mathfrak{so}(2n)$ nilpotent orbit closure with partition $(2^n)$. Finally, adding $\mathrm{O5}^+$ planes, the resulting non-simply laced orthosymplectic quiver is obtained. This quiver has a unitary counterpart on the right whose Coulomb branch is the $\mathfrak{sl}(n)$ nilpotent orbit closure of partition $(2^{\frac{n}{2}})$. }
    \label{height2max}
\end{figure}

\begin{figure}[p]
    \centering
        \scalebox{0.811}{
\begin{tikzpicture}
	\begin{pgfonlayer}{nodelayer}
		\node [style=none] (53) at (0.5, 0.75) {O$3^+$};
		\node [style=none] (54) at (0.5, -4.75) {O$5^+$};
		\node [style=none] (107) at (0, -9.75) {=};
		\node [style=none] (108) at (0, 1.25) {};
		\node [style=none] (109) at (0, -0.25) {};
		\node [style=none] (110) at (0, -4) {};
		\node [style=none] (111) at (0, -5.5) {};
		\node [style=miniU] (112) at (-3.5, -1.75) {};
		\node [style=miniBlue] (113) at (-2.5, -1.75) {};
		\node [style=miniBlue] (114) at (-4.5, -1.75) {};
		\node [style=none] (115) at (5.25, -1.75) {$\cdots$};
		\node [style=none] (116) at (-5.25, -1.75) {$\cdots$};
		\node [style=miniU] (117) at (6, -1.75) {};
		\node [style=miniBlue] (118) at (7, -1.75) {};
		\node [style=miniBlue] (119) at (-7, -1.75) {};
		\node [style=miniU] (120) at (-6, -1.75) {};
		\node [style=miniU] (121) at (-8, -1.75) {};
		\node [style=miniU] (122) at (8, -1.75) {};
		\node [style=none] (123) at (5.5, -1.75) {};
		\node [style=none] (124) at (-5, -1.75) {};
		\node [style=none] (125) at (-5.5, -1.75) {};
		\node [style=miniBlue] (126) at (-3.5, -1.75) {};
		\node [style=miniBlue] (127) at (6, -1.75) {};
		\node [style=miniBlue] (128) at (8, -1.75) {};
		\node [style=miniBlue] (129) at (-6, -1.75) {};
		\node [style=miniBlue] (130) at (-8, -1.75) {};
		\node [style=miniU] (131) at (-4.5, -1.75) {};
		\node [style=miniU] (132) at (-2.5, -1.75) {};
		\node [style=miniU] (133) at (7, -1.75) {};
		\node [style=miniU] (134) at (-7, -1.75) {};
		\node [style=miniU] (135) at (-3.5, -1.75) {};
		\node [style=miniBlue] (136) at (-2.5, -1.75) {};
		\node [style=miniBlue] (137) at (-4.5, -1.75) {};
		\node [style=miniU] (138) at (6, -1.75) {};
		\node [style=miniU] (139) at (-6, -1.75) {};
		\node [style=miniBlue] (140) at (7, -1.75) {};
		\node [style=miniBlue] (141) at (-7, -1.75) {};
		\node [style=miniU] (142) at (8, -1.75) {};
		\node [style=miniU] (143) at (-8, -1.75) {};
		\node [style=miniU] (144) at (3.5, -1.75) {};
		\node [style=miniBlue] (145) at (4.5, -1.75) {};
		\node [style=none] (146) at (5, -1.75) {};
		\node [style=none] (147) at (-8, -2.25) {\small{$2$}};
		\node [style=none] (148) at (-7, -2.25) {\small{$2$}};
		\node [style=none] (149) at (-6, -2.25) {\small{$4$}};
		\node [style=miniU] (150) at (-4.5, -1.75) {};
		\node [style=miniU] (151) at (4.5, -1.75) {};
		\node [style=miniBlue] (152) at (-3.5, -1.75) {};
		\node [style=miniBlue] (153) at (3.5, -1.75) {};
		\node [style=miniU] (154) at (-2.5, -1.75) {};
		\node [style=miniU] (155) at (2.5, -1.75) {};
		\node [style=flavourRed] (156) at (-3.5, -0.75) {};
		\node [style=flavourRed] (157) at (3.5, -0.75) {};
		\node [style=none] (158) at (3.5, -0.25) {\small{$1$}};
		\node [style=none] (159) at (-3.5, -0.25) {\small{$1$}};
		\node [style=none] (160) at (-1.75, -1.75) {\dots};
		\node [style=none] (161) at (-2, -1.75) {};
		\node [style=none] (162) at (-1.5, -1.75) {};
		\node [style=none] (163) at (-4.5, -2.25) {\small{$2k$}};
		\node [style=none] (164) at (-3.5, -2.25) {\small{$2k$}};
		\node [style=none] (165) at (3.5, -2.25) {\small{$2k$}};
		\node [style=none] (166) at (4.5, -2.25) {\small{$2k$}};
		\node [style=none] (167) at (2.5, -2.25) {\small{$2k{+}1$}};
		\node [style=none] (168) at (-2.5, -2.25) {\small{$2k{+}1$}};
		\node [style=none] (169) at (-3.5, -2.5) {};
		\node [style=none] (170) at (3.5, -2.5) {};
		\node [style=none] (171) at (0, -3.15) {\small{$2n{-}4k{-}1$ nodes}};
		\node [style=none] (172) at (6, -2.25) {\small{$4$}};
		\node [style=none] (173) at (7, -2.25) {\small{$2$}};
		\node [style=none] (174) at (8, -2.25) {\small{$2$}};
		\node [style=miniBlue] (175) at (-1, -1.75) {};
		\node [style=none] (176) at (1.75, -1.75) {\dots};
		\node [style=none] (177) at (2, -1.75) {};
		\node [style=none] (178) at (1.5, -1.75) {};
		\node [style=none] (179) at (-1, -2.25) {\small{$2k$}};
		\node [style=miniBlue] (180) at (1, -1.75) {};
		\node [style=miniU] (181) at (0, -1.75) {};
		\node [style=none] (182) at (0, -2.25) {\small{$2k{+}1$}};
		\node [style=none] (183) at (1, -2.25) {\small{$2k$}};
		\node [style=miniBlue] (184) at (-2.5, -7.5) {};
		\node [style=none] (185) at (3.75, -7.5) {\dots};
		\node [style=miniU] (186) at (4.5, -7.5) {};
		\node [style=miniBlue] (187) at (5.5, -7.5) {};
		\node [style=none] (188) at (3.5, -7.5) {};
		\node [style=none] (189) at (4, -7.5) {};
		\node [style=none] (190) at (-2.5, -8) {\small{$2k$}};
		\node [style=none] (191) at (4.5, -8) {\small{$2$}};
		\node [style=none] (192) at (5.5, -8) {\small{$2$}};
		\node [style=miniU] (193) at (-3.5, -7.5) {};
		\node [style=none] (194) at (-3.5, -7.4) {};
		\node [style=none] (195) at (-2.5, -7.4) {};
		\node [style=none] (196) at (-3.5, -7.6) {};
		\node [style=none] (197) at (-2.5, -7.6) {};
		\node [style=none] (198) at (-3.325, -7.075) {};
		\node [style=none] (199) at (-2.825, -7.5) {};
		\node [style=none] (200) at (-3.325, -7.925) {};
		\node [style=none] (201) at (-3.5, -8) {\small{$2k{+}1$}};
		\node [style=miniBlue] (202) at (-2.5, -7.5) {};
		\node [style=miniU] (203) at (5.5, -7.5) {};
		\node [style=miniBlue] (204) at (4.5, -7.5) {};
		\node [style=miniU] (205) at (-1.5, -7.5) {};
		\node [style=none] (206) at (-1.5, -8) {\small{$2k+1$}};
		\node [style=none] (207) at (-0.75, -7.5) {\dots};
		\node [style=none] (208) at (-1, -7.5) {};
		\node [style=none] (209) at (-0.5, -7.5) {};
		\node [style=miniBlue] (210) at (0, -7.5) {};
		\node [style=miniU] (211) at (1, -7.5) {};
		\node [style=none] (212) at (0, -8) {\small{$2k$}};
		\node [style=none] (213) at (1, -8) {\small{$2k{+}1$}};
		\node [style=miniBlue] (214) at (2, -7.5) {};
		\node [style=none] (215) at (2, -8) {\small{$2k$}};
		\node [style=flavourRed] (216) at (2, -6.5) {};
		\node [style=none] (217) at (2, -6) {1};
		\node [style=miniU] (218) at (3, -7.5) {};
		\node [style=none] (219) at (3, -8) {\small{$2k$}};
		\node [style=none] (220) at (-3.5, -8.25) {};
		\node [style=none] (221) at (2, -8.25) {};
		\node [style=none] (222) at (-0.75, -9) {\small{$n{-}2k$ nodes}};
		\node [style=gauge3] (223) at (-1.75, -11.5) {};
		\node [style=gauge3] (224) at (1.75, -11.5) {};
		\node [style=gauge3] (225) at (2.75, -11.5) {};
		\node [style=gauge3] (226) at (-2.75, -11.5) {};
		\node [style=none] (227) at (3.5, -11.5) {\dots};
		\node [style=none] (228) at (3.25, -11.5) {};
		\node [style=none] (229) at (3.75, -11.5) {};
		\node [style=none] (230) at (-3.5, -11.5) {\dots};
		\node [style=none] (231) at (-3.75, -11.5) {};
		\node [style=none] (232) at (-3.25, -11.5) {};
		\node [style=gauge3] (233) at (4.25, -11.5) {};
		\node [style=gauge3] (234) at (-4.25, -11.5) {};
		\node [style=gauge3] (235) at (-5.25, -11.5) {};
		\node [style=gauge3] (236) at (5.25, -11.5) {};
		\node [style=none] (237) at (5.25, -12) {1};
		\node [style=none] (238) at (4.25, -12) {2};
		\node [style=flavour2] (239) at (-1.75, -10.5) {};
		\node [style=flavour2] (240) at (1.75, -10.5) {};
		\node [style=none] (241) at (-5.25, -12) {\small{$1$}};
		\node [style=none] (242) at (-4.25, -12) {\small{$2$}};
		\node [style=none] (243) at (-1.75, -10) {\small{$1$}};
		\node [style=none] (244) at (1.75, -10) {\small{$1$}};
		\node [style=gauge3] (245) at (-0.75, -11.5) {};
		\node [style=gauge3] (246) at (0.75, -11.5) {};
		\node [style=none] (247) at (0, -11.5) {\dots};
		\node [style=none] (248) at (-0.25, -11.5) {};
		\node [style=none] (249) at (0.25, -11.5) {};
		\node [style=none] (250) at (-2.75, -12) {\small{$k{-}1$}};
		\node [style=none] (251) at (-1.75, -12) {\small{$k$}};
		\node [style=none] (252) at (-0.75, -12) {\small{$k$}};
		\node [style=none] (253) at (0.75, -12) {\small{$k$}};
		\node [style=none] (254) at (1.75, -12) {\small{$k$}};
		\node [style=none] (255) at (2.75, -12) {\small{$k{-}1$}};
		\node [style=none] (256) at (1.75, -12.3) {};
		\node [style=none] (257) at (-1.75, -12.3) {};
		\node [style=none] (258) at (0, -12.8) {\small{$n{-}2k{+}1$ nodes}};
		\node [style=gauge3] (259) at (-1.75, 3.25) {};
		\node [style=gauge3] (260) at (1.75, 3.25) {};
		\node [style=gauge3] (261) at (2.75, 3.25) {};
		\node [style=gauge3] (262) at (-2.75, 3.25) {};
		\node [style=none] (263) at (3.5, 3.25) {\dots};
		\node [style=none] (264) at (3.25, 3.25) {};
		\node [style=none] (265) at (3.75, 3.25) {};
		\node [style=none] (266) at (-3.5, 3.25) {\dots};
		\node [style=none] (267) at (-3.75, 3.25) {};
		\node [style=none] (268) at (-3.25, 3.25) {};
		\node [style=gauge3] (269) at (4.25, 3.25) {};
		\node [style=gauge3] (270) at (-4.25, 3.25) {};
		\node [style=gauge3] (271) at (-5.25, 3.25) {};
		\node [style=gauge3] (272) at (5.25, 3.25) {};
		\node [style=none] (273) at (5.25, 2.75) {\small{$1$}};
		\node [style=none] (274) at (4.25, 2.75) {2};
		\node [style=flavour2] (275) at (-1.75, 4.25) {};
		\node [style=flavour2] (276) at (1.75, 4.25) {};
		\node [style=none] (277) at (-5.25, 2.75) {\small{$1$}};
		\node [style=none] (278) at (-4.25, 2.75) {\small{$2$}};
		\node [style=none] (279) at (-1.75, 4.75) {\small{$1$}};
		\node [style=none] (280) at (1.75, 4.75) {\small{$1$}};
		\node [style=gauge3] (281) at (-0.75, 3.25) {};
		\node [style=gauge3] (282) at (0.75, 3.25) {};
		\node [style=none] (283) at (0, 3.25) {\dots};
		\node [style=none] (284) at (-0.25, 3.25) {};
		\node [style=none] (285) at (0.25, 3.25) {};
		\node [style=none] (286) at (-2.75, 2.75) {\small{$2k$}};
		\node [style=none] (287) at (-1.75, 2.75) {\scriptsize{$2k+1$}};
		\node [style=none] (288) at (-0.75, 2.75) {\scriptsize{$2k+1$}};
		\node [style=none] (289) at (0.75, 2.75) {\scriptsize{$2k+1$}};
		\node [style=none] (290) at (1.75, 2.75) {\scriptsize{$2k+1$}};
		\node [style=none] (291) at (2.75, 2.75) {\scriptsize{$2k$}};
		\node [style=none] (292) at (1.75, 2.45) {};
		\node [style=none] (293) at (-1.75, 2.45) {};
		\node [style=none] (294) at (0, 1.95) {\small{$2n{-}4k{-}1$ nodes}};
	\end{pgfonlayer}
	\begin{pgfonlayer}{edgelayer}
		\draw [style=->] (108.center) to (109.center);
		\draw [style=->] (110.center) to (111.center);
		\draw (114) to (112);
		\draw (112) to (113);
		\draw (117) to (118);
		\draw (120) to (119);
		\draw (119) to (121);
		\draw (118) to (122);
		\draw (125.center) to (120);
		\draw (124.center) to (114);
		\draw (123.center) to (117);
		\draw (145) to (146.center);
		\draw (145) to (144);
		\draw (152) to (156);
		\draw (157) to (153);
		\draw (154) to (161.center);
		\draw (155) to (153);
		\draw [style=brace1] (170.center) to (169.center);
		\draw (175) to (162.center);
		\draw (180) to (178.center);
		\draw (155) to (177.center);
		\draw (181) to (175);
		\draw (181) to (180);
		\draw (186) to (187);
		\draw (189.center) to (186);
		\draw (194.center) to (195.center);
		\draw (196.center) to (197.center);
		\draw (198.center) to (199.center);
		\draw (199.center) to (200.center);
		\draw (205) to (202);
		\draw (208.center) to (205);
		\draw (210) to (209.center);
		\draw (210) to (211);
		\draw (214) to (211);
		\draw (216) to (214);
		\draw (214) to (218);
		\draw (218) to (188.center);
		\draw [style=brace1] (221.center) to (220.center);
		\draw (223) to (226);
		\draw (224) to (225);
		\draw (228.center) to (225);
		\draw (229.center) to (233);
		\draw (226) to (232.center);
		\draw (231.center) to (234);
		\draw (239) to (223);
		\draw (224) to (240);
		\draw (234) to (235);
		\draw (233) to (236);
		\draw (223) to (245);
		\draw (246) to (224);
		\draw (248.center) to (245);
		\draw (249.center) to (246);
		\draw [style=brace1] (256.center) to (257.center);
		\draw (259) to (262);
		\draw (260) to (261);
		\draw (264.center) to (261);
		\draw (265.center) to (269);
		\draw (262) to (268.center);
		\draw (267.center) to (270);
		\draw (275) to (259);
		\draw (260) to (276);
		\draw (270) to (271);
		\draw (269) to (272);
		\draw (259) to (281);
		\draw (282) to (260);
		\draw (284.center) to (281);
		\draw (285.center) to (282);
		\draw [style=brace1] (292.center) to (293.center);
	\end{pgfonlayer}
\end{tikzpicture}
}
\caption{The even $n$ case. The Coulomb branch of the top quiver is the $\mathfrak{sl}(2n)$ nilpotent orbit closure with partition $(2^{2k+1},1^{2n-4k-2})$. With the addition of $\mathrm{O3}^+$ planes, the following orthosymplectic quiver is the $\mathfrak{so}(2n)$ nilpotent orbit closure with partition $(2^{2k},1^{2n-4k})$. Finally, adding $\mathrm{O5}^+$ planes, the resulting non-simply laced orthosymplectic quiver is obtained. This quiver has a unitary counterpart on the right whose Coulomb branch is the $\mathfrak{sl}(n)$ nilpotent orbit closure of partition $(2^{k},1^{n-2k})$. }
    \label{evencase}
\end{figure}

\begin{figure}[t]
    \centering
    \scalebox{0.811}{
\begin{tikzpicture}
	\begin{pgfonlayer}{nodelayer}
		\node [style=none] (53) at (0.5, 0.75) {O$3^+$};
		\node [style=none] (54) at (0.5, -4.75) {O$5^+$};
		\node [style=none] (107) at (0, -9.75) {=};
		\node [style=none] (108) at (0, 1.25) {};
		\node [style=none] (109) at (0, -0.25) {};
		\node [style=none] (110) at (0, -4) {};
		\node [style=none] (111) at (0, -5.5) {};
		\node [style=gauge3] (223) at (-1.75, -11.5) {};
		\node [style=gauge3] (224) at (1.75, -11.5) {};
		\node [style=gauge3] (225) at (2.75, -11.5) {};
		\node [style=gauge3] (226) at (-2.75, -11.5) {};
		\node [style=none] (227) at (3.5, -11.5) {\dots};
		\node [style=none] (228) at (3.25, -11.5) {};
		\node [style=none] (229) at (3.75, -11.5) {};
		\node [style=none] (230) at (-3.5, -11.5) {\dots};
		\node [style=none] (231) at (-3.75, -11.5) {};
		\node [style=none] (232) at (-3.25, -11.5) {};
		\node [style=gauge3] (233) at (4.25, -11.5) {};
		\node [style=gauge3] (234) at (-4.25, -11.5) {};
		\node [style=gauge3] (235) at (-5.25, -11.5) {};
		\node [style=gauge3] (236) at (5.25, -11.5) {};
		\node [style=none] (237) at (5.25, -12) {1};
		\node [style=none] (238) at (4.25, -12) {2};
		\node [style=flavour2] (239) at (-1.75, -10.5) {};
		\node [style=flavour2] (240) at (1.75, -10.5) {};
		\node [style=none] (241) at (-5.25, -12) {\small{$1$}};
		\node [style=none] (242) at (-4.25, -12) {\small{$2$}};
		\node [style=none] (243) at (-1.75, -10) {\small{$1$}};
		\node [style=none] (244) at (1.75, -10) {\small{$1$}};
		\node [style=gauge3] (245) at (-0.75, -11.5) {};
		\node [style=gauge3] (246) at (0.75, -11.5) {};
		\node [style=none] (247) at (0, -11.5) {\dots};
		\node [style=none] (248) at (-0.25, -11.5) {};
		\node [style=none] (249) at (0.25, -11.5) {};
		\node [style=none] (250) at (-2.75, -12) {\small{$k{-}1$}};
		\node [style=none] (251) at (-1.75, -12) {\small{$k$}};
		\node [style=none] (252) at (-0.75, -12) {\small{$k$}};
		\node [style=none] (253) at (0.75, -12) {\small{$k$}};
		\node [style=none] (254) at (1.75, -12) {\small{$k$}};
		\node [style=none] (255) at (2.75, -12) {\small{$k{-}1$}};
		\node [style=none] (256) at (1.75, -12.3) {};
		\node [style=none] (257) at (-1.75, -12.3) {};
		\node [style=none] (258) at (0, -12.8) {\small{$n{-}2k{+}1$ nodes}};
		\node [style=gauge3] (259) at (-1.75, 3.25) {};
		\node [style=gauge3] (260) at (1.75, 3.25) {};
		\node [style=gauge3] (261) at (2.75, 3.25) {};
		\node [style=gauge3] (262) at (-2.75, 3.25) {};
		\node [style=none] (263) at (3.5, 3.25) {\dots};
		\node [style=none] (264) at (3.25, 3.25) {};
		\node [style=none] (265) at (3.75, 3.25) {};
		\node [style=none] (266) at (-3.5, 3.25) {\dots};
		\node [style=none] (267) at (-3.75, 3.25) {};
		\node [style=none] (268) at (-3.25, 3.25) {};
		\node [style=gauge3] (269) at (4.25, 3.25) {};
		\node [style=gauge3] (270) at (-4.25, 3.25) {};
		\node [style=gauge3] (271) at (-5.25, 3.25) {};
		\node [style=gauge3] (272) at (5.25, 3.25) {};
		\node [style=none] (273) at (5.25, 2.75) {\small{$1$}};
		\node [style=none] (274) at (4.25, 2.75) {\small{$2$}};
		\node [style=flavour2] (275) at (-1.75, 4.25) {};
		\node [style=flavour2] (276) at (1.75, 4.25) {};
		\node [style=none] (277) at (-5.25, 2.75) {\small{$1$}};
		\node [style=none] (278) at (-4.25, 2.75) {\small{$2$}};
		\node [style=none] (279) at (-1.75, 4.75) {\small{$1$}};
		\node [style=none] (280) at (1.75, 4.75) {\small{$1$}};
		\node [style=gauge3] (281) at (-0.75, 3.25) {};
		\node [style=gauge3] (282) at (0.75, 3.25) {};
		\node [style=none] (283) at (0, 3.25) {\dots};
		\node [style=none] (284) at (-0.25, 3.25) {};
		\node [style=none] (285) at (0.25, 3.25) {};
		\node [style=none] (286) at (-2.75, 2.75) {\small{$2k$}};
		\node [style=none] (287) at (-1.75, 2.75) {\small{$2k{+}1$}};
		\node [style=none] (288) at (-0.75, 2.75) {\small{$2k{+}1$}};
		\node [style=none] (289) at (0.75, 2.75) {\small{$2k{+}1$}};
		\node [style=none] (290) at (1.75, 2.75) {\small{$2k{+}1$}};
		\node [style=none] (291) at (2.75, 2.75) {\small{$k$}};
		\node [style=none] (292) at (1.75, 2.45) {};
		\node [style=none] (293) at (-1.75, 2.45) {};
		\node [style=none] (294) at (0, 1.95) {\small{$2n{-}4k{-}1$ nodes}};
		\node [style=miniU] (295) at (-3.5, -1.5) {};
		\node [style=miniBlue] (296) at (-2.5, -1.5) {};
		\node [style=miniBlue] (297) at (-4.5, -1.5) {};
		\node [style=none] (298) at (5.25, -1.5) {\dots};
		\node [style=none] (299) at (-5.25, -1.5) {\dots};
		\node [style=miniU] (300) at (6, -1.5) {};
		\node [style=miniBlue] (301) at (7, -1.5) {};
		\node [style=miniBlue] (302) at (-7, -1.5) {};
		\node [style=miniU] (303) at (-6, -1.5) {};
		\node [style=miniU] (304) at (-8, -1.5) {};
		\node [style=miniU] (305) at (8, -1.5) {};
		\node [style=none] (306) at (5.5, -1.5) {};
		\node [style=none] (307) at (-5, -1.5) {};
		\node [style=none] (308) at (-5.5, -1.5) {};
		\node [style=miniBlue] (309) at (-3.5, -1.5) {};
		\node [style=miniBlue] (310) at (6, -1.5) {};
		\node [style=miniBlue] (311) at (8, -1.5) {};
		\node [style=miniBlue] (312) at (-6, -1.5) {};
		\node [style=miniBlue] (313) at (-8, -1.5) {};
		\node [style=miniU] (314) at (-4.5, -1.5) {};
		\node [style=miniU] (315) at (-2.5, -1.5) {};
		\node [style=miniU] (316) at (7, -1.5) {};
		\node [style=miniU] (317) at (-7, -1.5) {};
		\node [style=miniU] (318) at (-3.5, -1.5) {};
		\node [style=miniBlue] (319) at (-2.5, -1.5) {};
		\node [style=miniBlue] (320) at (-4.5, -1.5) {};
		\node [style=miniU] (321) at (6, -1.5) {};
		\node [style=miniU] (322) at (-6, -1.5) {};
		\node [style=miniBlue] (323) at (7, -1.5) {};
		\node [style=miniBlue] (324) at (-7, -1.5) {};
		\node [style=miniU] (325) at (8, -1.5) {};
		\node [style=miniU] (326) at (-8, -1.5) {};
		\node [style=miniU] (327) at (3.5, -1.5) {};
		\node [style=miniBlue] (328) at (4.5, -1.5) {};
		\node [style=none] (329) at (5, -1.5) {};
		\node [style=none] (330) at (-8, -2) {\small{$2$}};
		\node [style=none] (331) at (-7, -2) {\small{$2$}};
		\node [style=none] (332) at (-6, -2) {\small{$4$}};
		\node [style=miniU] (333) at (-4.5, -1.5) {};
		\node [style=miniU] (334) at (4.5, -1.5) {};
		\node [style=miniBlue] (335) at (-3.5, -1.5) {};
		\node [style=miniBlue] (336) at (3.5, -1.5) {};
		\node [style=miniU] (337) at (-2.5, -1.5) {};
		\node [style=miniU] (338) at (2.5, -1.5) {};
		\node [style=flavourRed] (339) at (-3.5, -0.5) {};
		\node [style=flavourRed] (340) at (3.5, -0.5) {};
		\node [style=none] (341) at (3.5, 0) {\small{$1$}};
		\node [style=none] (342) at (-3.5, 0) {\small{$1$}};
		\node [style=none] (343) at (-1.75, -1.5) {\dots};
		\node [style=none] (344) at (-2, -1.5) {};
		\node [style=none] (345) at (-1.5, -1.5) {};
		\node [style=none] (346) at (-4.5, -2) {\small{$2k$}};
		\node [style=none] (347) at (-3.5, -2) {\small{$2k$}};
		\node [style=none] (348) at (3.5, -2) {\small{$2k$}};
		\node [style=none] (349) at (4.5, -2) {\small{$2k$}};
		\node [style=none] (350) at (2.5, -2) {\small{$2k{+}1$}};
		\node [style=none] (351) at (-2.5, -2) {\small{$2k{+}1$}};
		\node [style=none] (352) at (-3.5, -2.25) {};
		\node [style=none] (353) at (3.5, -2.25) {};
		\node [style=none] (354) at (0, -2.9) {\small{$2n{-}4k{-}1$ nodes}};
		\node [style=none] (355) at (6, -2) {\small{$4$}};
		\node [style=none] (356) at (7, -2) {\small{$2$}};
		\node [style=none] (357) at (8, -2) {\small{$2$}};
		\node [style=miniBlue] (358) at (-1, -1.5) {};
		\node [style=none] (359) at (1.75, -1.5) {\dots};
		\node [style=none] (360) at (2, -1.5) {};
		\node [style=none] (361) at (1.5, -1.5) {};
		\node [style=none] (362) at (-1, -2) {\small{$2k{+}1$}};
		\node [style=miniBlue] (363) at (1, -1.5) {};
		\node [style=miniU] (364) at (0, -1.5) {};
		\node [style=none] (365) at (0, -2) {\small{$2k$}};
		\node [style=none] (366) at (1, -2) {\small{$2k{+}1$}};
		\node [style=miniU] (367) at (-1, -1.5) {};
		\node [style=miniU] (368) at (1, -1.5) {};
		\node [style=miniBlue] (369) at (0, -1.5) {};
		\node [style=miniBlue] (370) at (-2.75, -7.25) {};
		\node [style=none] (371) at (3.5, -7.25) {\dots};
		\node [style=miniU] (372) at (4.25, -7.25) {};
		\node [style=miniBlue] (373) at (5.25, -7.25) {};
		\node [style=none] (374) at (3.25, -7.25) {};
		\node [style=none] (375) at (3.75, -7.25) {};
		\node [style=none] (376) at (-2.75, -7.75) {\small{$2k{+}1$}};
		\node [style=none] (377) at (4.25, -7.75) {\small{$2$}};
		\node [style=none] (378) at (5.25, -7.75) {\small{$2$}};
		\node [style=miniU] (379) at (-3.75, -7.25) {};
		\node [style=none] (380) at (-3.75, -7.15) {};
		\node [style=none] (381) at (-2.75, -7.15) {};
		\node [style=none] (382) at (-3.75, -7.35) {};
		\node [style=none] (383) at (-2.75, -7.35) {};
		\node [style=none] (384) at (-3.575, -6.825) {};
		\node [style=none] (385) at (-3.075, -7.25) {};
		\node [style=none] (386) at (-3.575, -7.675) {};
		\node [style=none] (387) at (-3.75, -7.75) {\small{$2k$}};
		\node [style=miniBlue] (388) at (-2.75, -7.25) {};
		\node [style=miniU] (389) at (5.25, -7.25) {};
		\node [style=miniBlue] (390) at (4.25, -7.25) {};
		\node [style=miniU] (391) at (-1.75, -7.25) {};
		\node [style=none] (392) at (-1.75, -7.75) {\small{$2k$}};
		\node [style=none] (393) at (-1, -7.25) {\dots};
		\node [style=none] (394) at (-1.25, -7.25) {};
		\node [style=none] (395) at (-0.75, -7.25) {};
		\node [style=miniBlue] (396) at (-0.25, -7.25) {};
		\node [style=miniU] (397) at (0.75, -7.25) {};
		\node [style=none] (398) at (-0.25, -7.75) {\small{$2k$}};
		\node [style=none] (399) at (0.75, -7.75) {\small{$2k{+}1$}};
		\node [style=miniBlue] (400) at (1.75, -7.25) {};
		\node [style=none] (401) at (1.75, -7.75) {\small{$2k$}};
		\node [style=flavourRed] (402) at (1.75, -6.25) {};
		\node [style=none] (403) at (1.75, -5.75) {1};
		\node [style=miniU] (404) at (2.75, -7.25) {};
		\node [style=none] (405) at (2.75, -7.75) {\small{$2k$}};
		\node [style=none] (406) at (-3.75, -8) {};
		\node [style=none] (407) at (1.75, -8) {};
		\node [style=none] (408) at (-1, -8.55) {\small{$n{-}k{-}1$ nodes}};
		\node [style=miniBlue] (409) at (-3.75, -7.25) {};
		\node [style=miniBlue] (410) at (-1.75, -7.25) {};
		\node [style=miniU] (411) at (-2.75, -7.25) {};
	\end{pgfonlayer}
	\begin{pgfonlayer}{edgelayer}
		\draw [style=->] (108.center) to (109.center);
		\draw [style=->] (110.center) to (111.center);
		\draw (223) to (226);
		\draw (224) to (225);
		\draw (228.center) to (225);
		\draw (229.center) to (233);
		\draw (226) to (232.center);
		\draw (231.center) to (234);
		\draw (239) to (223);
		\draw (224) to (240);
		\draw (234) to (235);
		\draw (233) to (236);
		\draw (223) to (245);
		\draw (246) to (224);
		\draw (248.center) to (245);
		\draw (249.center) to (246);
		\draw [style=brace1] (256.center) to (257.center);
		\draw (259) to (262);
		\draw (260) to (261);
		\draw (264.center) to (261);
		\draw (265.center) to (269);
		\draw (262) to (268.center);
		\draw (267.center) to (270);
		\draw (275) to (259);
		\draw (260) to (276);
		\draw (270) to (271);
		\draw (269) to (272);
		\draw (259) to (281);
		\draw (282) to (260);
		\draw (284.center) to (281);
		\draw (285.center) to (282);
		\draw [style=brace1] (292.center) to (293.center);
		\draw (297) to (295);
		\draw (295) to (296);
		\draw (300) to (301);
		\draw (303) to (302);
		\draw (302) to (304);
		\draw (301) to (305);
		\draw (308.center) to (303);
		\draw (307.center) to (297);
		\draw (306.center) to (300);
		\draw (328) to (329.center);
		\draw (328) to (327);
		\draw (335) to (339);
		\draw (340) to (336);
		\draw (337) to (344.center);
		\draw (338) to (336);
		\draw [style=brace1] (353.center) to (352.center);
		\draw (358) to (345.center);
		\draw (363) to (361.center);
		\draw (338) to (360.center);
		\draw (364) to (358);
		\draw (364) to (363);
		\draw (372) to (373);
		\draw (375.center) to (372);
		\draw (380.center) to (381.center);
		\draw (382.center) to (383.center);
		\draw (384.center) to (385.center);
		\draw (385.center) to (386.center);
		\draw (391) to (388);
		\draw (394.center) to (391);
		\draw (396) to (395.center);
		\draw (396) to (397);
		\draw (400) to (397);
		\draw (402) to (400);
		\draw (400) to (404);
		\draw (404) to (374.center);
		\draw [style=brace] (407.center) to (406.center);
	\end{pgfonlayer}
\end{tikzpicture}
}
\caption{The odd $n$ case. The Coulomb branch of the top quiver is the $\mathfrak{sl}(2n)$ nilpotent orbit closure with partition $(2^{2k+1},1^{2n-4k-2})$. With the addition of $\mathrm{O3}^+$ planes, the following orthosymplectic quiver is the $\mathfrak{so}(2n)$ nilpotent orbit closure with partition $(2^{2k-1},1^{2n-4k+2})$. Finally, adding $\mathrm{O5}^+$ planes, the resulting non-simply laced orthosymplectic quiver is obtained. This quiver has a unitary counterpart on the right whose Coulomb branch is the $\mathfrak{sl}(n)$ nilpotent orbit closure of partition $(2^{k},1^{n-2k})$.}
    \label{oddcase}
\end{figure}
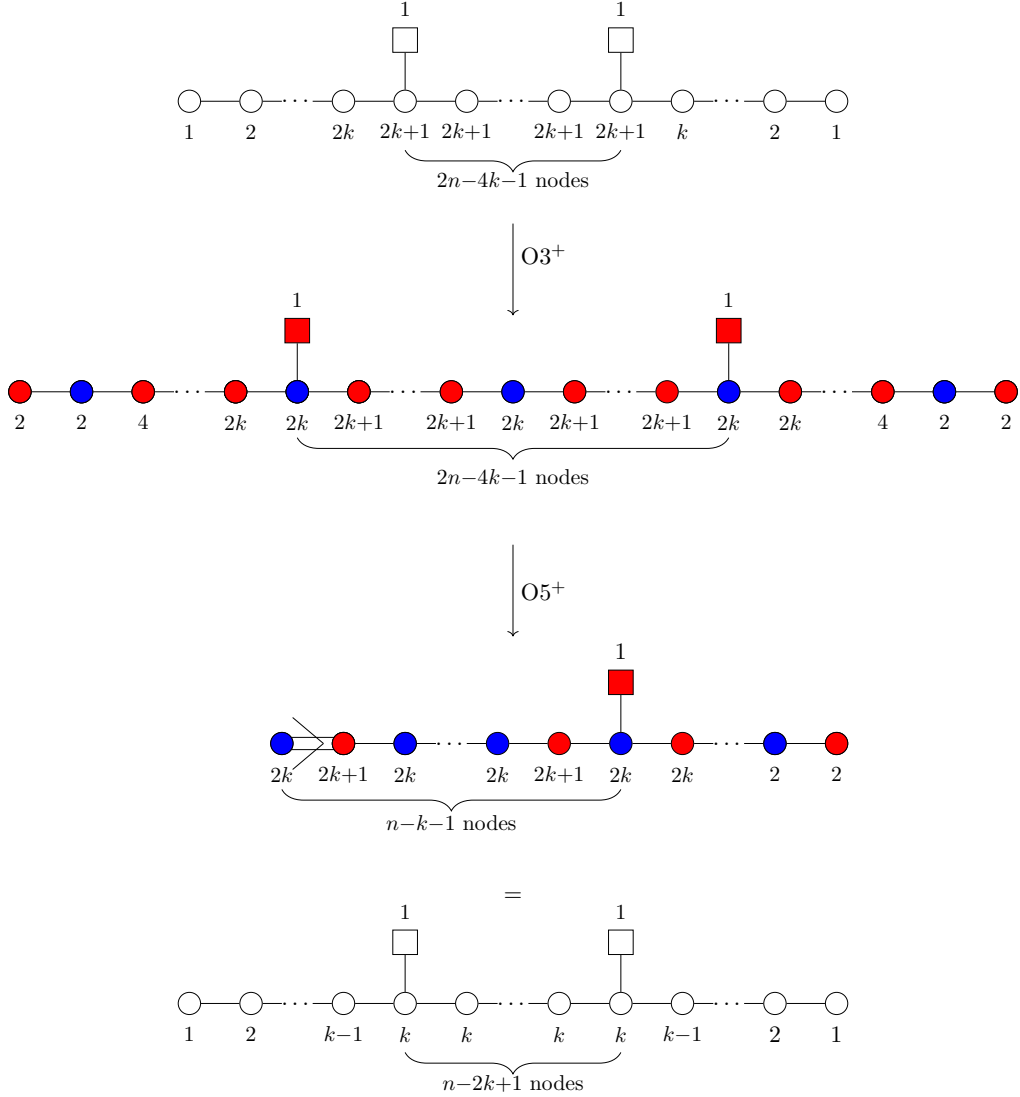

\clearpage

\subsection{Kraft-Procesi transitions}
\begin{figure}[p]
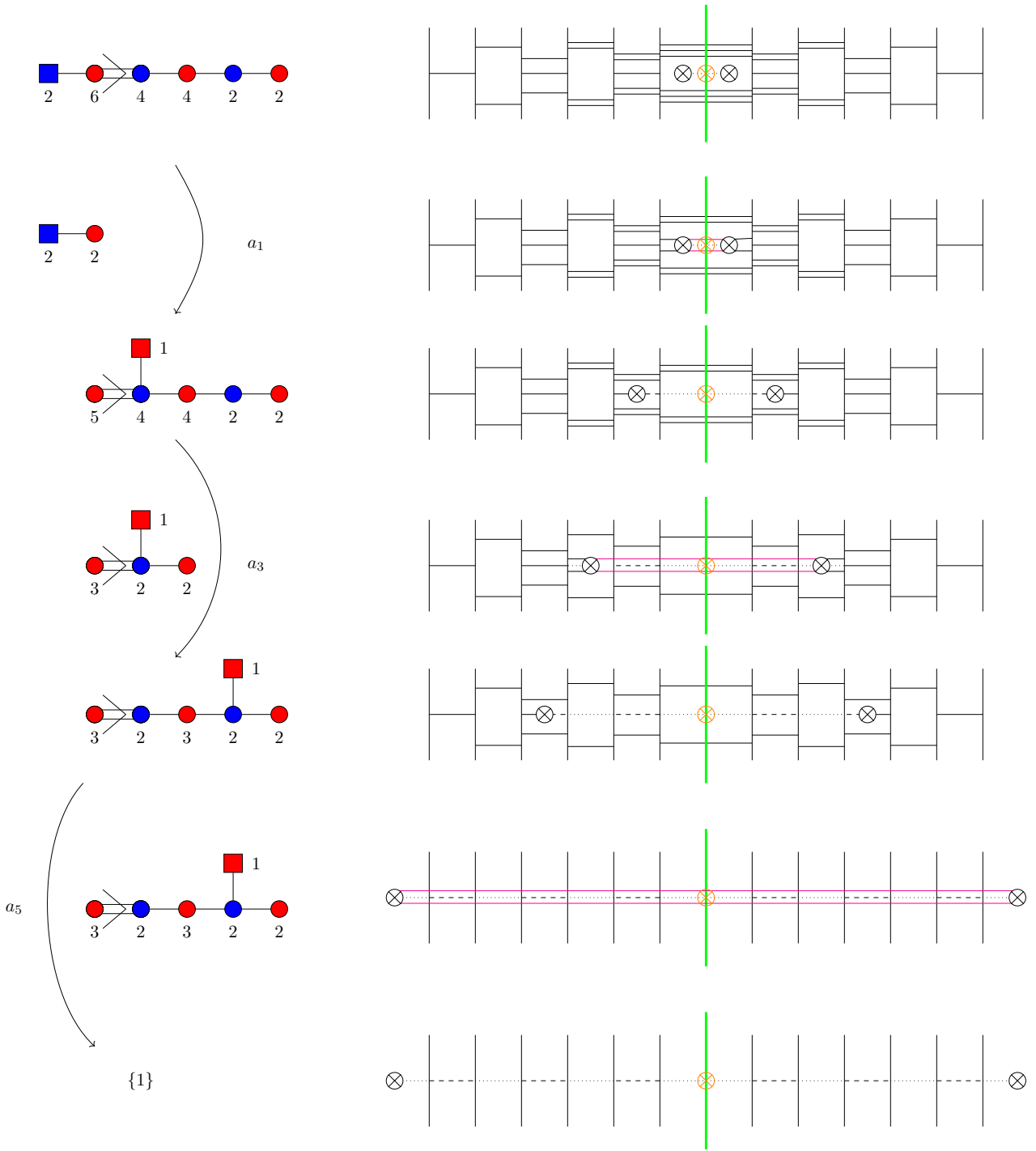

 \hspace*{-1cm}\scalebox{0.75}{


}
\caption{Deriving the Hasse diagram via quiver subtraction and Kraft-Procesi transitions. The left-hand side details the quiver subtraction algorithm; while the right-hand side shows the corresponding Kraft-Procesi transitions in the brane configurations. The notation for the branes and O3 planes follows \cite{Cabrera:2017njm}. The branes colored in magenta correspond to the subtracted figures on the left side.}
\label{KPT}
\end{figure}

In Section \ref{Hasse}, the different phases of a quiver theory are studied via the Hasse diagram. The phase diagram can be explicitly derived via Kraft-Procesi transitions \cite{Cabrera:2017njm} in a brane configuration. For concreteness, consider the brane configuration at the top of Figure \ref{KPT}, which displays a Higgs branch phase of the D3-D5-NS5 system. The corresponding magnetic quiver for this Higgs branch is displayed on the left and describes the closure of the maximal height 2 orbit of $\mathfrak{sl}(6)$. A Kraft-Procesi transition is realised by moving a full D3 brane onto the O3 plane such that the D3 can split on the NS5 branes. The D3 segment, which is solely suspended between NS5 branes, gives rise to an electric theory that characterises the KP transition via its Higgs branch. In other words, this is the transverse slice and the associated magnetic quiver is depicted next to the arrow in Figure \ref{KPT}. Moving this D3 brane segment along the NS5 branes implies that this modulus has left the Higgs branch and entered the Coulomb branch. The remaining brane configuration, displayed at the centre of Figure \ref{KPT}, has an associated magnetic quiver which accounts for the symplectic leaf below the top. The next KP transition proceeds as before, segments of full D3 branes are aligned and moved onto the O3 plane such that the resulting D3 can split on the NS5 brane. The Higgs branch of the electric theory characterises the transverse slice and Figure \ref{KPT} displays the associated magnetic quiver. After moving this D3 onto the Coulomb branch, the remaining parts of the Hasse diagram are obtained by repeating KP transition until all D3 branes are moved onto the Coulomb branch. Since there are no Higgs branch moduli left, the magnetic quiver is trivial. This is the trivial symplectic leaf of the Hasse diagram.

The Kraft-Procesi transitions are almost identical to those in \cite{Cabrera:2017njm} and, therefore, are only shown for one example. The only complication is that the presence of both O3 and O5 planes fixes the ON plane.

\section{Conclusion and outlook}
\label{conclusion}
In this article, the folding of orthosymplectic quivers, both framed and unframed, has been studied. In the case of framed orthosymplectic quivers, whose Coulomb branches are  $\mathfrak{so}(2n)$ nilpotent orbit closures of height 2, folding gives non-simply laced quivers whose Coulomb branches are $\mathfrak{sl}(n)$ nilpotent orbit closures of height 2. Amongst unframed orthosymplectic quivers, the magnetic quivers of the 5d $E_n$ families discussed in \cite{Bourget:2020gzi} have been treated. The Coulomb branches of the folded quivers have known unitary quiver counterparts, and in some cases are magnetic quivers for known 5d $\mathcal{N}=1$ or 4d $\mathcal{N}=2$ theories.
The brane systems studied alternatively allow for $\widetilde{\mathrm{O}5}^+$ and O$5^-$ orientifold planes, giving rise to orthosymplectic quivers of B and D type. We leave the study of these quivers for future work.
It would furthermore be interesting to study five-brane webs including O$5$ and O$7^+$ orientifold planes, which are expected to yield the unframed non-simply laced orthosymplectic magnetic quivers studied in this work.

\section*{Acknowledgements}
We would like to thank Satoshi Nawata, Dan Xie and Gabi Zafrir for helpful discussions. This work is supported by STFC grants ST/P000762/1 and ST/T000791/1.
M.S. is supported by the National Natural Science Foundation of China (grant no. 11950410497), and the China Postdoctoral Science Foundation (grant no. 2019M650616). Z.Z. is thankful for the kind hospitality of YMSC and Tsinghua University where part of this work is completed.

\appendix

\section{Hall Littlewood polynomials and star shaped quivers}
\label{app:HL}

We can calculate the Coulomb branch Hilbert series of many 3d ${\cal N}=4$ unframed unitary and/or orthosymplectic star shaped quivers ${\mathsf{Q}}(J)$, that are characterised by a central node with gauge group $J$, by gluing resolved Slodowy slices. The procedure depends upon being able to identify the linear quivers that form the legs of the star shaped quiver as Slodowy slices; these may be transverse to orbits in the GNO dual group $J^\vee$, or in a twist related simply laced dual $J'$, as discussed later, see Table \ref{tab:apx1}. This correspondence between linear quivers and Slodowy slices draws on the Barbasch-Vogan map and related dualities, as elaborated in \cite{Cabrera:2018ldc, Hanany:2019tji}. The pairings for low rank Classical groups are tabulated in \cite{Cabrera:2018ldc}, and can in principle be extended to higher rank.

Recall that, for any Lie group $G$, nilpotent orbits are labeled by homomorphisms $\rho : \mathrm{SU}(2) \rightarrow G$, which correspond to partitions. The Slodowy slice $\slice{\rho}^G$ transverse to the closure of the nilpotent orbit $\orbit{\rho}$ defines a decomposition $G \to F(\rho) \otimes \mathrm{SU}(2)$, where $F(\rho)$ is the centralizer of the image of $\rho$ in $G$. This partition $\rho$ thus determines a fugacity map ${{\bf x} \to ({\bf y}},t) $, where the $ {\bf x}$ and $ {\bf y}$ are Cartan subalgebra (CSA) fugacities for $G$ and the subgroup $  F(\rho) \subseteq  {{G}}$, respectively, and $t$ is a CSA fugacity for the $\mathrm{SU}(2)$. This fugacity map is unique up to Weyl group transformations. The Slodowy slice transforms under $F(\rho)$.

The Hilbert series for a Slodowy slice can be found from the  branching of the adjoint representation of $G$ under $\rho$: 
\begin{equation}
\label{eq:eq:HL1}
\begin{aligned}
\chi _{[adjoint]} ^G ({\bf x})&= \mathop  \bigoplus \limits_{[n],{\bf [m]}} {a_{n,{\bf m}}}\left( {\chi _{ [n]}^{\mathrm{SU}(2)} \otimes \chi _{\bf [m]}^F} ({\bf y}) \right),
\end{aligned}
\end{equation}
where the $ {a_{n,{\bf m}}}$ are branching coefficients. The HS is obtained from \eqref{eq:eq:HL1} by (i) replacing the $\mathrm{SU}(2)$ characters by highest weight fugacities \cite{RudolphHWG}, $\chi_{[n]}^{\mathrm{SU}(2)} \to t^n$, (ii) symmetrising the representations of $F(\rho)$ under a grading by $t^2$, and (iii) taking a quotient by the Casimirs of $G$. This leads to the refined and unrefined HS:
\begin{equation}
\label{eq:eq:HL2}
\begin{aligned}
g_{HS}^{{{\cal S}_{{\cal N},\rho} ^G}}({\bf y},t) &= PE\left[ {\mathop  \bigoplus \limits_{[n], \bf [m]}  a_{n,\bf m} {~} \chi _{\bf [m]}^F({\bf y}){t^{n + 2}} - \sum\limits_{i = 1}^r {{t^{2{d_i}}}} } \right],\\
g_{HS}^{{{{\cal S}}_{{{\cal N},\rho}} ^G}}(1,t) &= PE\left[ {\sum\limits_n {{a_n}{t^{n + 2}} - \sum\limits_{i = 1}^r {{t^{2{d_i}}}} } } \right],
\end{aligned}
\end{equation}
where the $d_i$ are the degrees of the Casimirs of $G$. Further details on the construction of slices can be found in \cite{Cabrera:2018ldc}.

Developing the methods of \cite{CremonesiHall}, we now introduce \emph{resolved} Slodowy slices, $ \slice{\rho,{\bf [n]}}^{{G}} $, which carry background charges ${\bf [n]}$ parameterised by the Dynkin labels of ${{G}}$, as described in \cite[App.\ B]{Hanany:2019tji}. These are related to uncharged slices (which effectively carry singlet charges) by a quotient of Hall Littlewood polynomials \cite{Hanany:2015hxa} under the fugacity map for $\rho$:
\begin{equation}
\ghs{ \slice{\rho,{\bf [n]}}^{{G}}}({\bf y},t)
\equiv {\ghs{ \slice{\rho}^{{G}}}}    ({\bf y},t) {~}
{\left. { \frac{ HL_{{\bf{[n]}}}^{{G}}({\bf x},t)}{ HL_{{\bf{[0]}}}^{{G}}({\bf x},t)} } \right|_{\rho:{~} {\bf x} \to ( {\bf y}, t )    }}.
\label{eq:HL3}
\end{equation}
The above considerations are quite general and apply to any Lie group $G$. 

Returning to our star shaped quiver ${\mathsf{Q}}(J)$, let us assume that we can identify its $k$ legs with a set of Slodowy slices $ \slice{\rho(i)}^{{J}^\vee} $, where $i \in \{1,\ldots, k \}$. The partitions $ \rho (i) $ determine fugacity maps ${{\bf x} \to ({\bf y}(i)},t) $. Given such a set of slices, we can apply charges $\bf [n]$ to these slices and carry out gluing \cite{CremonesiHall} by summation over the weight lattice of $J^\vee$. We obtain the following Hilbert series:
\begin{equation}
\ghs{{\mathsf{Q}}(J)}({\bf y}(1),\dots,{\bf y}(k),t)= 
\sum\limits_{\left[ \bf n \right] \in { \Gamma_{J^\vee / W^\vee}}}
\underbrace{
 {P_{\left[ \bf n \right]}^{J ^\vee}} {~}
{t^{2 \Delta [ {\bf [n]}]}}
 }_{\text{central term}}
 {~}
 \prod\limits_{i=1}^{k} T_{\rho(i),[\bf n]}(J).
\label{eq:HL4}
\end{equation}
Here, the symmetry factors ${P_{\left[ \bf n \right]}^{J ^\vee}}$ match the $P (t;m)$ terms that appear in the monopole formula \eqref{monopoleeqn}, being related by the map between the magnetic weight lattice charges $\bf m$ in the orthogonal basis of ${J ^\vee}$, and the weight lattice charges $\bf [n]$ in the Dynkin label or $\omega$-basis \cite{Feger:2012bs,Feger:2019tvk}. The term $\Delta[\bf[n]]$ equals the conformal dimension contribution $\Delta_{vec}(\bf m)$, as in Figure \ref{conformalD}.\footnote{Alternatively, $\Delta[\bf[n]]$ can be found directly, either from the weight map \cite{Hanany:2016gbz} associated to the nilcone of $J ^{\vee}$, as ${\Delta [\bf[ n]]}=- [\bf n] \cdot \omega ({\cal N}) $, or from the Cartan matrix ${\bf A}^{\vee}$ and Weyl vector ${\bf 1}$, as $\Delta[{\bf[n]}] = - 2 {\bf [n]} \cdot {{\bf A} ^{\vee} }^{-1} \cdot {\bf 1}$.} Each leg, described by the remaining terms, corresponds to the Coulomb branch of a theory of type $T(J)$ carrying external charges, expressed as a Slodowy slice:
\begin{equation}
T_{\rho(i),[\bf n]}(J) = {t^{- \Delta [ {\bf [n]}]}}{~} 
\underbrace{\ghs{ \slice{{\rho (i),[\bf n]}}^{J^\vee }}({\bf y}(i),t)} _{\text{slices}}.
\label{eq:HL4a}
\end{equation}
The Hilbert series $g_{HS}^{{\mathsf{Q}}(J)}$ transforms in the product group $\mathop  \otimes \limits_i F( \rho(i) )$ (before possible symmetry enhancement) and matches that of the Coulomb branch of the star shaped quiver $\mathsf{Q}(J)$.

\paragraph{Selection Rule.}
Not every collection of slices yields a well formed Hilbert series wherein all the fields (other than the singlet at its origin) have positive conformal dimension, $ \forall {\bf{n}} \ne \left[ {\bf{0}} \right]:\Delta \left[ {\left[ {\bf{n}} \right]} \right] > 0$. The selection rule can be formulated in terms of the weights associated with each slice by the partitions $\rho(i)$ and their contribution to the overall charges carried by $t$. Thus, each fugacity map ${{\bf x} \to ({\bf y}(i)},t) $ incorporates a weight map $\omega(i)$ that assigns R-charges to the CSA fugacities $\bf{x}$, viz $\omega(i):{~}\{x_1,\ldots,x_r\} \to \{t^{\omega_1(i)},\ldots,t^{\omega_r(i)}\}$ \cite{Cabrera:2018ldc}. Collecting terms that contribute to conformal dimension (via exponents of $t$) from \eqref{eq:HL4} and \eqref{eq:HL4a} we find a selection rule that requires:
\begin{equation}
    \omega(\mathsf{Q}) \equiv  - 2{~} \omega (\text{reg.}) + \sum\limits_{i = 1}^k\left( {\omega (\text{reg.}) - \omega (i)}\right)  \mathop  > \limits_{\text{strict}} {\bf{0}},
\label{eq:puncselection}
\end{equation}
where $\omega (\text{reg.})$ is the weight map associated with the zero dimensional regular slice, and the inequality requires that all the entries of the weight vector $ \omega(\mathsf{Q})$ should be greater than zero.

For example, consider the quiver with $E_7$ global symmetry in Table \ref{generalfamily}. We can read from \cite[Fig.\ 25]{Bourget:2020xdz} that this comprises slices to orbits with $D$-partitions $\rho=(1^6),(1^6)$ and $(3,1^3)$. Using the nilpotent orbit data in \cite[App.\ B]{Hanany:2016gbz}, we find that these correspond to weights $[0,0,0],[0,0,0] \text{ and } [2,1,1]$, and that the weight map for the regular slice is $[4,3,3]$. We obtain,
\begin{equation}
\begin{aligned}
\omega(\mathsf{Q})& =-2[4,3,3]+ 2([4,3,3]-[0,0,0])+([4,3,3]-[2,1,1])\\
& =[2,2,2].
\end{aligned}
\end{equation}
This is strictly greater than $[0,0,0]$, so $\omega(\mathsf{Q})$ is the weight vector of a good quiver.

\paragraph{Integer and Fractional Integer Lattices.}
The regular summation over resolved slices in \eqref{eq:HL4} is carried out over the entire weight or Dynkin label lattice ${\left[ \bf n \right] \in { \Gamma_{J^\vee / W^\vee}}}$ of the GNO dual group $J^\vee$ of the central quiver node. This corresponds to the combination of integer and half integer lattices as in \cite{Bourget:2020xdz}. When $J^\vee$ is special orthogonal, the summation can be restricted to correspond to the integer lattice by the simple expedient of restricting the summation over the weight lattice to exclude irreps from spinor lattices, i.e.\ to exclude those ${\left[ \bf n \right]}$ where the sum of spinor Dynkin labels is odd.

\paragraph{Non-Simply Laced Fixtures.}
The above formulae \eqref{eq:HL4} to \eqref{eq:puncselection} can be extended to calculate refined Coulomb branch Hilbert series from quivers with one or more non-simply laced legs by the expedient of applying a multiple $[ m(i) \bf n]$ of the Dynkin label charges $[\bf n]$ to each leg, according to the multiplicity $m(i)$ of the non-simply laced link $i$ in the quiver diagram \cite{Hanany:2015hxa}. Thus, \eqref{eq:HL4a} becomes:
\begin{equation}
T_{\rho(i),[ m(i) \bf n]}(J) = {t^{- \Delta [[  m(i) {\bf n}]]}}{~} 
\ghs{ \slice{{\rho (i),[ m(i) \bf n]}}^{J^\vee }}({\bf y}(i),t),
\label{eq:HL4b}
\end{equation}
and the selection rule \eqref{eq:puncselection} is modified to reflect the higher Dynkin label charges on the non-simply laced quiver legs:
\begin{equation}
    \omega(\mathsf{Q}) \equiv  - 2{~} \omega (\text{reg.}) + \sum\limits_{i = 1}^k m(i)  \left( {\omega (\text{reg.}) - \omega (i)}\right)  \mathop  > \limits_{\text{strict}} {\bf{0}}.
\label{eq:puncselectionNSL}
\end{equation}
In a sense, the folding of $m$ magnetic quiver legs $T_{\rho,[\bf n]}(J)$ corresponds to the creation of a single leg $T_{\rho,[m {\bf n}]}(J)$ carrying charges in the $m \text{-th}$ symmetrisation of the charges carried by the original quiver leg.

\paragraph{Twisted Fixtures.}
The foregoing applies to magnetic quivers for standard fixtures, where all the slices are transverse to orbits from the same symmetry group $J^\vee$. Quivers can also be constructed for ``twisted" fixtures, where some of the punctures are represented by slices from a different symmetry group ${J'}$ (being always simply laced and generally of different rank), which are ``twisted" to fit the weight lattice of the $J^\vee$ symmetry group. Compatibility requirements between the lattices of $J^\vee$ and ${J'}$ mean that only certain pairs of groups can be related by such twists.

To accommodate twisted quivers, equations \eqref{eq:HL4} and \eqref{eq:HL4a} require modification. We set the terms for a twisted puncture to contain a conformal dimension contribution and slice drawn from ${{J'}}$:
\begin{equation}
\begin{aligned}
T_{\rho(i),[\bf n']}({{J'}}) = {t^{- \Delta' [ {\bf [n']}]}}{~} \ghs{ \slice{{\rho (i),[\bf n']}}^{{{J'}}}}({\bf y}(i),t).
\end{aligned}
\end{equation}
\begin{enumerate}
  \item We identify the partition $\rho (i)$ from amongst the orbits of ${J'}$, based on the quiver diagram, so that the resolved slices $\slice{{\rho (i),[\bf n']}}^{{{J'}}}$ belong to the twisted group.
  \item We also need to identify the ``twisted" map $[\bf n] \to [\bf n']$ between the Dynkin labels of $J^\vee$ and those of ${J'}$, and use this to express the $\Delta' [ {\bf [n']}]$ contribution to conformal dimension in terms of $[\bf n]$.
\end{enumerate}
Various types of twist are encountered when dealing with mixed unitary and orthosymplectic quivers. These are tabulated in Table \ref{tab:apx1}, along with the maps between the Dynkin labels of $J'$ and ${J}^\vee$ that are necessary in order to carry out the summation over the ${J}^\vee$ lattice.

\begin{table}[t]
\centering
\begin{tabular}{c|c|c|c|c}
\toprule
$\text{Type}$ & $ J'$ & $ J$ & ${J}^\vee$ & Dynkin Labels of $ J'$ \\
\midrule
Standard & ${G_r}^\vee$ &${G_r}$ &${G_r}^\vee$ &$[n_1, \ldots, n_r]$ \\
Twisted $A^{I}_{\mathrm{odd}}$ & ${A_{2r-1}}$ &${D_r}$ &${D_r}$ & $[n_1, \ldots,M, A,M, \ldots, n_1]$\\
Twisted $A^{II}_{\mathrm{odd}}$ & ${A_{2r-1}}$ &${C_r}$ &${B_r}$ & $[n_1, \ldots,n_{r-1}, n_r,n_{r-1}, \ldots, n_1]$\\
Twisted $A_{\mathrm{even}}$ & ${A_{2r}}$ &${B_r}$ &${C_r}$ & $[n_1, \ldots,n_{r}, n_{r}, \ldots, n_1]$\\
Twisted $D$ & ${D_{r+1}}$ &${B_r}$ &${C_r}$ & $[n_1, \ldots,n_{r}, n_r]$\\
\bottomrule
\end{tabular}
\caption{Types of Twisted Puncture. The maps express the Dynkin labels $[n']$ of $ J'$ in terms of the Dynkin labels $[n]$ of ${J}^\vee$. Here $M \equiv \text{Min}[n_{r-1},n_r]$ and $A \equiv \text{Abs}[n_{r-1} - n_r]$. The notation for twisted $A_{\mathrm{odd}}$ punctures is adapted from \cite{onishchik2004lectures,bump2004lie}. }
\label{tab:apx1}
\end{table}

The twisted $A_{2r-1}$ fixtures analysed in Sections \ref{Example} and \ref{unframed} contain a $J'=A_{2r-1}$ linear quiver affixed to a star shaped quiver with $ J = C_r$ and $ J^\vee = B_r$. For example, consider the quiver in Table \ref{Entable} whose Coulomb branch is the minimal orbit of $E_6$. We can read from \cite[Tab.\ 20]{Bourget:2020xdz} that this comprises two slices to orbits with the $B$-partitions $\rho=(1^5)$ and a slice to the orbit with $A_3$ partition $(3,1)$. Using the nilpotent orbit data in \cite[App.\ B]{Hanany:2016gbz}, we find these correspond to $B_2$ weights $[0,0]$ and $A_3$ weights $[2,2,2]$, respectively. The weight map for the regular slice of $B_2$ is $[4,3]$ and that for $A_3$ is $[3,4,3]$. So,
\begin{equation}
\begin{aligned}
\omega(\mathsf{Q})& =\underbrace{-2[4,3]+ 2([4,3]-[0,0])}_{\text{Dynkin Labels } [n_1,n_2]}+\underbrace{([3,4,3]-[2,2,2])}_{\text{Dynkin Labels } [n_1,n_2,n_1]}\\
& =\underbrace{[1,2,1]}_{\text{Dynkin Labels } [n_1,n_2,n_1]}\\
& =\underbrace{[2,2]}_{\text{Dynkin Labels } [n_1,n_2]}.
\end{aligned}
\end{equation}
Application of the selection rule shows $\omega(\mathsf{Q})$ is strictly positive and therefore the weight vector is that of a good star shaped quiver. The summation is carried out over the $B_2$ Dynkin label lattice $[n_1,n_2]$. Only $A_3$ slices with Dynkin labels of the form $[n_1,n_2,n_1]$ contribute to the summation. 

Alternatively, consider the quiver in Table \ref{Entable} whose Coulomb branch is the minimal orbit of $D_5$. This comprises two slices to orbits with the $D$-partitions $\rho=(1^4)$ and a slice to the orbit with $A_3$ partition $(3,1)$. Using the nilpotent orbit data in Appendix B of \cite{Hanany:2016gbz}, we find these correspond to $D_2$ weights $[0,0]$ and $A_3$ weights $[2,2,2]$, respectively. The weight map for the regular slice of $D_2$ is $[1,1]$ and that for $A_3$ is $[3,4,3]$. So,
\begin{equation}
\begin{aligned}
\omega(\mathsf{Q})& =
\underbrace{-2[1,1]+ 2([1,1]-[0,0])}_{\text{Dynkin Labels } [n_1,n_2]}
+\underbrace{([3,4,3]-[2,2,2])}_{\text{Dynkin Labels } [M,A,M]}\\
& =\underbrace{[1,2,1]}_{\text{Dynkin Labels }[M,A,M]}.
\end{aligned}
\end{equation}
The contribution to conformal dimension from $\omega(\mathsf{Q})$ is positive for $[n_1,n_2] \neq [0,0]$, and so application of the selection rule shows that the weight vector is that of a good star shaped quiver. The summation is carried out over the $D_2$ weight lattice $[n_1,n_2]$. Only $A_3$ slices with Dynkin labels of the form $[M,A,M]$ are involved, where $M=\text{Min}[n_1,n_2]$ and $A=\text{Abs}[n_1-n_2]$.

It is clearly possible to construct fixtures that combine features of twists, non-simply laced legs and/or sub-lattices. Further discussion of such fixtures is left for future work.

\providecommand{\href}[2]{#2}\begingroup\raggedright\endgroup

\end{document}